\documentclass[11pt]{article}
\usepackage{jheppub} 
 \usepackage[T1]{fontenc}
 \usepackage{booktabs}
 \usepackage{xcolor} 
 \usepackage{xspace}
 \usepackage{dcolumn}
 \usepackage{hyperref}
 \usepackage{caption}
 \usepackage[subrefformat=parens, position=top, skip=-15pt, margin=15pt, justification=justified, singlelinecheck=false] {subcaption}
 \usepackage{placeins}   
 \usepackage[group-digits=false]{siunitx}
 \usepackage{microtype}
 \usepackage{lmodern}
 \usepackage{hyperref}
\usepackage{multirow}

\pdfoutput=1

\allowdisplaybreaks[1]

\newcommand{\Pl}{\ell}

\newcommand{\fb}{{\ensuremath\unskip\,\text{fb}}\xspace}

\def\refeq#1{\mbox{(\ref{#1})}}
\def\reffi#1{\mbox{Fig.~\ref{#1}}}
\def\reffis#1{\mbox{Figs.~\ref{#1}}}
\def\refta#1{\mbox{Table~\ref{#1}}}
\def\reftas#1{\mbox{Tables~\ref{#1}}}
\def\refse#1{\mbox{Section~\ref{#1}}}

\def\citere#1{\mbox{Ref.~\cite{#1}}}
\def\citeres#1{\mbox{Refs.~\cite{#1}}}

\newcommand{\ie}{\emph{i.e.}\ }
\newcommand{\eg}{\emph{e.g.}\ }

\def\be{\begin{equation}}
\def\ee{\end{equation}}

\newcommand{\PH}{\ensuremath{\text{H}}\xspace}
\newcommand{\Pj}{\ensuremath{\text{j}}\xspace}
\newcommand{\Pp}{\ensuremath{\text{p}}\xspace}
\newcommand{\Pe}{\ensuremath{\text{e}}\xspace}

\newcommand{\Pb}{\ensuremath{\text{b}}\xspace}
\newcommand{\Pq}{\ensuremath{q}\xspace}
\newcommand{\Pt}{\ensuremath{\text{t}}\xspace}

\newcommand{\Pg}{\ensuremath{\text{g}}\xspace}

\newcommand{\PW}{\ensuremath{\text{W}}\xspace}
\newcommand{\PZ}{\ensuremath{\text{Z}}}

\newcommand{\Mt}{\ensuremath{m_\Pt}\xspace}
\newcommand{\MH}{\ensuremath{M_\PH}\xspace}
\newcommand{\MWOS}{\ensuremath{M_\PW^\text{OS}}\xspace}
\newcommand{\MW}{\ensuremath{M_\PW}\xspace}
\newcommand{\MZOS}{\ensuremath{M_\PZ^\text{OS}}\xspace}
\newcommand{\MZ}{\ensuremath{M_\PZ}\xspace}

\newcommand{\Gt}{\ensuremath{\Gamma_\Pt}\xspace}
\newcommand{\GH}{\ensuremath{\Gamma_\PH}\xspace}

\newcommand{\GZOS}{\ensuremath{\Gamma_\PZ^\text{OS}}\xspace}

\newcommand{\GWOS}{\ensuremath{\Gamma_\PW^\text{OS}}\xspace}

\newcommand{\GeV}{\ensuremath{\,\text{GeV}}\xspace}
\newcommand{\TeV}{\ensuremath{\,\text{TeV}}\xspace}

\newcommand{\sw}{s_{\mathrm{w}}}
\newcommand{\alphas}{\ensuremath{\alpha_\text{s}}\xspace}
\newcommand{\order}[1]{\ensuremath{\mathcal{O}{\left(#1\right)}}\xspace}

\newcommand{\GF}{\ensuremath{G_\mu}}

\newcommand{\ptsub}[1]{\ensuremath{p_{\text{T},#1}}\xspace}

\newcommand{\MVOS}{\ensuremath{M_{V}^\text{OS}}\xspace}%
\newcommand{\GVOS}{\ensuremath{\Gamma_{V}^\text{OS}}\xspace}%

\newcommand{\newc}{\newcommand}
\newc{\bi}{\begin{itemize}}
\newc{\ei}{\end{itemize}}
\newc{\benu}{\begin{enumerate}}
\newc{\eenu}{\end{enumerate}}
\newc{\bc}{\begin{center}}
\newc{\ec}{\end{center}}
\newc{\bfig}{\begin{figure}}
\newc{\efig}{\end{figure}}
\newc{\qbar}{\bar{q}}
\newc{\go}{\tilde{g}}
\newc{\PB}{\textsc{Powheg-Box}}

\newcommand{\recola}{{\sc Recola}\xspace}

\newcommand{\mocanlo}{{\sc MoCaNLO}\xspace}

\newcommand{\collier}{{\sc Collier}\xspace}

\newcommand{\rT}{{\mathrm{T}}}
\newcolumntype{.}{D{.}{.}{-1}}
\newcolumntype{d}[1]{D{.}{.}{#1}}

\colorlet{tableoverheadcolor}{gray!37.5}
\colorlet{tableheadcolor}{gray!25}
\colorlet{tablerowcolor}{gray!12.5}

\newlength{\width}
\newlength{\height}

\def\draftdate{\relax}
\def\mda{\relax}
\def\mua{\relax}
\def\mla{\relax}
\def\draft{
\def\thtystars{******************************}
\def\sixtystars{\thtystars\thtystars}
\typeout{}
\typeout{\sixtystars**}
\typeout{* Draft mode!
         For final version remove \protect\draft\space in source file *}
\typeout{\sixtystars**}
\typeout{}
\def\draftdate{\today}
\def\mua{\marginpar[\boldmath\hfil$\uparrow$]%
                   {\boldmath$\uparrow$\hfil}\color{black}%
                    \typeout{marginpar: $\uparrow$}\ignorespaces}
\def\mda{\color{red}\marginpar[\boldmath\hfil$\downarrow$]%
                   {\boldmath$\downarrow$\hfil}%
                    \typeout{marginpar: $\downarrow$}\ignorespaces}
\def\mla{\marginpar[\boldmath\hfil$\rightarrow$]%
                   {\boldmath$\leftarrow $\hfil}%
                    \typeout{marginpar: $\leftrightarrow$}\ignorespaces}
\def\Mua{\marginpar[\boldmath\hfil$\Uparrow$]%
                   {\boldmath$\Uparrow$\hfil}\color{black}%
                    \typeout{marginpar: $\uparrow$}\ignorespaces}
\def\Mda{\color{red}\marginpar[\boldmath\hfil$\Downarrow$]%
                   {\boldmath$\Downarrow$\hfil}%
                    \typeout{marginpar: $\downarrow$}\ignorespaces}
\def\Mla{\marginpar[\boldmath\hfil\textcolor{red}{$\Rightarrow$}]%
                   {\boldmath\textcolor{red}{$\Leftarrow $}\hfil}%
                    \typeout{marginpar: $\leftrightarrow$}\ignorespaces}
\overfullrule 5pt
\oddsidemargin 15mm
\marginparwidth 29mm
}

\title{\hfill ~\\[-58mm]
\phantom{h}\hfill\mbox{\small {}}
\\[1cm]
\vspace{13mm}   
NLO QCD and EW corrections to vector-boson scattering into
$\PW^+\PW^-$ at the LHC} 

\subheader{\today}

\author{Ansgar Denner,}
\author{Robert Franken,}
\author{Timo Schmidt,}
\author{Christopher Schwan}

\affiliation{Institut f\"ur Theoretische Physik und Astrophysik,  %
        Universit\"at W\"urzburg, \\%
        Emil-Hilb-Weg 22,  %
        97074 W\"urzburg, %
        Germany%
}
\emailAdd{denner@physik.uni-wuerzburg.de}
\emailAdd{robert.franken@physik.uni-wuerzburg.de}
\emailAdd{timo.schmidt@physik.uni-wuerzburg.de}
\emailAdd{christopher.schwan@physik.uni-wuerzburg.de}

\abstract{We present the full next-to-leading-order electroweak and
  QCD corrections to vector-boson scattering into a pair of off-shell
  opposite-sign W bosons decaying into leptons of different flavour at
  the LHC. We include full leading-order predictions for the
  irreducible background.  Explicitly, we investigate the process
  $\Pp\Pp \to \Pe^+ \nu_e\mu^-\bar\nu_\mu\Pj\Pj + X$ at leading orders
  $\order{\alpha^6}$, $\order{\alphas\alpha^5}$,
  $\order{\alphas^2\alpha^4}$, supplemented by the loop-induced
  $\order{\alphas^4\alpha^4}$ contribution, and at
  next-to-leading orders $\order{\alpha^7}$ and
  $\order{\alphas\alpha^6}$ in two setups providing fiducial cross
  sections as well as differential distributions.  We take full
  account of photon-induced next-to-leading-order contributions, which
  prove to be non negligible. With $-11.4\%$ and $-6.7\%$ in the two
  setups, the electroweak corrections are smaller than for other
  vector-boson-scattering processes. This can be traced back to the
  presence of the Higgs-boson resonance in the fiducial phase space,
  whose effects we analyse within an additional unphysical, but
  manifestly gauge-invariant setup. The QCD corrections amount to
  $-5.1\%$ and $-21.6\%$ in the two setups. The large size of the
  latter correction, compared to other vector-boson scattering
  processes, is explained by a very restrictive definition of its
  fiducial phase space.}

\begin{document} 

\maketitle

\newpage

\section{Introduction}
Vector-boson scattering (VBS) processes allow for important tests of
the electroweak (EW) sector and the Standard Model (SM) as a whole.
Especially the scattering of vector bosons into a pair of $\PW$
bosons provides the possibility to investigate triple and quartic
gauge couplings and the couplings of EW gauge bosons to the scalar
Higgs sector. Owing to strong gauge cancellations in the SM, VBS is
very sensitive to possible deviations from the SM in the EW sector and an
experimentally interesting testing ground.

In recent years, VBS processes have been observed by both ATLAS and
CMS in leptonically decaying same-sign $\PW$
\cite{Aad:2014zda,Khachatryan:2014sta,Aaboud:2016ffv,Sirunyan:2017ret,Aaboud:2019nmv},
$\PW\PZ$ \cite{Aaboud:2018ddq,Sirunyan:2019ksz,CMS:2020gfh} and
$\PZ\PZ$ \cite{Sirunyan:2017fvv, Aad:2020zbq,Sirunyan:2020alo} final states
(denoted for short as same-sign W, WZ and ZZ scattering). Very
recently, also opposite-sign $\PW$ scattering could be measured by the
CMS experiment \cite{CMS:2022woe}. Besides, opposite-sign $\PW$
scattering is an important part of the EW Higgs-boson production
and decay, which has also been studied at the LHC
\cite{CMS:2018zzl,Atlas:2018xbv}.

On the theoretical side, next-to-leading-order (NLO) QCD corrections
to all VBS processes and its irreducible background and
especially to opposite-sign W scattering have been calculated  \cite{Jager:2006zc,
  Melia:2011dw,Greiner:2012im} and matched to parton showers
\cite{Jager:2013mu,Rauch:2016upa} over the last two decades, whereas
NLO EW corrections to massive VBS processes have only become available
recently for same-sign $\PW$
\cite{Biedermann:2016yds,Biedermann:2017bss}, $\PW\PZ$
\cite{Denner:2019tmn} and $\PZ\PZ$ scattering
\cite{Denner:2020zit,Denner:2021hsa}. In this article, we continue
this series of studies and provide EW and QCD corrections to the scattering
of opposite-sign $\PW$ bosons decaying into a pair of oppositely
charged leptons of different generations and two corresponding
neutrinos, leading to the final state $\Pe^+\nu_\Pe\mu^- \bar\nu_\mu
\Pj\Pj$.  We exclude jets with bottom quarks in the final state, since a major
experimental problem in opposite-sign W scattering is the background
from $\Pt\bar\Pt$ production, whose cross section is typically 
orders of magnitude larger than VBS cross sections
\cite{ATLAS:2019hau} and needs to be suppressed by a $\Pb$-jet veto.

The final states of VBS scattering processes receive contributions 
from triple vector-boson production where one of the vector bosons
decays hadronically and two decay leptonically. While we are not aware
of existing NLO QCD and EW calculations for these processes, there are
calculations for off-shell triple W-boson production in the fully leptonic
channel \cite{Campanario:2008yg,Schonherr:2018jva,Dittmaier:2019twg,Amoroso:2020lgh}.

At leading order (LO), the process $\Pp\Pp\to\Pe^+\nu_\Pe\mu^-
\bar\nu_\mu \Pj\Pj+X$ receives tree-level contributions of the orders
$\order{\alpha^6}$, $\order{\alphas\alpha^5}$, and
$\order{\alphas^2\alpha^6}$. We also include the loop-induced process
of $\order{\alphas^4\alpha^4}$ with two gluons in the initial and
final states in our calculation.
VBS scattering is part of the EW contribution of $\order{\alpha^6}$,
the order $\order{\alphas^2\alpha^6}$ is the QCD-induced background,
and the $\order{\alphas\alpha^5}$ consists of interferences of EW and
QCD contributions and gluon--photon-induced contributions. The
QCD-induced background constitutes the largest part of the cross
section, followed by the EW contribution, the loop-induced process and
the interference.

In this paper, we focus on the NLO contributions of orders
$\order{\alpha^7}$ and $\order{\alphas\alpha^6}$. These result from EW
and QCD corrections to the LO EW contribution and from non-separable
EW corrections to the LO interference. For simplicity, we call the 
$\order{\alphas\alpha^6}$ contributions QCD corrections for short. 
In our previous investigations, we found the EW corrections to VBS 
processes to be typically
$-15\%$, resulting from EW logarithms enhanced by high intrinsic
scales of VBS. For opposite-sign W scattering a sizeable fraction of the
cross section results from the intermediate Higgs-boson resonance at
the scale of the Higgs-boson mass, leading to smaller corrections.  We
fully include photon-induced corrections, which involve new partonic
channels with VBS signatures at NLO.

This paper is structured as follows: In \refse{sec:process}, the
process and details of the validation of our calculation are
described. Section \ref{sec:results} summarises the numerical input
parameters, the event-selection criteria, and the integrated cross
sections in two physical setups. Special attention is devoted to the
Higgs resonance, which we investigate in an additional unphysical
setup. We conclude with a presentation of differential distributions
for the physical setups. In \refse{sec:conclusion}, we give a
brief summary.

\section{Description of the calculation}
\label{sec:process}

In this article we investigate the process
\begin{equation}\label{eq:LOprocess}
\Pp\Pp \to \Pe^+ \nu_\Pe \mu^- \bar \nu_\mu \Pj\Pj+X 
\end{equation}
at the LHC.  We only consider the final state with
  $\Pe^+\mu^-$ and do not take into account the charge-conjugate final state. Since we
  treat the leptons as massless and use the same cuts for electrons
  and muons, the $\Pe^-\mu^+$ final state yields identical results. The two jets
  $\Pj$ in the final state result from 
  clustering of quarks, antiquarks, gluons and photons. However,
  unclustered photons are not considered as jets. As a consequence, we
  only have to take into account partonic processes with at least two
  strongly-interacting partons in the final state.

\subsection{Leading order}
\label{sec:LO_process}
Just like all other processes with $2\Pj + 4\ell$ final states, the
cross section of \refeq{eq:LOprocess} consists of three components at
LO: a purely EW one of $\order{\alpha^6}$ which includes VBS, one of
$\order{\alphas \alpha^5}$ from interference,
(anti)quark--photon-induced and gluon--photon-induced
contributions, and a QCD-induced one of $\order{\alpha^2_{\rm s}
  \alpha^4}$.

\begin{figure}
\centering
\begin{subfigure}{0.3\textwidth}
\captionsetup{skip=0pt}
\caption{}
\centering
\includegraphics[page=1,width=1.\linewidth]{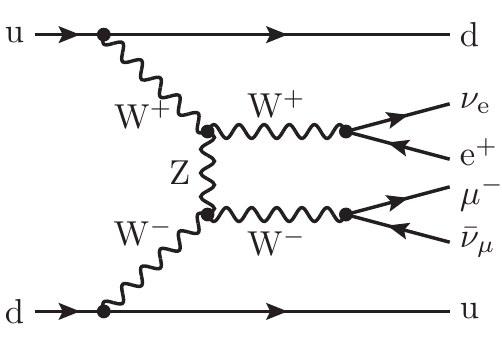}
\end{subfigure}
\begin{subfigure}{0.3\textwidth}
\captionsetup{skip=0pt}
\caption{}
\centering
 \includegraphics[page=2,width=1.\linewidth]{axodraw-diagrams/diagrams.pdf}
\end{subfigure}
\begin{subfigure}{0.3\textwidth}
\captionsetup{skip=0pt}
\caption{}
\centering
 \includegraphics[page=4,width=1.\linewidth]{axodraw-diagrams/diagrams.pdf}
\end{subfigure}
\par
\begin{subfigure}{0.3\textwidth}
\captionsetup{skip=0pt}
\caption{}
\centering
\includegraphics[page=5,width=1.\linewidth]{axodraw-diagrams/diagrams.pdf}
\end{subfigure}
\begin{subfigure}{0.3\textwidth}
\captionsetup{skip=0pt}
\caption{}
\centering
 \includegraphics[page=6,width=1.\linewidth]{axodraw-diagrams/diagrams.pdf}
\end{subfigure}
\begin{subfigure}{0.3\textwidth}
\captionsetup{skip=0pt}
\caption{}
\centering
 \includegraphics[page=7,width=1.\linewidth]{axodraw-diagrams/diagrams.pdf}
\end{subfigure}
\par
\begin{subfigure}{0.3\textwidth}
\captionsetup{skip=0pt}
\caption{}
\centering
\includegraphics[page=9,width=1.\linewidth]{axodraw-diagrams/diagrams.pdf}
\end{subfigure}
\begin{subfigure}{0.3\textwidth}
\captionsetup{skip=0pt}
\caption{}
\centering
 \includegraphics[page=8,width=1.\linewidth]{axodraw-diagrams/diagrams.pdf}
\end{subfigure}
\begin{subfigure}{0.3\textwidth}
\captionsetup{skip=0pt}
\caption{}
\centering
 \includegraphics[page=12,width=1.\linewidth]{axodraw-diagrams/diagrams.pdf}
\end{subfigure}
\caption{Examples of LO diagrams of $\order{g^6}$. The first row (a)
  -- (c) shows signal diagrams containing VBS subprocesses, the second
  row
  diagrams with two (d), one (e), or no (f) resonant vector bosons, and
  the third row triple-vector-boson production (g), an $s$-channel
  diagram (h), and a photon-induced (i) diagram.}
\label{fig:Born_a6}
\end{figure}
A sample of Feynman diagrams at order%
\footnote{We denote the weak coupling constant by $g=e/\sw$ and the strong
  coupling constant by $g_{\rm s} $.}
 $\order{g^6}$ is shown in
\reffi{fig:Born_a6}.  
VBS appears at LO as a subprocess in quark-induced partonic processes
as scattering of a pair of electroweak gauge bosons emitted from
different quark lines into a pair of opposite charged $\PW$ bosons
that decay leptonically into two opposite charged leptons and their
corresponding neutrinos (in the following simply referred to as $\PW^+
\PW^-$ final state) [\reffi{fig:Born_a6} (a) -- (c)].
Besides diagrams with triple and quartic gauge couplings, VBS
includes diagrams with an $s$-channel Higgs
exchange [\reffi{fig:Born_a6} (c)]. In each partonic channel, all of
these sub-diagrams contribute.

In contrast to the similar process $\Pp\Pp \to \Pe^+ \Pe^- \mu^+ \mu^-
\Pj\Pj + X$ ($\PZ \PZ$ scattering), which we investigated in
\citeres{Denner:2020zit,Denner:2021hsa}, the two neutrinos in the
final state are only visible as missing transverse momentum in the
experiment. It is hence impossible to construct a physical cut to suppress
contributions of resonant Higgs production and decay, such as an
invariant-mass cut on the four-lepton system. As we will discuss in
the course of this paper, the presence of the Higgs resonance changes
the characteristics of $\PW^+ \PW^-$ scattering compared to other VBS
processes.

Besides diagrams containing the VBS subprocess, the order $\order{g^6}$ receives contributions
from ($t$-channel) diagrams with two, one or no resonant vector bosons
[\reffi{fig:Born_a6} (d), (e), and (f)], $s$-channel diagrams
[\reffi{fig:Born_a6} (g) and (h)], and photon-induced contributions
[\reffi{fig:Born_a6} (i)]. In partonic processes that involve
$s$-channel sub-diagrams, in particular, triple-vector-boson production
appears [\reffi{fig:Born_a6} (g)]. 
\begin{figure}
\centering
\begin{subfigure}{0.3\textwidth}
\captionsetup{skip=0pt}
\caption{}
\centering
\includegraphics[page=13,width=1.\linewidth]{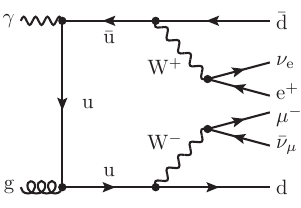}
\end{subfigure}
\begin{subfigure}{0.3\textwidth}
\captionsetup{skip=0pt}
\caption{}
\centering
 \includegraphics[page=11,width=1.\linewidth]{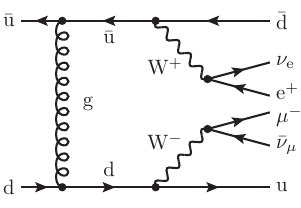}
\end{subfigure}
\begin{subfigure}{0.3\textwidth}
\captionsetup{skip=0pt}
\caption{}
\centering
 \includegraphics[page=28,width=1.\linewidth]{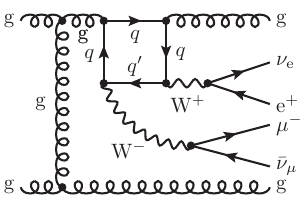}
\end{subfigure}
\caption{Examples of LO diagrams of $\order{g_\mathrm s g^5}$, a gluon--photon-induced contribution (a), $\order{g_\mathrm s^2 g^4}$, a $t$-channel gluon exchange (b), and  $\order{g_\mathrm s^4 g^4}$, a loop-induced contribution~(c).}
\label{fig:Born_rest}
\end{figure}
Contributions of the irreducible background emerge via
gluon--photon-induced diagrams at $\order{g_\mathrm{s}g^5}$
[\reffi{fig:Born_rest} (a)] and diagrams of $\order{g^2_{\mathrm s}
  g^4}$ [\reffi{fig:Born_rest} (b)], which are characterised by a $t$-
or $s$-channel gluon exchange. An example diagram for the loop-induced channel
is shown in \reffi{fig:Born_rest} (c).

At $\order{\alpha^6}$, there are 60 quark-induced partonic channels
(not counting $\Pq\Pq'$ and $\Pq'\Pq$ initial states separately and
omitting bottom quarks), which we divide in 5 subclasses. Specifically, 36 partonic
channels include Feynman diagrams with VBS topology, but no
triple-vector-boson production, 4 include both VBS and triple-W
production, 8 VBS and WWZ production, 4 triple-W production but no
VBS, and 8 WWZ production but no VBS. For further details concerning
the structure of the process, we refer to our paper on $\PZ \PZ$
scattering \cite{Denner:2020zit}. The partonic processes for
opposite-sign W scattering are obtained from those for $\PZ\PZ$
scattering upon replacing two oppositely-charged leptons by two
neutrinos, without touching the external quarks, gluons, and photons.
Therefore, in particular, the counting of partonic processes
is the same in both cases.

In our calculation we neglect quark mixing and use a unit CKM matrix.
Similarly to $\PZ \PZ$ scattering, this affects the suppressed quark-induced
$s$-channel contributions by about 5\% \cite{Denner:2021hsa}, hence
its effects are negligible. Furthermore, we exclude the contributions
of initial- or final-state bottom quarks. Those of initial-state
bottom quarks are suppressed owing to their parton-distribution
functions (PDFs). Thus, for  $\PZ \PZ$
scattering, the contributions of bottom quarks have been found to be
below $3\%$ \cite{Denner:2021hsa}. In contrast to $\PZ \PZ$
scattering, the contribution of final-state bottom quarks would be
overwhelming for  $\PW^+\PW^-$ scattering owing to contributions of 
$\Pt \bar\Pt$ production to the identical final state.
In this respect, we stress the importance of
experimental bottom-jet vetoes when measuring VBS cross sections to
avoid contamination from $\Pt \bar\Pt$ production.
Our calculation is based on the assumption of a perfect bottom-jet veto.

Further contributions to the hadronic process \refeq{eq:LOprocess} originate
from photon-induced partonic processes with 
$\gamma\gamma$, $\Pq\gamma$, or $\Pg\gamma$ as initial states which are
included in this computation. The process $\gamma\gamma \to
\Pq\Pq+4\Pl$ appears at $\order{\alpha^6}$, the processes $\Pg\gamma
\to \Pq\Pq+4\Pl$ and $\Pq\gamma \to \Pq\Pg+4\Pl$ at
$\order{\alphas\alpha^5}$ as non-interference processes. Finally,
at $\order{\alphas^4 \alpha^4}$ a loop-induced
channel $\Pg\Pg \to \Pg\Pg+4\Pl$ with four external gluons opens up,
which has been shown to be non-negligible for $\PZ \PZ$ scattering,
despite of its higher order, owing to the enhanced gluon PDFs.

\subsection{Real emission}

\begin{figure}
\centering
\includegraphics{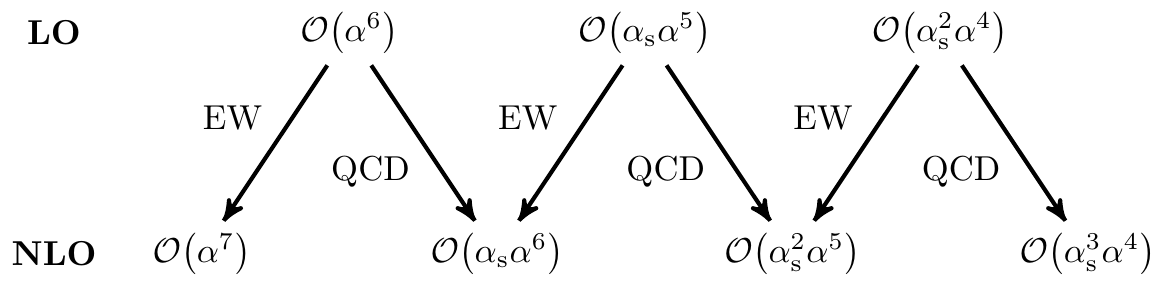}
\caption{LO and NLO contributions to $\PW^+ \PW^-$ scattering and its
  irreducible background. In this paper, we discuss both the EW and
  QCD NLO corrections to $\order{\alpha^6}$ and the non-separable EW
  corrections to the interference process for several experimental
  setups.} 
\label{fig:All_Orders}
\end{figure}

Real EW and QCD corrections to all LO processes lead
to four contributions of orders $\order{\alpha^7}$,
$\order{\alphas\alpha^6}$, $\order{\alphas^2\alpha^5}$, and
$\order{\alphas^3 \alpha^4}$ at NLO (see \reffi{fig:All_Orders}). 
In this paper, we discuss only
the first two orders, which are the NLO corrections to the VBS
signal process. The order $\order{\alpha^7}$ are pure EW
corrections to the EW LO contribution, whereas the
$\order{\alphas\alpha^6}$ contains both QCD corrections to the EW LO
process as well as EW corrections to the LO interference.

\begin{figure}
\centering
\begin{subfigure}{0.3\textwidth}
\captionsetup{skip=0pt}
\caption{}
\centering
\includegraphics[page=20,width=1.\linewidth]{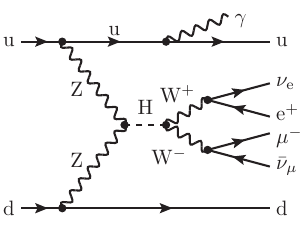}
\end{subfigure}
\begin{subfigure}{0.3\textwidth}
\captionsetup{skip=0pt}
\caption{}
\centering
 \includegraphics[page=22,width=1.\linewidth]{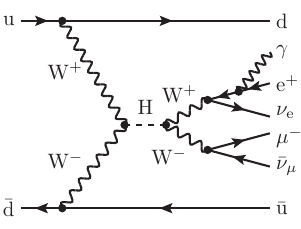}
\end{subfigure}
\begin{subfigure}{0.3\textwidth}
\captionsetup{skip=0pt}
\caption{\label{fig:ggamma-induced}}
\centering
\includegraphics[page=21,width=1.\linewidth]{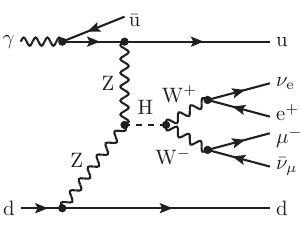}
\end{subfigure}
\par
\begin{subfigure}{0.3\textwidth}
\captionsetup{skip=0pt}
\caption{}
\centering
\includegraphics[page=23,width=1.\linewidth]{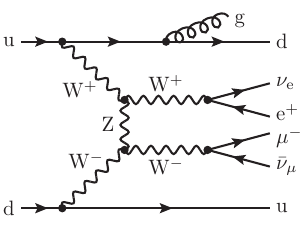}
\end{subfigure}
\begin{subfigure}{0.3\textwidth}
\captionsetup{skip=0pt}
\caption{ \label{fig:gluon_emission}}
\centering
 \includegraphics[page=27,width=1.\linewidth]{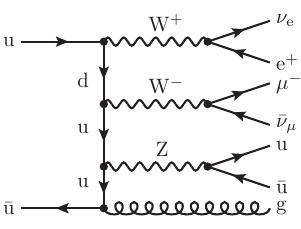}
\end{subfigure}
\begin{subfigure}{0.3\textwidth}
\captionsetup{skip=0pt}
\caption{\label{fig:non-VBS_Higgs}}
\centering
 \includegraphics[page=31,width=1.\linewidth]{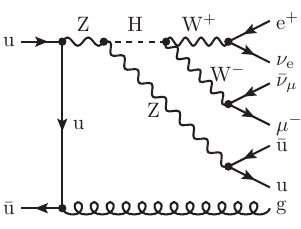}
\end{subfigure}
\par
\begin{subfigure}{0.3\textwidth}
\captionsetup{skip=0pt}
\caption{}
\centering
\includegraphics[page=30,width=1.\linewidth]{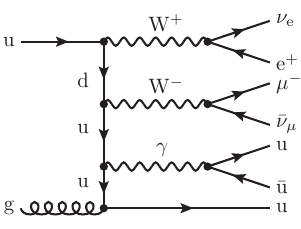}
\end{subfigure}
\begin{subfigure}{0.3\textwidth}
\captionsetup{skip=0pt}
\caption{}
\centering
\includegraphics[page=32,width=1.\linewidth]{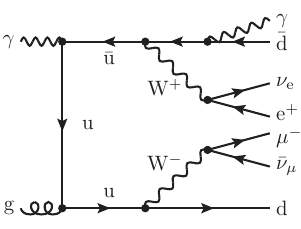}
\end{subfigure}
\begin{subfigure}{0.3\textwidth}
\captionsetup{skip=0pt}
\caption{}
\centering
 \includegraphics[page=24,width=1.\linewidth]{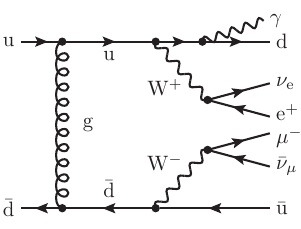}
\end{subfigure}
\caption{Examples of real-emission diagrams of $\order{g^7}$,
  $\order{g_\mathrm s g^6}$, and $\order{g_\mathrm s^2 g^5}$: In the
  first row emission of a real photon from a quark (a) or a lepton
  line (b) and initial-state photon splitting (c).  In the second row
  gluon emission from signal (d) and background (e,f) diagrams.  In
  the third row a gluon-induced process with final-state photon
  splitting (g), photon emission from a photon--gluon-induced diagram
  (h) and from a QCD-induced diagram (i). Note that diagrams of
  $\order{g_\mathrm s^2 g^5}$ only contribute via interferences within
  of the calculations of this article.}
\label{fig:real}
\end{figure}

Apart from the emission of a real photon from any charged particle, which does not
change the classification of the process [\reffi{fig:real} (a) and
(b)], at $\order{\alpha^7}$ additional photon-induced processes appear
in which the initial-state photon splits into a quark--antiquark pair
[\reffi{fig:real}~(c)], leading to an additional jet in the final
state. As a consequence, VBS topologies appear in photon-induced processes.

Real QCD corrections, on the other hand, always produce an additional jet,
either from emission of a real gluon from a quark line [\reffi{fig:real}
(d), (e), and (f)], or from a gluon splitting [\reffi{fig:real} (g)]
giving rise to an additional quark or antiquark in the final state.
Gluon splitting also leads to diagrams of $\order{g_{\mathrm s} g^6}$
with VBS topologies, \eg diagram \reffi{fig:real}~(c) 
with the initial-state photon replaced by a gluon, which contribute at
$\order{\alphas\alpha^6}$. Figure
\ref{fig:real} (h) illustrates a contribution to the
photon--gluon-induced process, and \reffi{fig:real}~(i) a
QCD-induced real-correction diagram that contributes at
$\order{\alphas\alpha^6}$ via the interference with diagram \reffi{fig:real}~(b).%

At the orders
$\order{\alphas^2\alpha^5}$ and $\order{\alphas^3\alpha^4}$, which we
do not discuss in this article, VBS never appears as a subprocess.

To summarise, three types of partonic processes constitute the real corrections at
$\order{\alpha^7}$: $\Pq\Pq \to \Pq\Pq\gamma+4\Pl$, $\gamma\gamma \to
\Pq\Pq\gamma+4\Pl$ and $\Pq\gamma \to \Pq\Pq\Pq+4\Pl$. At
$\order{\alphas\alpha^6}$, the partonic channels $\Pq\Pq \to
\Pq\Pq\Pg+4\Pl$, $\gamma\gamma \to \Pq\Pq\Pg+4\Pl$ and $\Pg\Pq \to
\Pq\Pq\Pq+4\Pl$ contribute as QCD corrections to the EW process,
$\Pq\Pq\to \Pq\Pq\gamma+4\Pl$ and $\Pq\gamma \to\Pq\Pq\Pq+4\Pl$ as EW
corrections to the interference, and $\Pq\gamma \to \Pq\Pg\gamma+4\Pl$
and $\Pg \gamma \to \Pq \Pq \gamma+4\Pl$ as EW corrections to the
non-interference $\order{\alphas\alpha^5}$ processes $\Pq\gamma \to \Pq\Pg+4\Pl$
and $\Pg \gamma \to \Pq \Pq+4\Pl$.

At $\order{\alphas\alpha^6}$, partonic channels of the type shown in
\reffi{fig:real} (g) contain both QCD and QED singularities.
Therefore,  strictly speaking, QCD and EW corrections cannot be
separated in a physically meaningful way, and counting these
contributions as QCD corrections is merely a convention.   

The real emission of massless particles leads to IR singularities in
the fiducial phase space. We use the Catani--Seymour dipole
subtraction method \cite{Catani:1996vz} and its generalisation to QED
\cite{Dittmaier:1999mb,Dittmaier:2008md}, just as in our previous works.
The previously mentioned $\order{\alphas\alpha^6}$ contributions from
diagrams  where a photon splits into a quark--antiquark pair
[\reffi{fig:real} (g)] lead to an additional complication. Their QED
singularities would cancel against a virtual correction to a $\Pj\gamma+4\Pl$
final state that is not part of our signal phase space requiring two
final-state jets.
As already in $\PW \PZ$ and $\PZ \PZ$ scattering,
this singularity can be absorbed into a photon-to-jet conversion
function, which is related to the non-perturbative hadronic vacuum
polarisation \cite{Denner:2019zfp}, while treating the photon
as a jet in this case.

\subsection{Virtual corrections}

\begin{figure}
\centering
\begin{subfigure}{0.3\textwidth}
\captionsetup{skip=0pt}
\caption{}
\centering
\includegraphics[page=14,width=1.\linewidth]{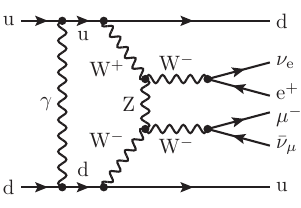}
\end{subfigure}
\begin{subfigure}{0.3\textwidth}
\captionsetup{skip=0pt}
\caption{}
\centering
 \includegraphics[page=16,width=1.\linewidth]{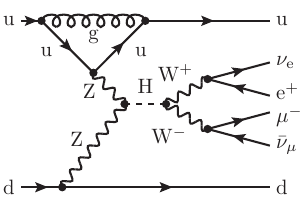}
\end{subfigure}
\begin{subfigure}{0.3\textwidth}
\captionsetup{skip=0pt}
\caption{}
\centering
 \includegraphics[page=19,width=1.\linewidth]{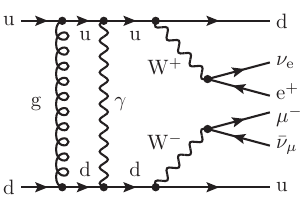}
\end{subfigure}
\caption{Examples of loop  diagrams of $\order{g^8}$ and
  $\order{g_\mathrm s^2 g^6}$. The first one (a) is a pure EW
  diagram, the second one (b) a QCD-induced contribution, and the third
  one (c) a mixed EW--QCD one.} 
\label{fig:One-Loop}
\end{figure}

Virtual corrections emerge from purely EW one-loop diagrams of order
$\order{g^8}$ and 
from one-loop diagrams of order
$\order{g_{\rm s}^2g^6}$ involving a virtual gluon
 (see \reffi{fig:One-Loop} for sample diagrams).
Interfering these with tree-level diagrams of orders $\order{g^6}$ or
$\order{g_{\rm s}^2 g^4}$ leads to contributions at orders
$\order{\alpha^7}$ and $\order{\alphas\alpha^6}$. As is the case for the real
corrections, the virtual ones of order $\order{\alpha^7}$ consist only of EW
corrections to the EW diagrams [\reffi{fig:One-Loop}~(a)], whereas the
order $\order{\alphas\alpha^6}$ contains both the interference of
$\order{g_{\rm s}^2 g^6}$ [\reffi{fig:One-Loop}~(b) and (c)] with
$\order{g^6}$ and the one of $\order{g^8}$ [\reffi{fig:One-Loop}~(a)]
with $\order{g_{\rm s}^2 g^4}$.  As for all VBS
processes, it is impossible to separate QCD and EW corrections at
$\order{\alphas\alpha^6}$ on the basis of Feynman diagrams, because
diagrams of order $\order{g_{\rm s}^2 g^6}$ with mixed EW and QCD
particle content [\reffi{fig:One-Loop}~(c)] can be viewed as QCD
corrections to the EW $\order{g^6}$ diagram or vice versa as EW
corrections to the QCD $\order{g_{\rm s}^2 g^4}$ diagram.

The virtual corrections contain both ultraviolet and infrared (IR)
singularities. The IR singularities cancel against those of
the real corrections or are absorbed in the PDFs. The ultraviolet
singularities are treated by renormalisation. We remark that diagrams
like the one in \reffi{fig:One-Loop} (c) 
give rise to IR singularities of both QCD
and EW origin.

\subsection{Details and validation}
Our results have been produced using the Monte Carlo program BBMC and
the matrix-element generator \recola
\cite{Actis:2012qn,Actis:2016mpe}. BBMC is a multi-channel integrator
which is able to compute multiparticle processes at NLO accuracy. For
fast and stable integration, it uses mappings similar to those in
\citeres{Berends:1994pv,Denner:1999gp,Dittmaier:2002ap}, which map the
final-state phase space according to the resonance structure of the
process. \recola is a general tree-level and one-loop matrix-element
provider which relies on the \collier library
\cite{Denner:2014gla,Denner:2016kdg}, which provides one-loop scalar
\cite{'tHooft:1978xw,Beenakker:1988jr,Dittmaier:2003bc,Denner:2010tr}
and tensor integrals
\cite{Passarino:1978jh,Denner:2002ii,Denner:2005nn} numerically. The
combination of \recola and \collier has been shown to be working
reliably for many high-multiplicity processes and in particular for
VBS.

BBMC has been checked against \mocanlo, an independent multi-channel
integrator, for same-sign W
\cite{Biedermann:2016yds,Biedermann:2017bss} and ZZ~scattering
\cite{Denner:2020zit,Denner:2021hsa}. For opposite-sign W scattering,
representative channels at LO, as well as $\order{\alpha^7}$ and
$\order{\alphas\alpha^6}$ have been compared for both codes. All
results were confirmed to agree within statistical uncertainties at the
percent level for all VBS processes.

Both in BBMC and \mocanlo the Catani--Seymour dipole formalism
\cite{Catani:1996vz,Dittmaier:1999mb,Dittmaier:2008md,Denner:2019vbn} is used to
subtract (and add back) IR divergences. Since opposite-sign W
scattering and ZZ scattering share related partonic processes,
as pointed out in \refse{sec:LO_process},
in their calculations the same dipole structures appear. Each dipole that was not
already utilised in same-sign W scattering was checked against
\mocanlo at the phase-space-point level for the real-subtracted  as well as
the integrated-dipole contributions in the course of our calculations
for ZZ scattering. Since these dipoles are process-independent
quantities, those checks remain valid for opposite-sign W scattering.

For the treatment of the vector-boson resonances we use the
complex-mass scheme
\cite{Denner:1999gp,Denner:2005fg,Denner:2006ic,Denner:2019vbn}. 

\section{Numerical Results}
\label{sec:results}
\subsection{Input parameters and event selection}

\subsubsection*{Input parameters}
Our setup is designed for an LHC run at a
centre-of-mass (CM) energy of $13\TeV$. We use the NLO NNPDF3.1luxQED PDF set with
$\alphas(\MZ) = 0.118$ \cite{Ball:2014uwa,Bertone:2017bme}
via LHAPDF~\cite{Andersen:2014efa,Buckley:2014ana} in the
$N_\text{F}=5$ fixed-flavour scheme in both the LO and NLO predictions. The
initial-state collinear splittings are treated by ${\overline{\rm
    MS}}$ redefinition of the PDFs.

As in our previous works, 
the central renormalisation and factorisation scales are chosen as the
geometric average of the transverse momenta of the tagging jets
\begin{equation}
\label{eq:defscale}
\mu_{\rm ren}^{\rm central} = \mu_{\rm fac}^{\rm central} = \sqrt{p_{\rm T, j_1}\, p_{\rm T, j_2}},
\end{equation}
where $\Pj_1$ and $\Pj_2$ are the two hardest identified jets (see
definition below) ordered according to transverse momentum. Based on
this central scale, we perform a 7-point scale variation of both the
renormalisation and factorisation scale, \ie we calculate the
observables for the pairs
\begin{equation}
(\mu_\mathrm{ren}/\mu_{\rm ren}^{\rm central},\mu_\mathrm{fact}/\mu_{\rm fact}^{\rm central}) = (0.5,0.5),
(0.5,1),(1,0.5),(1,1),(1,2),(2,1),(2,2)
\end{equation}
 of renormalisation and factorisation scales
and use the resulting envelope to estimate the perturbative (QCD) uncertainty.

We employ the $G_\mu$ scheme \cite{Denner:2000bj} to define the
electromagnetic coupling, which fixes the EW coupling $\alpha$ using the Fermi
constant $G_\mu$ as input parameter via
\begin{equation}
  \alpha = \frac{\sqrt{2}}{\pi} G_\mu \MW^2 \left( 1 - \frac{\MW^2}{\MZ^2} \right)  \qquad \text{with}  \qquad   {\GF    = 1.16638\times 10^{-5}\GeV^{-2}}.
\end{equation}
The masses and widths of the massive gauge bosons are taken from the PDG review 2020 \cite{ParticleDataGroup:2020ssz},
\begin{subequations}
    \begin{equation}
\begin{aligned}
                \MZOS &=  91.1876\GeV,      & \quad \quad \quad \GZOS &= 2.4952\GeV, \\
                \MWOS &=  80.379\GeV,       & \GWOS &= 2.085\GeV,
\end{aligned}
\end{equation}
and furthermore we set
\begin{equation}
\begin{aligned}
                  \Mt   &=  173.0\GeV,      & \quad \quad \quad \Gt &= 0 \GeV, \\
                M_{\rm H} &=  125.0\GeV,    &  \GH  &=  4.07 \times 10^{-3}\GeV.
\end{aligned}
\end{equation}
\label{eqn:ParticleMassesAndWidths}
\end{subequations}
The bottom quark is assumed to be massless, and no partonic channels
with initial-state and/or final-state bottom quarks are included.
Without any resonant top quarks in the considered processes, we set
the top-quark width to zero. The Higgs-boson width is taken
from \citere{Heinemeyer:2013tqa}.  From the measured on-shell (OS)
values of the masses and widths of the weak vector bosons $V=\PW,\PZ$,
we obtain the corresponding pole quantities used in the calculation
via \cite{Bardin:1988xt}
\begin{equation}
        M_V = \frac{\MVOS}{\sqrt{1+(\GVOS/\MVOS)^2}}\,,\qquad  
\Gamma_V = \frac{\GVOS}{\sqrt{1+(\GVOS/\MVOS)^2}}.
\end{equation}

Our calculation is performed in the 5-flavour scheme assuming
  a perfect bottom-jet veto.  The contribution of the remaining
  channels with bottom quarks only in the initial state has
  been neglected. It is PDF-suppressed and does not involve any VBS
  contributions. We verified numerically that this contribution
  is in relative terms below $10^{-7}$ at $\order{\alpha^6}$ and at the level of  $10^{-4}$
  at $\order{\alphas^2\alpha^4}$ for the setups considered
  in this paper.

\subsubsection*{Event selection with VBS cuts}

The event selection used for this analysis is inspired by the CMS and
ATLAS measurements of opposite-sign $\PW$-boson-pair production
\cite{CMS:2020mxy, ATLAS:2019rob} and our previous works on VBS. It is
similar to the one employed for the observation of this process
by CMS \cite{CMS:2022woe}. At
the LHC, different final states of opposite-sign W scattering have
been probed, either with two charged leptons of opposite charge and
same flavour or different flavour  and at least two QCD jets.
We restrict our analysis to the different-flavour channel. 
QCD partons (quarks, antiquarks, gluons), leptons and photons are
clustered simultaneously using 
the anti-$k_\text{T}$ algorithm \cite{Cacciari:2008gp} with a
resolution parameter $R=0.4$ for all pairs of QCD partons,
leptons and photons. Pairs of QCD partons are recombined to QCD
partons, QCD partons and photons to QCD partons, and
leptons and photons to (dressed) leptons. Unrecombined
photons are not treated as jets.
Only partons with rapidity $|y| < 5$ are considered for recombination,
while particles with larger $|y|$ are assumed to be lost in the beam
pipe.  The rapidity $y$ and the transverse momentum $p_{\rm T}$ of a
particle are defined as
\begin{equation}
y = \frac 12 \ln \frac{E + p_z}{E-p_z}, \qquad 
p_{\rm T} = \sqrt{p_x^2 + p_y^2}, 
\end{equation}
where $E$ is the energy of the particle, $p_z$ the component of its momentum along
the beam axis, and $p_x,p_y$ the components perpendicular to the beam axis.
The result of the clustering are jets, dressed leptons, and
  photons. All cuts are applied to these objects and the distributions
  shown below are based on them. In the following, leptons have to be
  understood as dressed leptons throughout.

Each of the two charged leptons $\ell$ has to fulfil
\begin{align}
 \ptsub{\Pl} >  25\GeV, \qquad |y_\Pl| < 2.4,
\end{align}
and together they must satisfy
\begin{align}
 \ptsub{\Pl^+\Pl^-} > 30\GeV, \qquad M_{\Pl^+\Pl^-} > 20\GeV,
\end{align}
where $\ptsub{\Pl^+\Pl^-}$ is the (vectorial) sum of the transverse momenta of the
charged leptons and $M_{\Pl^+\Pl^-}$ is the invariant mass of the
charged-lepton pair. The
missing transverse momentum is required to fulfil 
\begin{equation}
p_{\rm T,miss} > 20 \GeV
\end{equation}
and is computed as the transverse part of the sum of the two neutrino momenta.
After jet clustering, jets that fulfil the conditions
\begin{align}
 \ptsub{\Pj} >  30\GeV, \qquad |y_\Pj| < 4.5,\qquad\Delta R_{\Pj\Pl} > 0.4 \label{eq:deltaR}
\end{align}
are called identified jets, where the distance $\Delta R_{ij}$ is defined as
\begin{align}
\Delta R_{ij} = \sqrt{(\Delta \phi_{ij})^2 + (\Delta y_{ij})^2}
\end{align}
with the azimuthal-angle difference $\Delta \phi_{ij} = \min(|\phi_i -
\phi_j|, 2\pi - |\phi_i - \phi_j|)$ and the rapidity difference
$\Delta y_{ij} = y_i - y_j$. Note that the condition
\eqref{eq:deltaR} demands a minimal distance between the jet and any
of the charged leptons.  The two identified jets with highest
transverse momenta, called hardest, leading, or tagging jets, must obey
\begin{align}
\label{eq:vbscuts}
 M_{\Pj_1 \Pj_2} >  500\GeV, \qquad |\Delta y_{\Pj_1\Pj_2}| > 2.5.
\end{align}

\subsubsection*{Event selection with Higgs-search cuts}
Since the VBS channels contain contributions consisting of a Higgs
production and decay subprocess, we consider a second set of
event-selection criteria which is inspired by the different-flavour event selection
in the CMS Higgs search \cite{CMS:2018zzl}.

The clustering of QCD partons, leptons, and photons is done in
exactly the same way as for the VBS setup above, \ie using the
  anti-$k_\mathrm T$ algorithm with a resolution parameter of $R =
  0.4$ for partons within $|y| < 5$.
The two tagging jets are required to fulfil
\begin{align}\label{eq:tagging}
\ptsub{\Pj_{1,2}} > 30\GeV, \qquad |y_{\Pj_{1,2}}| < 4.7, \qquad \Delta R_{\Pj_{1,2}\Pl} > 0.4
\end{align}
as well as the typical VBS topology cuts
\begin{align}
\qquad |\Delta y_{\Pj_1\Pj_2}| > 3.5, \qquad M_{\Pj_1\Pj_2} > 400\GeV.
\end{align}
Additionally, we introduce a jet veto for any third
jet resulting from the clustering and fulfilling the
  conditions \refeq{eq:tagging}, \ie events are only kept if the third jet obeys
\begin{align}\label{eq:jet_veto}
\ptsub{\Pj_3} < 30 \GeV.
\end{align}

The (dressed) charged leptons have to fulfil
\begin{align}
\ptsub{\Pl}^{\mathrm{lead}} > 25\GeV, \qquad \ptsub{\Pl}^{\mathrm{trail}} > 10\GeV,
\end{align}
where $\ptsub{\Pl}^{\mathrm{lead}}$ and $\ptsub{\Pl}^{\mathrm{trail}}$ are the transverse momenta of the leading (harder) and trailing (softer) charged lepton. The minimum distance and the rapidity of the charged leptons are chosen to be
\begin{align}
\Delta R_{\Pl^+ \Pl^-} > 0.4, \qquad |y_\Pl| < 2.4,
\end{align}
their pair invariant mass is required to exceed
\begin{align}
M_{\Pl^+\Pl^-} > 12\GeV,
\end{align}
and the sum of their transverse momenta must respect
\begin{align}
\ptsub{\Pl^+\Pl^-} > 30\GeV.
\end{align}

The missing transverse momentum has to fulfil
\begin{align}
p_{\rm T,miss}  > 20 \GeV,
\end{align}
as in the VBS setup. Additionally, we require the transverse mass of
the lepton system $M_\mathrm T$ to be within a range below the Higgs
mass,
\begin{align}
60\GeV < M_\mathrm T < 125\GeV,
\end{align}
where $M_\mathrm T$ is defined as
\begin{align}
M_\mathrm T = \sqrt{2\ptsub{\Pl^+\Pl^-}p_{\rm T,miss}[1 - \cos \Delta\phi(\Pl^+\Pl^-, \nu\bar\nu)]}
\end{align}
and $\Delta\phi(\Pl^+\Pl^+,\nu\bar\nu)$ is the azimuthal angle between the
sum of the charged-lepton momenta and the sum of the neutrino momenta.

Furthermore the rapidities of the charged leptons are bounded by the
rapidities of the two tagging jets. To this end, we use the {\em Zeppenfeld
variable} $z_{\Pl\Pj_1\Pj_2}$, 
also called {\em centrality}, defined as
\begin{align}\label{eq:def_zepp}
z_{\Pl\Pj_1\Pj_2} = \frac{y_\Pl - \frac{y_{\Pj_1}+y_{\Pj_2}}{2}}{\Delta y_{\Pj_1\Pj_2}},
\end{align}
and require
\begin{align}\label{eq:cut_zepp}
-0.5 < z_{\Pl\Pj_1\Pj_2} < 0.5.
\end{align}

\subsection{Cross sections}
We start with presenting LO results in \refta{tab:LO}. 
\begin{table}
\sisetup{group-digits=false}
\centering
\begin{tabular}{cccccc}
\toprule
Order
    & $\order{           \alpha^6 }$
    & $\order{ \alphas   \alpha^5 }$
    & $\order{ \alphas^2   \alpha^4 }$
    & $\order{ \alphas^4   \alpha^4 }$ 
    & Sum\\
\midrule
\multicolumn{6}{l}{VBS setup} \\
\midrule 
$\sigma_{\mathrm{LO}} [\si{\femto\barn}]$
    & \num{ 2.6988 +- 0.0003}
    & \num{ 0.06491 +- 0.00002}
    & \num{ 6.9115 +- 0.0009}
    & \num{ 0.1952 +- 0.0008}
    & \num{ 9.8704 +- 0.0012}
    \\
$\textrm{fraction}\ [\si{\percent}]$
    & \num{27.3}
    & \num{0.7}
    & \num{70.0}
    & \num{2.0}
    & \num{100}
    \\
\midrule 
\multicolumn{3}{l}{Higgs setup} \\
\midrule
$\sigma_{\mathrm{LO}} [\si{\femto\barn}]$
    & \num{ 1.5322 +- 0.0002}
    & \num{ 0.008996 +- 0.000005}
    & \num{ 1.6923 +- 0.0003}
    & \num{ 0.1057 +- 0.0007}
    & \num{ 3.3392 +- 0.0008}
    \\
$\textrm{fraction}\ [\si{\percent}]$
    & \num{45.9}
    & \num{0.3}
    & \num{50.7}
    & \num{3.2}
    & \num{100}
    \\
\bottomrule
\end{tabular}
\caption{LO cross sections and contributions of individual orders
  $\order{\alpha^6}$, $\order{\alphas\alpha^5}$,
  $\order{\alphas^2\alpha^4}$, and $\order{ \alphas^4 \alpha^4 }$ for
  $\Pp\Pp \to \Pe^+\nu_e\mu^-\bar\nu_\mu\Pj\Pj+X$ at $13\TeV$ CM energy
  at the central scale. No contributions with external bottom quarks are
  included. Each contribution is given in $\fb$ and as fraction
  relative to the sum of the four contributions in percent. The
  digits in parentheses indicate integration errors.} 
\label{tab:LO}
\end{table}
For the VBS setup, the EW cross section of $\order{\alpha^6}$ is
$2.7\fb$ and thus the largest cross section among all VBS processes at the LHC.
Whilst the interference is almost negligible with $0.065\fb$, the
QCD-induced background is very large with $6.9\fb$, due to processes with two
gluons in the initial state that contribute for
opposite-sign $\PW$~scattering as opposed to same-sign
$\PW$~scattering.  The EW and QCD cross-section fractions
are $27.3\%$ and $70.0\%$ and thus comparable to
the situation of $\PZ\PZ$ scattering, which features, up to the leptons,
the same partonic processes as $\PW^+\PW^-$ scattering. There we found
$32.2\%$ versus $59.5\%$ with strict VBS cuts \cite{Denner:2020zit}.
The contribution of the loop-induced gluon channel amounts to
$0.19\fb$ or $2\%$.  Including the 7-point scale variations, the full
LO cross section is obtained as
\begin{align}
\sigma_\mathrm{LO}^{\mathrm{VBS}} = 9.871(1)^{+30\%}_{-21\%}\fb.
\end{align}
The sizeable scale dependence results from the dominant contribution
of order  $\order{\alphas^2\alpha^4}$.

For the Higgs setup the overall picture does not change drastically at
LO. Owing to stricter cuts, the EW cross section is reduced to
$1.5\fb$ which is almost of the same size as the QCD cross section
with $1.7\fb$, \ie the relative contribution of VBS is increased.
While the fraction of the interference contribution drops, the one of
the loop-induced gluon channel grows. Together with the 7-point scale variation for
this setup the LO result is
\begin{align}
\sigma_{\mathrm{LO}}^\mathrm{Higgs} = 3.3392(8)^{+25\%}_{-17\%}\fb.
\end{align}
The scale dependence is somewhat reduced since the relative
contribution of the QCD-induced process is smaller.

To investigate the contribution of VBS to the $\order{\alpha^6}$ cross section
and to compare this with its corresponding
background channels of the same initial and final states, we split up
the cross sections into further sub-contributions in \refta{ta:xs_contributions}. 
\begin{table}
\sisetup{group-digits=false}
\centering
\begin{tabular}{ccccc}
\toprule
Contribution & $\order{\alpha^6}$ & $\order{\alphas\alpha^5}$  & $\order{\alphas^2\alpha^4}$ & sum \\
\midrule             
\multicolumn{5}{l}{VBS setup } \\
\midrule 
$\sigma(4\Pq, \text{VBS})[\fb]$        & \num{2.6988 +- 0.0003}   & \num{0.05439 +- 0.00002}     & \num{2.2315 +- 0.0003}     & \num{4.9846 +- 0.0004} \\
$\sigma(4\Pq, \text{non-VBS})[\fb]$    & \num{1.4734(9)e-4}       & --                           & \num{0.008641+-0.000003}   & \num{0.008788+-0.000003} \\ 
$\sigma(\gamma\gamma/\Pg\gamma/\Pg\Pg)[\fb]$              & \num{6.832(2)e-6}        & \num{0.010605+-0.000002}     & \num{4.6820 +- 0.0008}     & \num{4.6926 +- 0.0008} \\
$\sigma(\mathrm{total})[\fb]$          & \num{2.6988 +- 0.0003}   & \num{0.06500 +- 0.00002}     & \num{6.9221 +- 0.0009}     & \num{9.6860 +- 0.0009} \\
\midrule             
\multicolumn{5}{l}{Higgs setup } \\
\midrule 
$\sigma(4\Pq, \text{VBS})[\fb]$        & \num{1.5322 +- 0.0002}   & \num{0.007490 +- 0.000005}     & \num{0.39866 +- 0.00007}     & \num{1.9384 +- 0.0002} \\
$\sigma(4\Pq, \text{non-VBS})[\fb]$    & \num{1.850(2)e-5}       & --                           & \num{0.0012729+-0.0000006}   & \num{0.00129138+-0.00000006} \\ 
$\sigma(\gamma\gamma/\Pg\gamma/\Pg\Pg)[\fb]$              & \num{7.764(4)e-7}        & \num{0.0015062+-0.0000004}     & \num{1.2923 +- 0.0003}     & \num{1.2938 +- 0.0003} \\
$\sigma(\mathrm{total})[\fb]$          & \num{1.5322 +- 0.0002}   & \num{0.008996 +- 0.000005}   & \num{1.6923 +- 0.0003}     & \num{3.2335 +- 0.0003} \\
\bottomrule
\end{tabular}
\caption{Division of LO cross sections into contributions of partonic
  channels with specific subprocesses. Note that no non-VBS processes
  containing two quarks in both the initial and final state contribute
  at order $\order{\alphas\alpha^5}$.} 
\label{ta:xs_contributions}
\end{table}
In contrast to ZZ~scattering, there is no remarkable difference
between processes with internal $\PZ\PZ \to \PW\PW$ scattering
compared to those with internal $\PW\PW \to \PW\PW$ scattering; hence
partonic channels with these subprocesses are combined and
labelled as ($4\Pq$, VBS).
Channels without VBS contributions, denoted by ($4\Pq$, non-VBS) and
containing triple vector-boson production diagrams, are
negligible at all considered orders.  Photon-induced processes
($\gamma\gamma$) are negligible at $\order{\alpha^6}$, whereas
processes containing gluons in the initial or final state
($\Pg\Pg$) become dominant at
\order{\alphas^2\alpha^4}. Photon--gluon-induced processes
($\Pg\gamma$) at $\order{\alphas\alpha^5}$ remain at the per-mille
level.  Both setups do not show remarkable differences at LO; however,
the Higgs setup suppresses the QCD background of $4\Pq$-VBS processes
very efficiently.

We move to the discussion of NLO contributions starting with the scale
dependence.  The LO scale dependence of the EW contribution and the
NLO scale dependence of the orders $\order{\alpha^7}$ and
$\order{\alphas\alpha^6}$ is shown in \refta{tab:scales}. We emphasise
that the scale uncertainties for the EW contribution within the VBS
setup are significantly reduced when taking the NLO QCD contributions
into account, whereas this reduction is not observed within the Higgs
setup.
\begin{table}
\sisetup{group-digits=false}
\centering
\begin{tabular}{ccccc}
\toprule
Order
    & $\order{ \alpha^6 }$
    & $\order{ \alpha^6 } + \order{\alpha^7}$
    & $\order{ \alpha^6 } + \order{\alphas \alpha^6}$
    & \!$\order{ \alpha^6 } + \order{\alpha^7} + \order{\alphas\alpha^6}$\! \\
\midrule
\multicolumn{3}{l}{VBS setup } \\
\midrule 
$\sigma_{\mathrm{central}} [\si{\femto\barn}]$
    & \num{ 2.6988 +- 0.0003}
    & \num{ 2.391 +- 0.001}
    & \num{ 2.563 +- 0.003}
    & \num{ 2.255 +- 0.004}
    \\
$\sigma_\mathrm{min} [\si{\fb}]$
    & \num{2.5069 +- 0.0003}
    & \num{ 2.232 +- 0.001}
    & \num{ 2.545 +- 0.004}
    & \num{ 2.200 +- 0.004}
    \\
$\delta \sigma_\mathrm{min} [\%]$ 
    & \num{-7.1}
    & \num{-6.7}
    & \num{-0.7}
    & \num{-2.5}
    \\
$\sigma_\mathrm{max} [\si{\fb}]$
    & \num{2.9187 +- 0.0003}
    & \num{2.573 +- 0.001}
    & \num{2.581 +- 0.004}
    & \num{2.285 +- 0.003} 
    \\
$\delta \sigma_\mathrm{max} [\%]$ 
    & \num{+8.2}
    & \num{+7.6}
    & \num{+0.7}
    & \num{+1.3}
    \\
\midrule 
\multicolumn{3}{l}{Higgs setup } \\
\midrule
$\sigma_{\mathrm{central}} [\si{\femto\barn}]$
    & \num{ 1.5322 +- 0.0002}
    & \num{ 1.429 +- 0.001}
    & \num{ 1.202 +- 0.002}
    & \num{ 1.099 +- 0.002}
    \\
$\sigma_\mathrm{min} [\si{\fb}]$
    & \num{1.4418 +- 0.0002}
    & \num{1.349 +- 0.001}
    & \num{1.130 +- 0.002}
    & \num{1.014 +- 0.002} 
    \\
$\delta \sigma_\mathrm{min} [\%]$ 
    & \num{-5.9}
    & \num{-5.6}
    & \num{-6.0}
    & \num{-7.7}
    \\
$\sigma_\mathrm{max} [\si{\fb}]$
    & \num{1.6324 +- 0.0002}
    & \num{1.517 +- 0.001}
    & \num{1.248 +- 0.002}
    & \num{1.155 +- 0.002} 
    \\
$\delta \sigma_\mathrm{max} [\%]$ 
    & \num{+6.5}
    & \num{+6.1}
    & \num{+3.8}
    & \num{+5.1}
    \\
\bottomrule
\end{tabular}
\caption{Cross sections at LO and NLO with 7-point scale variation
  of $\mu_{\rm{fact}}$ and $\mu_{\rm{ren}}$ around the central value \refeq{eq:defscale}. We list
  the result for the central scale and those for the scales
  that show the largest deviation from the central value, both in
  absolute and relative quantities.} 
\label{tab:scales}
\end{table}
This is due to the comparably large QCD corrections in this setup and
their scale dependence. The large QCD corrections are related
  to the jet veto, as discussed below.

We present the NLO cross sections in \refta{tab:nlo}, dividing the
partonic channels with four quarks into a group that contains only VBS
as a subprocess, two in which triple-vector-boson production, either
WWW or WWZ, can occur, and another two in which both
triple-vector-boson production and VBS appear as subprocesses.
\begin{table}
\sisetup{group-digits=false}
\centering
\tabcolsep 5pt
\begin{tabular}{cccccc}
\toprule
Contribution  & $\sigma_{\rm LO}^{\alpha^6} [\fb]$ & $\Delta \sigma_{\rm NLO}^{\alpha^7} [\fb]$ & $\delta^{\alpha^7}[\%]$ & $\Delta \sigma_{\rm NLO}^{\alphas\alpha^6}[\fb]$ & $\delta^{\alphas\alpha^6}[\%]$ \\ 
\midrule
\multicolumn{4}{l}{VBS setup } \\ 
\midrule
VBS only     & \num{2.1695  +- 0.0003}  & \num{-0.2812+- 0.0008}   & \num{-13.0}  &
                                          \num{-0.146 +- 0.003}    & \num{-6.7}   \\
VBS + WWW    & \num{0.13783 +- 0.00003} & \num{-0.0164+-0.0002}    & \num{-11.9}  &
                                          \num{0.0071+-0.0004}     & \num{5.2}    \\
VBS + WWZ    & \num{0.39140 +- 0.00006} & \num{-0.0427+-0.0003}    & \num{-10.9}  &
                                          \num{-0.013+-0.001}      & \num{-3.3}   \\
WWW only     & \num{5.319(8)e-5}        & \num{-1.49(5)e-5}        & \num{-28.0}  &
                                          \num{0.01169+-0.00001}   & \num{2.2e+4} \\
WWZ only     & \num{9.415(3)e-5}        & \num{-2.72(3)e-5}        & \num{-28.8}  &
                                          \num{0.003907+-0.000002} & \num{4.1e+3} \\
$\gamma\gamma$/$\gamma\Pg$ & \num{6.832(2)e-6} & \num{ 0.03292+-0.00001}  & \num{+4.8e+5}&
                                          \num{-.0002+-0.0006}     & \num{-2.9e+3}\\
total        & \num{2.6988+-0.0003}     & \num{-0.3074+-0.0009}    & \num{-11.4}  &
                                          \num{-0.136+-0.003}      & \num{-5.1}   \\
\midrule
\multicolumn{4}{l}{Higgs setup } \\ 
\midrule
VBS only     & \num{1.1958  +- 0.0002}  & \num{-0.0913  +- 0.0012}  & \num{ -7.6}  &
                                          \num{-0.2400  +- 0.0013}  & \num{-20.0}  \\
VBS + WWW    & \num{0.06603 +- 0.00001} & \num{-0.0052 +- 0.0002}   & \num{ -7.8}  &
                                          \num{-0.0093 +- 0.0003}   & \num{-14.1}  \\
VBS + WWZ    & \num{0.27030 +- 0.00004} & \num{-0.0160 +- 0.0005}   & \num{ -5.9}  &
                                          \num{-0.0451 +- 0.0007}   & \num{-16.7}  \\
WWW only     & \num{6.28(2)e-6}         & \num{-1.8(1)e-7}          & \num{-28.9}  &
                                          \num{0.002508+- 0.000005} & \num{+4.0e+4}\\
WWZ only     & \num{1.223(2)e-5}        & \num{-3.30(8)e-6}         & \num{-27.0}  &
                                          \num{0.0006770+-0.0000007}& \num{+5.5e+3}\\
$\gamma\gamma$/$\gamma\Pg$ & \num{7.764(4)e-7} & \num{0.00916 +- 0.00002}  & \num{+1.2e+6}&
                                          \num{-0.039  +- 0.001}    & \num{-5.0e+5}\\
total        & \num{1.5322+-0.0002}     & \num{-0.1033 +- 0.0013}   & \num{-6.7}   &
                                          \num{-0.330 +- 0.002}     & \num{-21.6}  \\
\bottomrule
\end{tabular}
\caption{NLO corrections at $\order{\alpha^7}$ and
  $\order{\alphas\alpha^6}$ in relation to their LO
  counterparts of $\order{\alpha^6}$, classified after appearing
  subprocesses, and corresponding relative corrections $\delta^{\alpha^7} = \Delta
  \sigma_{\rm NLO}^{\alpha^7} / \sigma_{\rm LO}^{\alpha^6}$ and $\delta^{\alphas\alpha^6} = \Delta
  \sigma_{\rm NLO}^{\alphas\alpha^6} / \sigma_{\rm LO}^{\alpha^6}$.} 
\label{tab:nlo}
\end{table}
Comparing the NLO corrections to $\PW^+\PW^-$ scattering to other VBS
processes, we notice some important differences. At first, the overall
EW corrections are with $-11.4\%$ in the VBS setup only three quarters and
with $-6.7\%$ in the Higgs setup only half as large as the EW
corrections for the VBS processes in same-sign WW, WZ, or ZZ
scattering. We recapitulate this briefly in
\refta{tab:EW_corrections}, where we list the EW corrections of
$\order{\alpha^7}$ compared with the EW LO processes of
$\order{\alpha^6}$ for all massive VBS processes. 
\begin{table}
\sisetup{group-digits=false}
\centering
\begin{tabular}{cccccc}
\toprule
Process 
    & $\PW^+ \PW^+$
    & $\PW^+ \PZ$
    & $\PZ \PZ$
    & $\PW^+ \PW^-$ 
    & $\PW^+ \PW^-$  \\
    &&&
    & (VBS setup)
    & (Higgs setup) \\
\midrule
$\Delta \sigma_{\mathrm{NLO}}^{\alpha^7} [\si{\femto\barn}]$
    & \num{ -0.2169   +- 0.0003}
    & \num{ -0.04091  +- 0.00002}
    & \num{ -0.015573 +- 0.000005}
    & \num{ -0.307 +- 0.001}
    & \num{ -0.103 +- 0.001}
    \\[.8ex]
$\sigma_{\mathrm{LO}}^{\alpha^6} [\si{\femto\barn}]$
    & \num{ 1.4178   +- 0.0002}
    & \num{ 0.25511  +- 0.00001}
    & \num{ 0.097683 +- 0.000002}
    & \num{ 2.6988   +- 0.0003}
    & \num{ 1.5322   +- 0.0002}
    \\[.7ex]
$\delta^{\alpha^7} [\si{\percent}]$
    & \num{-15.3}
    & \num{-16.0}
    & \num{-15.9}
    & \num{-11.4}
    & \num{-6.7} \\
\bottomrule
\end{tabular}
\caption{NLO cross section of $\order{\alpha^7}$ and LO cross section
  of $\order{\alpha^6}$ in $\fb$ and as relative correction in
  percent for different VBS processes $\PW^+ \PW^+$ ($\Pp\Pp \to 
  \Pe^+\nu_e\mu^+\nu_\mu\Pj\Pj+X$) \cite{Biedermann:2017bss}, $\PW \PZ$ ($\Pp\Pp \to
  \Pe^+\nu_e\mu^+\mu^-\Pj\Pj+X$) \cite{Denner:2019tmn}, $\PZ \PZ$ ($\Pp\Pp \to
  \Pe^+\Pe^-\mu^+\mu^-\Pj\Pj+X$) \cite{Denner:2020zit} and $\PW^+ \PW^-$ ($\Pp\Pp \to
  \Pe^+\nu_e\mu^-\bar\nu_\mu\Pj\Pj+X$) normalized to the EW LO cross
  section, $\delta^{\alpha^7} = {\Delta
  \sigma_\mathrm{NLO}^{\alpha^7}}/{\sigma_\mathrm{LO}^{\alpha^6}}$.}
\label{tab:EW_corrections}
\end{table}
Next, the partonic channels without VBS receive significantly higher negative
EW corrections than those with VBS subprocesses in both setups, but because of
their small absolute size this does not lead to a sizeable effect in
the complete cross section.

The relatively small value of the EW
corrections for the total fiducial cross section of opposite-sign
W~scattering can be traced back to the presence of the Higgs resonance
in VBS subprocesses. As noted in \citere{Biedermann:2016yds},
the large EW corrections emerge from EW logarithms that become large
at high invariant masses of the four-lepton system. 
While typically a large di-jet invariant mass
  and/or a large rapidity separation is required for the tagging jets in VBS, the
  large EW corrections are not a result of  this event selection but an
  intrinsic feature of VBS processes at the LHC
  \cite{Biedermann:2016yds}. In fact, the  di-jet invariant-mass cut
  varies between $100\GeV$ and $500\GeV$ for the results for
  $\PW^+\PW^-$, $\PW^+\PZ$, and $\PZ\PZ$ listed in
  \refta{tab:EW_corrections}.
In contrast, the
Higgs resonance drives a considerable fraction of the cross section
towards low four-lepton invariant masses around $\MH$. This can 
be seen from the distribution in the four-lepton invariant mass shown
in \refse{sec:LO_distributions}.
In the Higgs setup, with cuts tailored to enhance the Higgs
contribution, this effect is even more pronounced. As a result, the EW
corrections are almost halved, from $-11.4\%$ in the VBS setup to
$-6.7\%$ in the Higgs setup. Note that the Higgs resonance is
  eliminated by the cuts in the setup for VBS into ZZ in
  \refta{tab:EW_corrections}.

Another
interesting effect at $\order{\alpha^7}$ are the NLO contributions of
photon-induced processes. While their absolute contribution remains at
the level of one percent with respect to the complete
$\order{\alpha^6}$ contribution, they are extremely large when
compared to the LO contributions of the photon-induced channels. This
is due to the appearance of new partonic channels with a photon and a
quark or antiquark in the initial state with internal VBS topology
(see \reffi{fig:ggamma-induced}) compared to partonic channels with
two photons in the initial state at LO.

At $\order{\alphas\alpha^6}$, we notice an even larger quantitative
difference between the two setups, this time with larger corrections
in the Higgs setup. This effect can be traced back to the
jet veto \refeq{eq:jet_veto}, which suppresses the
(positive) real corrections of an emitted final-state gluon, while it
leaves the (negative) virtual corrections unaffected. 
Furthermore, there are very large relative corrections to non-VBS
processes in both setups.  These can be explained as follows: In
non-VBS processes at LO, both final-state jets can only be produced
via an $s$-channel vector boson, which does not lead to the required
back-to-back jets with high invariant mass.  At NLO, however, one of the
incoming partons can emit a hard gluon in forward direction
(see \reffi{fig:gluon_emission}), providing one of the tagging
jets. The $s$-channel vector boson can become resonant leading to an enhancement of the
cross section. The enhancement of the $\order{\alphas\alpha^6}$
contributions owing to these radiative triple-vector-boson-production
diagrams is also seen in the VBS + triple-gauge-boson-production
channels, where the relative QCD corrections are more positive than in
the pure VBS channels.  The same effect was already observed in other
VBS processes \cite{Ballestrero:2018anz,Denner:2020zit}.  The
photon--gluon-induced corrections at $\order{\alphas\alpha^6}$ are
about $-2.5\%$ of the complete $\order{\alpha^6}$ in the Higgs setup
and compatible with zero in the VBS setup. The apparently large
relative corrections are again due to the smallness of the
corresponding LO contributions.

\subsection{The role of Higgs VBF in opposite-sign VBS}
\label{sec:higgs-subtraction}

We previously argued that the size of the EW corrections for
opposite-sign W-boson scattering is smaller than for all other VBS
processes, because the presence of an $s$-channel Higgs propagator
shifts a large fraction of events towards its resonance. Therefore it
lowers the effective four-fermion invariant mass $\langle M_{4\ell}
\rangle$ from high values, typical for genuine VBS, towards lower values
near $\MH$, which drives the size of the EW correction. 
Roughly speaking, our signal is not only opposite-sign W-boson
scattering, but also VBF Higgs-boson production with subsequent decay into four leptons.
In the Higgs setup the EW corrections are found to be even smaller, which
is consistent with this picture, because in this setup the cuts increase
the fraction of VBF into a Higgs boson.

We can further test this hypothesis by cutting out the Higgs resonance. This
should increase the size of the EW corrections to a similar
magnitude as found in other VBS processes.
To this end, we introduce a third setup based on the VBS setup, but additionally imposing an
unphysical invariant-mass cut on the four-lepton system, \ie requiring
\begin{equation}\label{eq:Higgs_cut}
 |M_{4\Pl} - \MH| > N \Gamma_\PH,
\end{equation}
where $\Gamma_\PH$ is the Higgs decay width and $N=20$. This method
has, in contrast to the diagram-removal techniques proposed in
\citeres{Frixione:2008yi,Hollik:2012rc}, the advantage to remain
manifestly gauge invariant and not spoil the important role of the
Higgs diagrams for the preservation of unitarity in the high-energy
limit. For those LO QCD and LO EW contributions in which the Higgs
resonance is absent (WWW only + WWZ only + $\gamma$/g) this additional
cut is almost negligible, changing their contribution to the fiducial
cross section only at the per-mille level from {$1.5417(9)\times
  10^{-4}\fb$ to $1.5401(3)\times 10^{-4}\fb$} at $\order{\alpha^6}$,
which is compatible with zero within the $2\sigma$ integration-error
range. Hence we expect this cut to mainly remove the influence
of the Higgs resonance on VBS contributions while leaving
contributions from other Feynman diagrams mostly unaffected. Taking a
Breit--Wigner (BW) resonance peak as a basis for an estimate, the cut
\refeq{eq:Higgs_cut}
results in an effective removal of a fraction
\begin{equation}
\frac{\sigma^\mathrm{cut}}{\sigma^\mathrm{BW}} \approx \frac{2}{\pi} \operatorname{arctan}(2N)
\end{equation}
of the resonance contribution leading to a removal of $98.4\%$ of the Higgs
contribution for $N=20$. 

In \refta{tab:subtracted} we present results for the VBS setup
modified by the cut \refeq{eq:Higgs_cut}.
\begin{table}
\sisetup{group-digits=false}
\centering
\begin{tabular}{cccccc}
\toprule
Contribution & $\sigma_{\rm LO}^{\alpha^6}[\fb]$ & $\Delta \sigma_{\rm NLO}^{\alpha^7}[\fb]$ & $\delta^{\alpha^7}[\%]$ & $\Delta \sigma_{\rm NLO}^{\alphas\alpha^6}[\fb]$ & $\delta^{\alphas\alpha^6}[\%]$   \\
\midrule 
\multicolumn{6}{l}{VBS setup with Higgs-resonance cut \refeq{eq:Higgs_cut}} \\
\midrule 
VBS only     & \num{1.6117  +- 0.0002}  & \num{-0.239  +- 0.002}  & \num{-14.8}  &
                                          \num{-0.043  +- 0.003}  & \num{- 2.7}  \\
VBS + WWW    & \num{0.11398 +- 0.00002} & \num{-0.0143 +- 0.0002}   & \num{-12.5}  &
                                          \num{+0.0080 +- 0.0005}   & \num{+07.1}  \\
VBS + WWZ    & \num{0.24916 +- 0.00004} & \num{-0.0324 +- 0.0003}   & \num{-13.0}  &
                                          \num{+0.0018 +- 0.0011}   & \num{+00.1}  \\
WWW only     & \num{5.303(2)e-5}        & \num{-1.43(2)e-5}         & \num{-27.0}  &
                                          \num{0.01110 +- 0.00002}  & \num{+2.1e+4}\\
WWZ only     & \num{9.415(2)e-5}        & \num{-2.80(2)e-5}         & \num{-29.7}  &
                                          \num{0.004021+-0.000003}  & \num{+4.3e+3}\\
$\gamma\gamma$/$\gamma\Pg$ & \num{6.832(4)e-6} & \num{0.02575 +- 0.00003}  & \num{+3.8e+5}&
                                          \num{0.0108  +- 0.0002}   & \num{+1.6e+5}\\
total        & \num{1.9750+-0.0002}     & \num{-0.260 +- 0.002}     & \num{-13.2}  &
                                          \num{-0.007 +- 0.003}     & \num{-00.4}  \\
\bottomrule
\end{tabular}
\caption{LO cross section of $\order{\alpha^6}$ and NLO cross section
  of $\order{\alpha^7}$ and $\order{\alphas\alpha^6}$ in $\fb$ and
  relative corrections in percent with an unphysical invariant-mass cut on the
  Higgs-boson resonance in the VBS setup, classified after appearing
  subprocesses and corresponding relative corrections $\delta^{\alpha^7} = \Delta
  \sigma_{\rm NLO}^{\alpha^7} / \sigma_{\rm LO}^{\alpha^6}$ and $\delta^{\alphas\alpha^6} = \Delta
  \sigma_{\rm NLO}^{\alphas\alpha^6} / \sigma_{\rm LO}^{\alpha^6}$.} 
\label{tab:subtracted}
\end{table}
We note that the relative EW corrections to VBS processes with
$-12.5\%$ to $-14.8\%$ for the VBS channels, giving an average of
$-14.5\%$, are comparable with those for same-sign WW, WZ and ZZ
scattering. The other effects that have already been discussed in the
previous section hold analogously for this setup. Besides, we remark
the accidental cancellation in the total $\order{\alphas\alpha^6}$
corrections between the various contributions.

\begin{table}
\sisetup{group-digits=false}
\centering
\begin{tabular}{cccccc}
\toprule
Contribution & $\sigma_{\rm LO}^{\alpha^6}[\fb]$ & fraction$[\%]$ & $\Delta \sigma_{\rm NLO}^{\alpha^7}[\fb]$ & $\delta^{\alpha^7}[\%]$ & $\Delta \sigma_{\rm NLO}^{\alphas\alpha^6}[\fb]$   \\
\midrule 
\multicolumn{5}{l}{VBS setup, Higgs-resonance contribution} \\
\midrule 
VBS only     & \num{0.5577  +- 0.0004}  & \num{25.7} &\num{-0.042  +- 0.002}   & \num{-7.5} &
                                                      \num{-0.103  +- 0.005}   \\
VBS + WWW    & \num{0.02385 +- 0.00004} & \num{17.3} &\num{-0.0023 +- 0.0002}  & \num{-9.5} &
                                                      \num{-0.0007 +- 0.0006}  \\
VBS + WWZ    & \num{0.14224 +- 0.00007} & \num{36.3} &\num{-0.0103 +- 0.0004}  & \num{-7.2} &
                                                      \num{-0.0148 +- 0.0015}  \\
WWW only     & \num{1.6(9)e-7}          &            &\num{-6(5)e-7}           &  &
                                                      \num{5.9(3)e-4}          \\
WWZ only     & \num{3(37)e-9}         &            &\num{+8(4)e-7}           &  &
                                                      \num{-1.14(3)e-4}        \\
$\gamma\gamma$/$\gamma\Pg$ & \num{-5(25)e-10} &            &\num{0.00717 +- 0.00003} &  &
                                                      \num{0.0110  +- 0.0006}  \\
total        & \num{0.7238+-0.0004}     & \num{26.8} &\num{-0.047 +- 0.002}    & \num{-6.5} &
                                                      \num{-0.129 +- 0.005}    \\
\bottomrule
\end{tabular}
\caption{LO cross section at $\order{\alpha^6}$ and NLO corrections at
  $\order{\alpha^7}$ and $\order{\alphas\alpha^6}$ in $\fb$ and 
  relative corrections in percent, classified after
  appearing subprocesses, for the contribution of the Higgs resonance to the
  cross section $\sigma_{\text{Higgs}} = \sigma_{\text{VBS setup}} - \sigma_{\text{
      VBS setup, Higgs cut}}$ and the fraction of the
  Higgs-resonance contribution to the fiducial cross section $\sigma_{\text{Higgs}} /
  \sigma_{\text{VBS setup}}$ at LO. We note the cancellations and
  correspondingly large relative combined integration errors in the
  subtraction for the contributions of the Higgs resonance to non-VBS processes, which
  lead to results that are compatible with zero.} 
\label{tab:resonance}
\end{table}
As a last step, we present the contributions of the Higgs resonance in
\refta{tab:resonance}, which are obtained by subtracting the results
of \reftas{tab:nlo} and \ref{tab:subtracted}. For contributions
involving VBS, its fraction ranges between 17\% and 36\%.
We note that the cut does not affect the non-VBS channels at LO,
leading to results that are compatible with zero within the
integration errors. 

There are, however, effects at NLO to discuss for
these channels. Non-negligible corrections show up at
$\order{\alphas\alpha^6}$ for the triple-vector-boson production
channels. These can be explained by the fact that also for these
channels, Feynman diagrams with Higgs bosons in the $s$ channel,
emitted from an $s$-channel W or Z boson, are present both at LO and
NLO (see \reffi{fig:non-VBS_Higgs}). Within the QCD corrections, those
channels get enhanced by emission of a gluon from an initial-state
quark, allowing the event to pass the VBS cuts, which leads to a tiny,
but sizeable change of the cross section when cutting out the
resonance. In contrast, this enhancement is absent for the EW
corrections and so is a difference in the cross section. The
relatively large difference for photon- and gluon-induced channels at
both $\order{\alpha^7}$ and $\order{\alphas\alpha^6}$ is due to the
opening up of new channels at NLO, in which VBS subprocesses appear (see
\reffi{fig:ggamma-induced}). The cut on the Higgs resonance evidently
affects those new channels while leaving the LO cross section without
VBS subprocesses untouched. While the total $\order{\alpha^7}$
corrections are only $-6.5\%$ and thus very close to the result in the
Higgs setup, the $\order{\alphas\alpha^6}$
corrections are with $-18\%$ almost as large as in the Higgs setup.

\subsection{Differential distributions}
\label{sec:distributions}
\subsubsection{LO distributions}
\label{sec:LO_distributions}
\begin{figure}
\begin{subfigure}{0.49\textwidth}
\centering
\includegraphics[width=1.\linewidth]{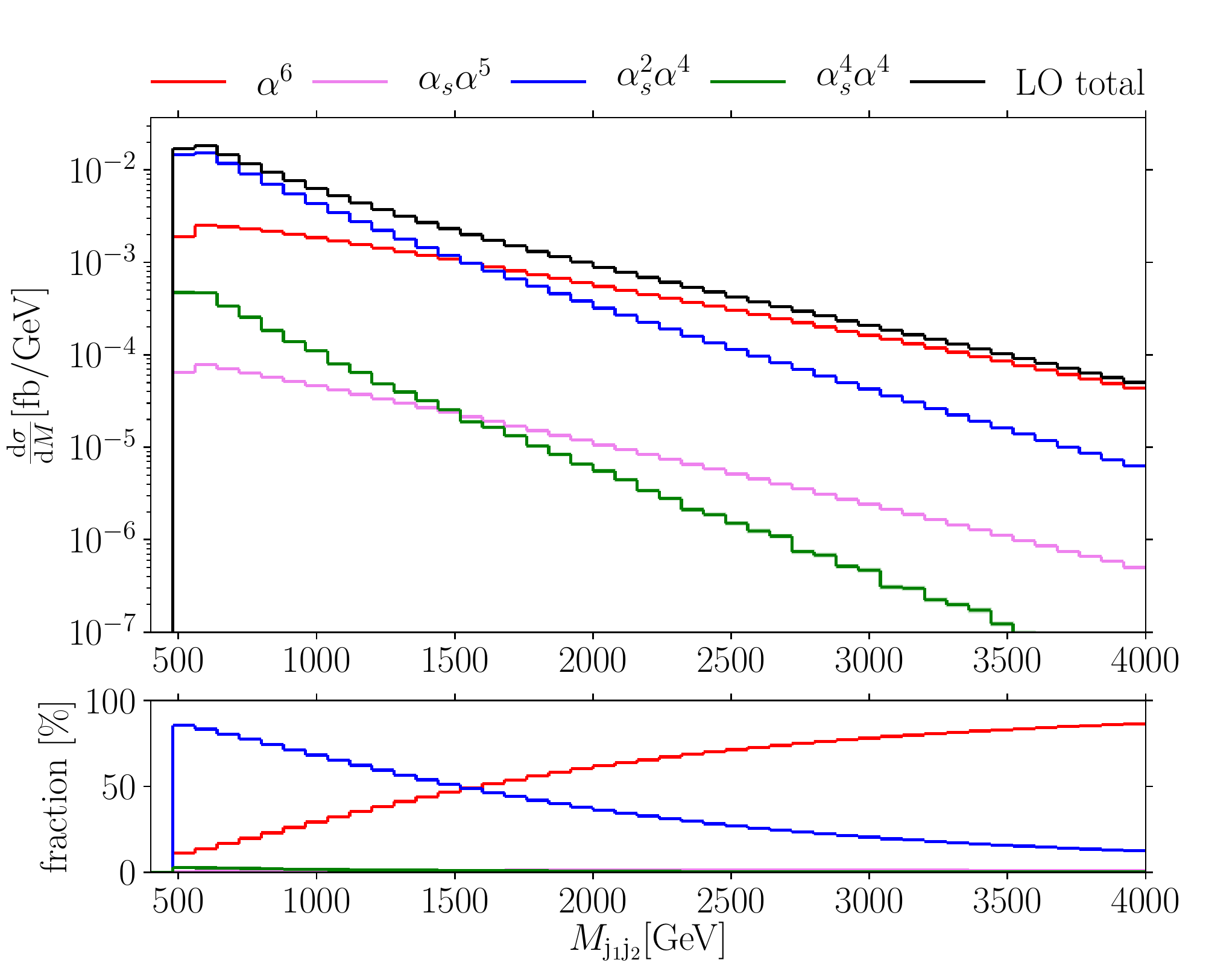}
\end{subfigure}
\begin{subfigure}{0.49\textwidth}
\centering
\includegraphics[width=1.\linewidth]{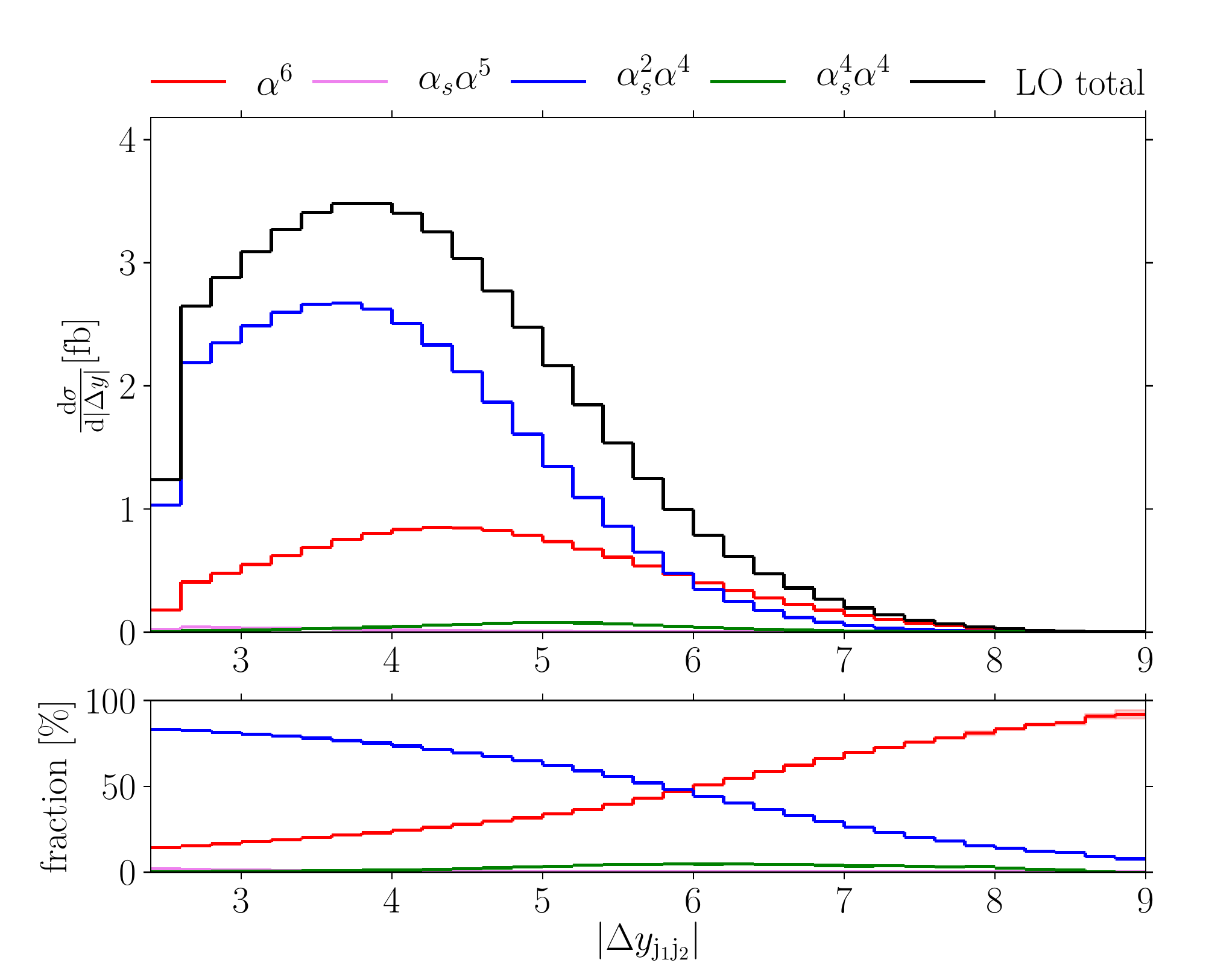}
\end{subfigure}%
\par
\begin{subfigure}{0.49\textwidth}
\centering
\includegraphics[width=1.\linewidth]{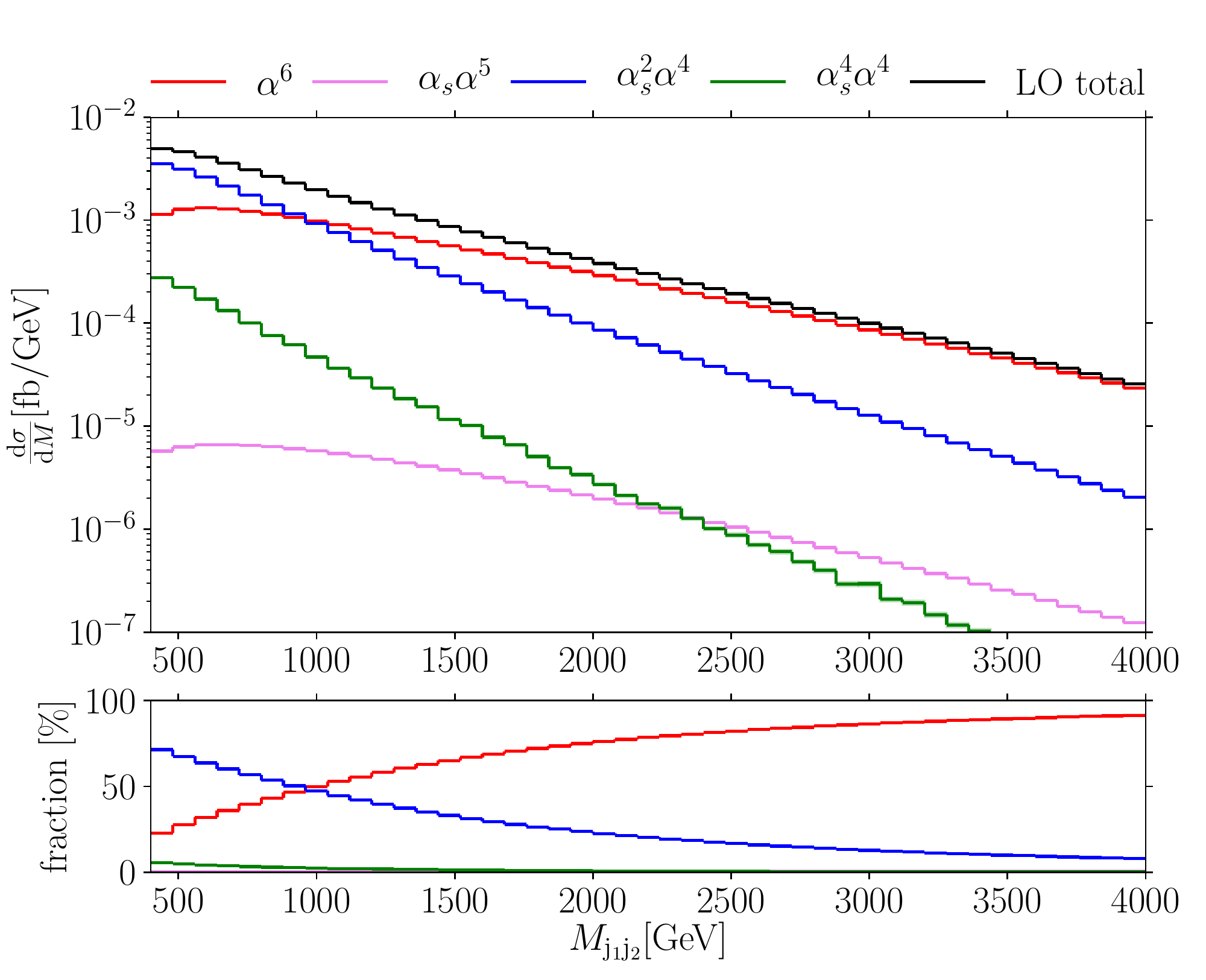}
\end{subfigure}
\begin{subfigure}{0.49\textwidth}
\centering
\includegraphics[width=1.\linewidth]{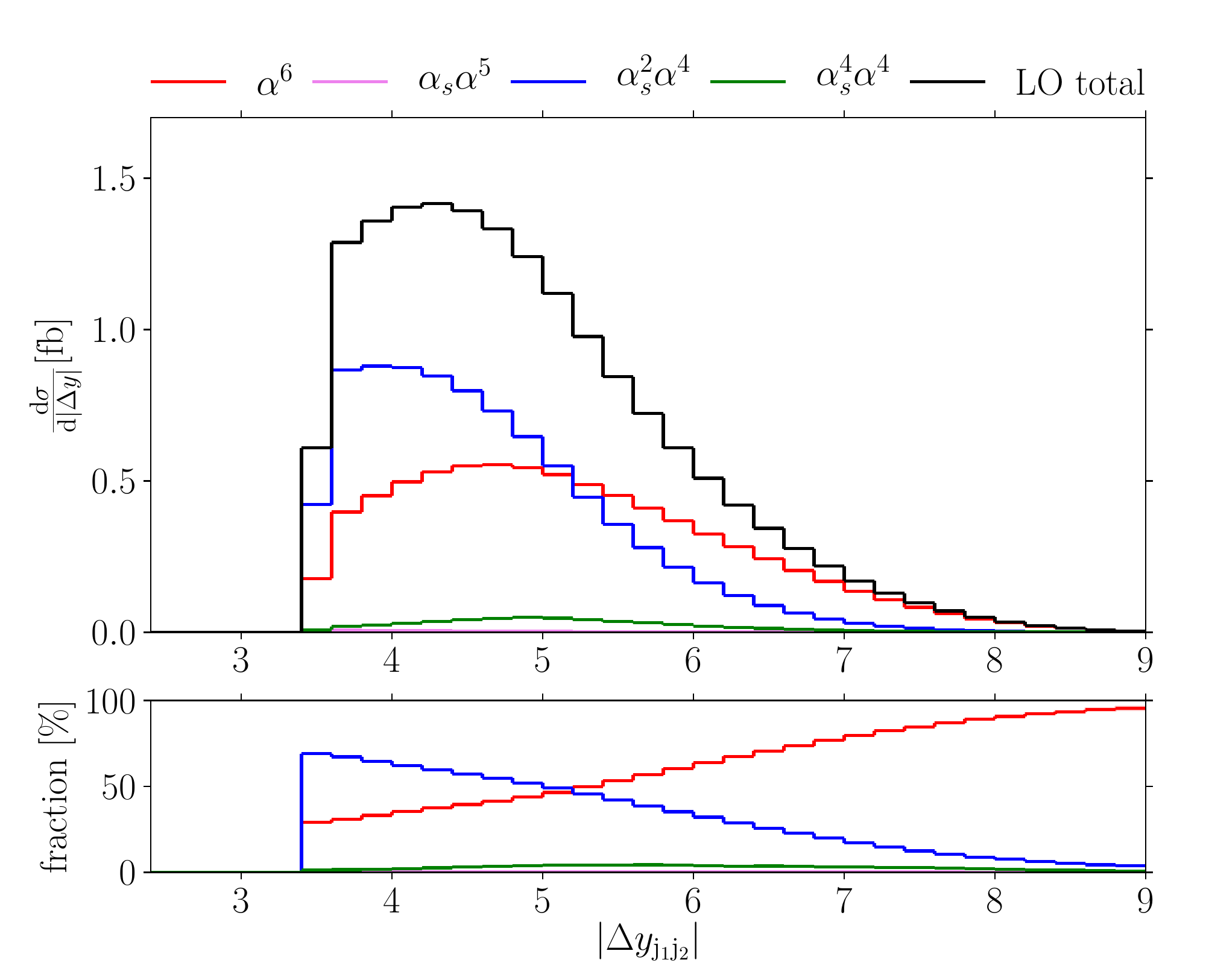}
\end{subfigure}
\caption{Differential distributions at LO in the invariant mass of the
  tagging jets (left) and the rapidity difference between the these
  jets (right) in the VBS (top) and the Higgs setup (bottom). The
  upper panels show the absolute EW contribution at
  $\order{\alpha^6}$, the interference at $\order{\alphas\alpha^5}$,
  the QCD contribution at $\order{\alphas^2\alpha^4}$, the
  loop-induced contribution at $\order{\alphas^4\alpha^4}$, and the
  sum of all contributions. The lower panels show the relative
  contributions normalised to the sum of all contributions. Shaded bands denote integration errors.}
\label{fig:LO_I}
\end{figure}

\begin{figure}
\begin{subfigure}{0.49\textwidth}
\centering
\includegraphics[width=1.\linewidth]{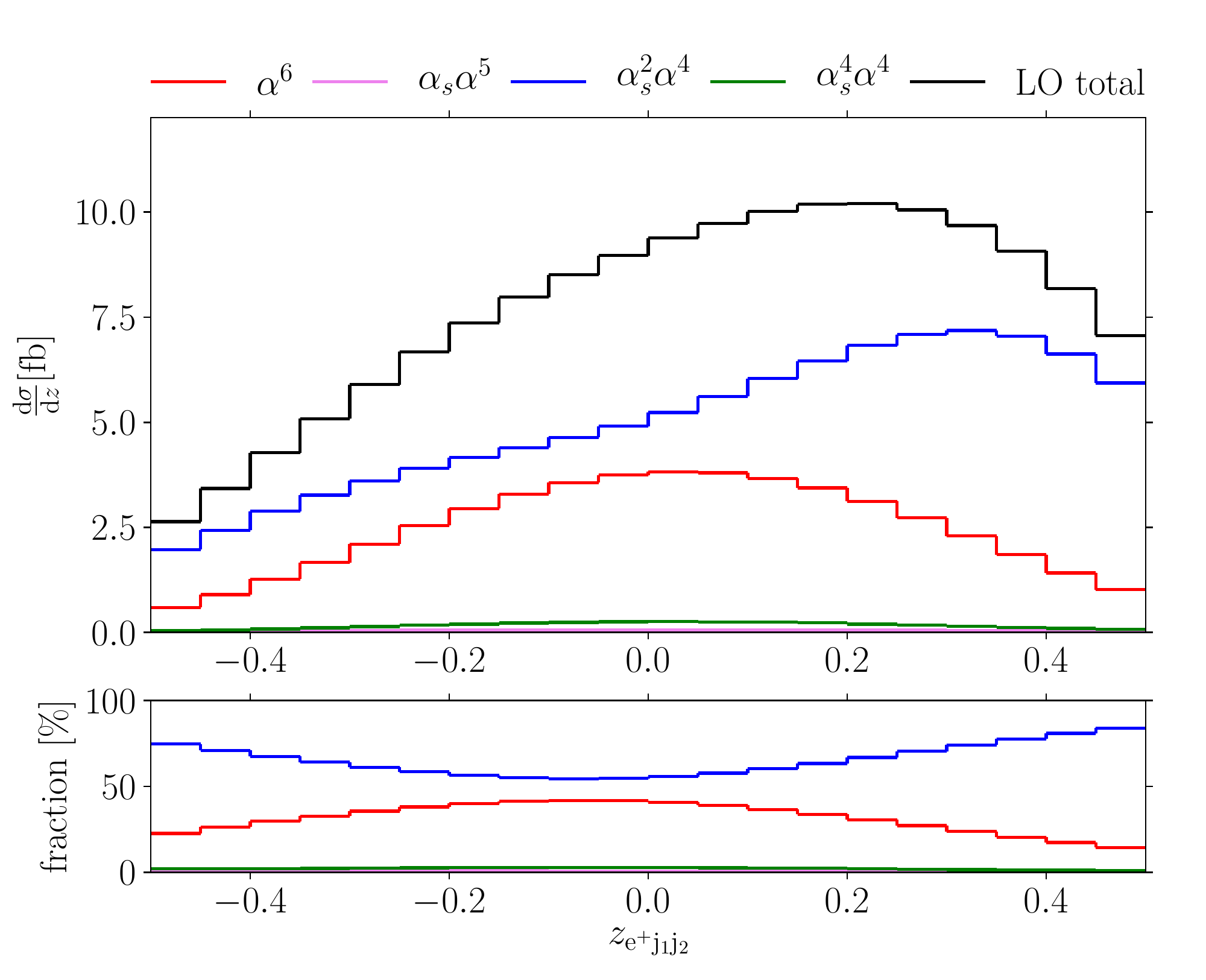}
\label{fig:zejj_VBS}
\end{subfigure}
\begin{subfigure}{0.49\textwidth}
\centering
\includegraphics[width=1.\linewidth]{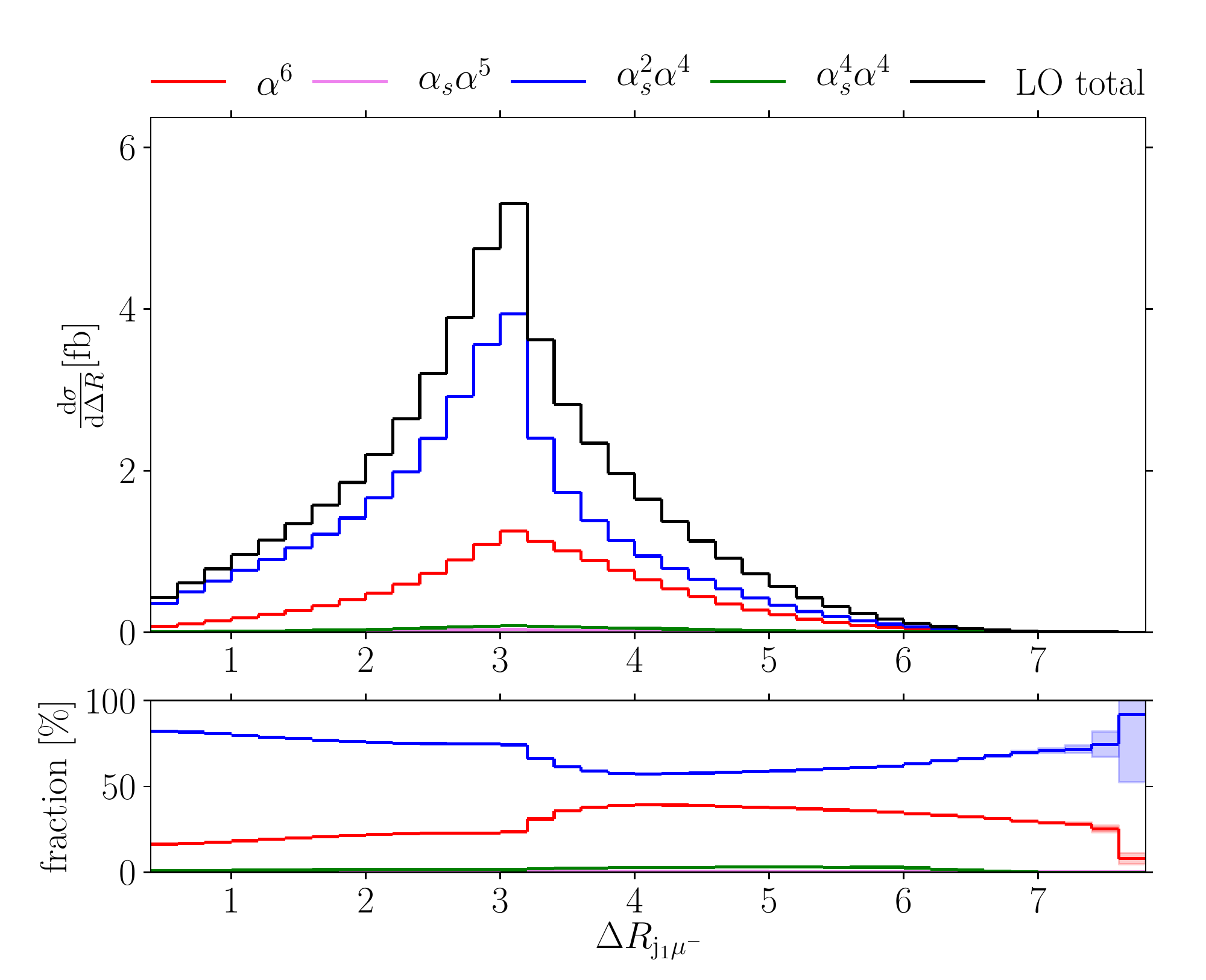}
\label{fig:drjmu_VBS} 
\end{subfigure}
\par
\begin{subfigure}{0.49\textwidth}
\centering
\includegraphics[width=1.\linewidth]{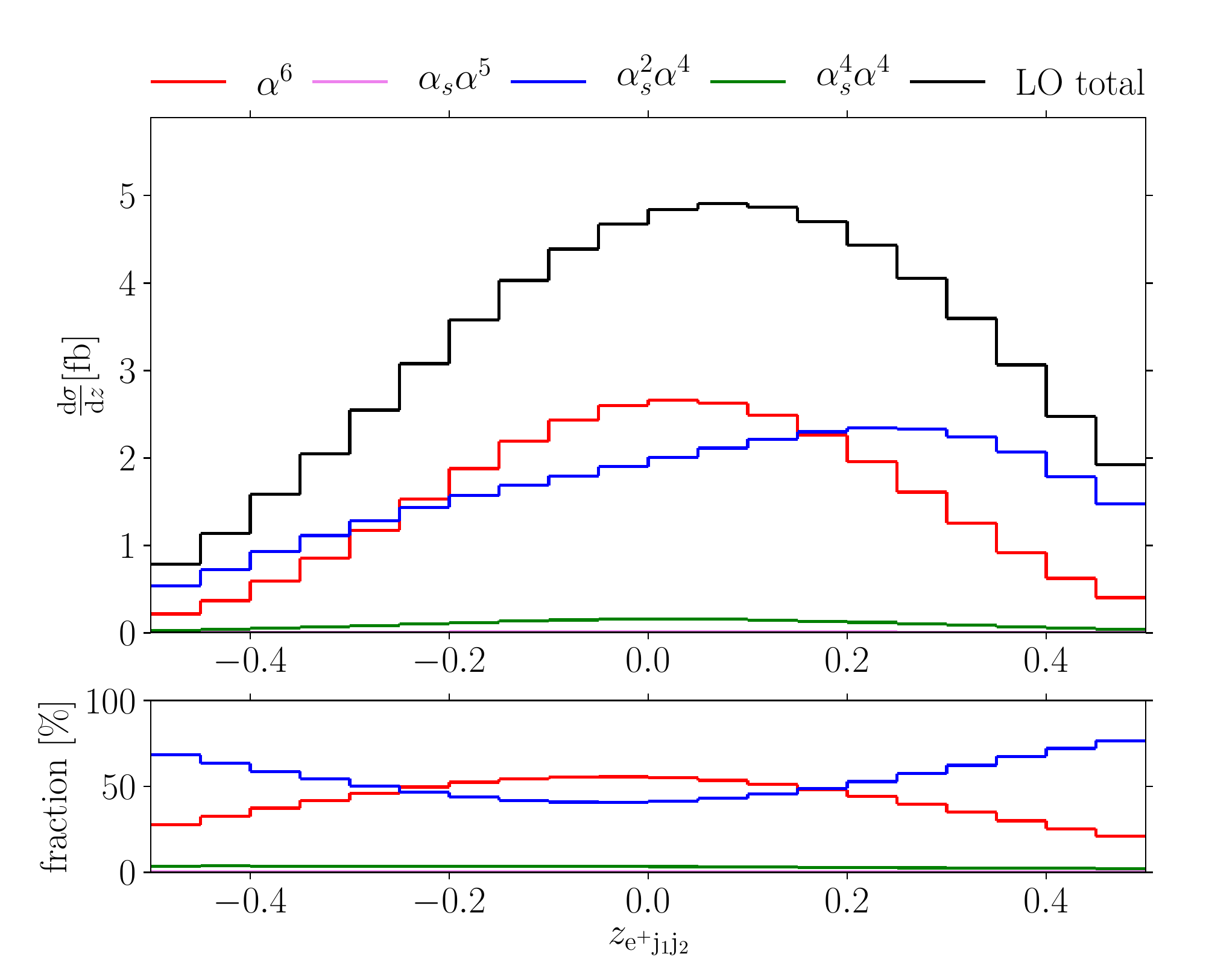}
\label{fig:zejj_Higgs}
\end{subfigure}
\begin{subfigure}{0.49\textwidth}
\centering
\includegraphics[width=1.\linewidth]{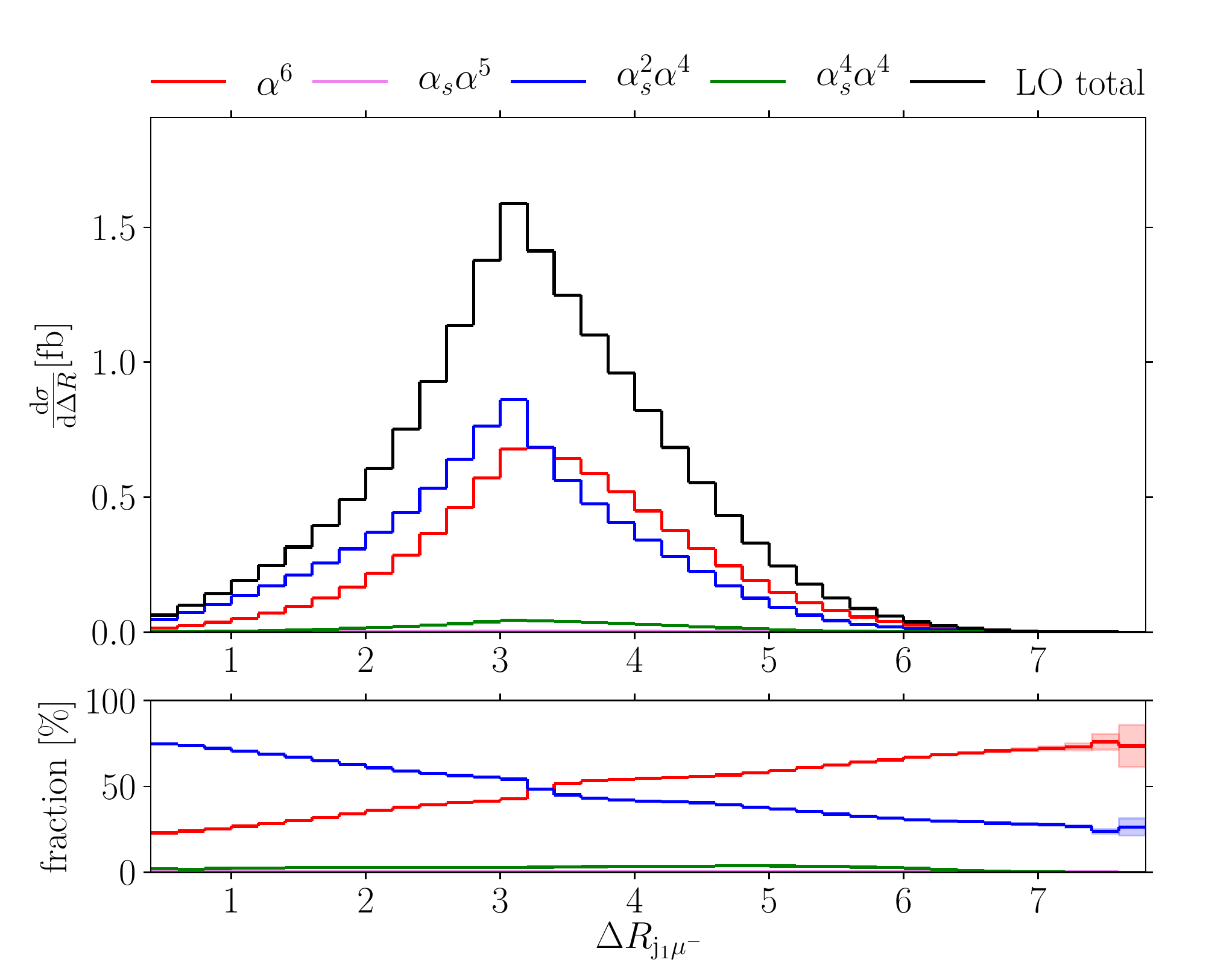}
\label{fig:drjmu_Higgs} 
\end{subfigure}
\caption{Differential distributions at LO in the Zeppenfeld variable
  (left) and the $\Delta R$ separation of the hardest jet and the muon (right) in the VBS (top) and the Higgs setup (bottom). The upper panels show the absolute EW contribution at $\order{\alpha^6}$, the interference at $\order{\alphas\alpha^5}$, the QCD contribution at $\order{\alphas^2\alpha^4}$, the loop-induced contribution at $\order{\alphas^4\alpha^4}$, and the sum of all contributions. The lower panels show the relative contributions normalised to the sum of all contributions. Shaded bands denote integration errors.}
\label{fig:LO_II}
\end{figure}

\begin{figure}
\begin{subfigure}{0.49\textwidth}
\centering
\includegraphics[width=1.\linewidth]{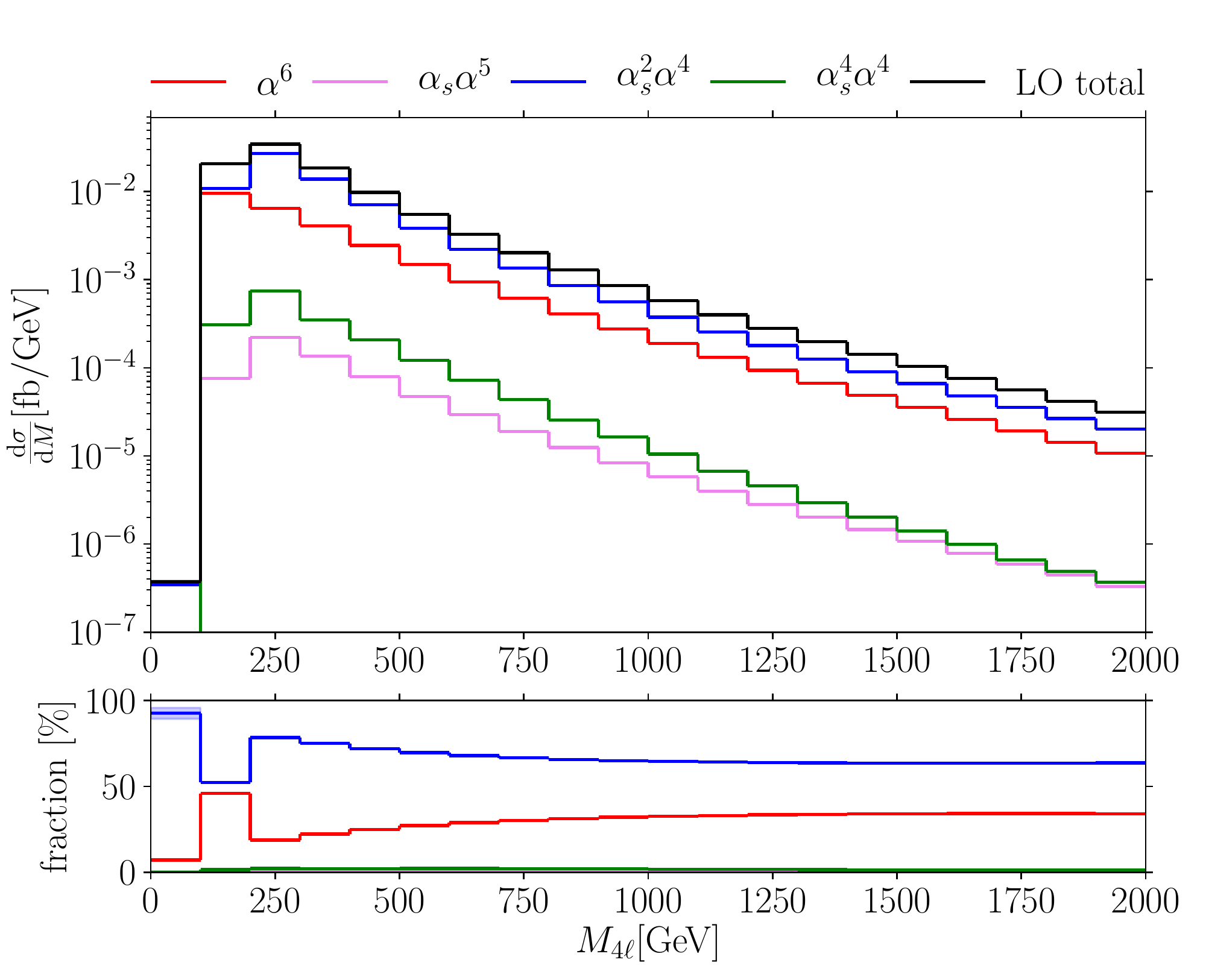}
\end{subfigure}
\begin{subfigure}{0.49\textwidth}
\centering
\includegraphics[width=1.\linewidth]{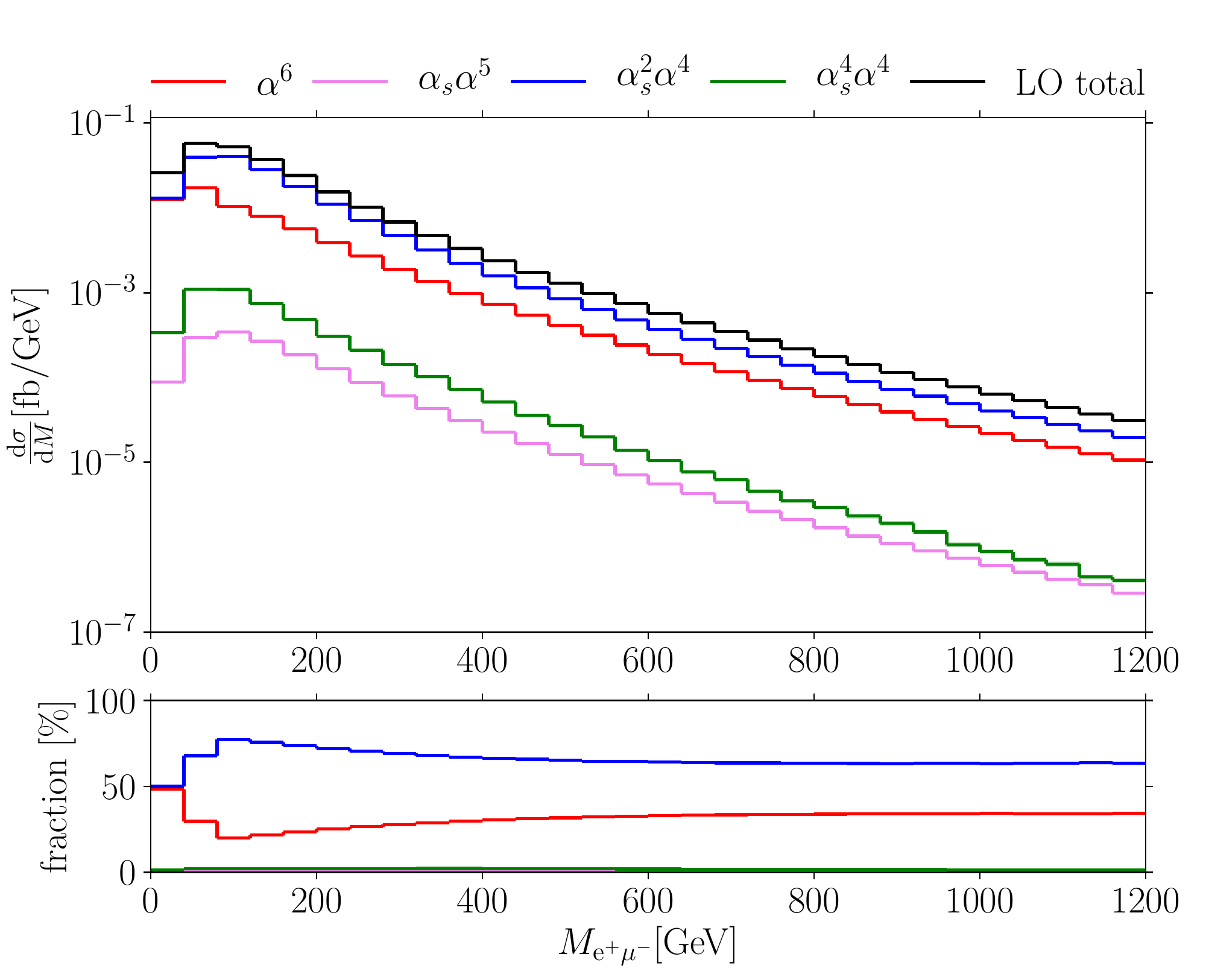}
\end{subfigure}%
\par
\begin{subfigure}{0.49\textwidth}
\centering
\includegraphics[width=1.\linewidth]{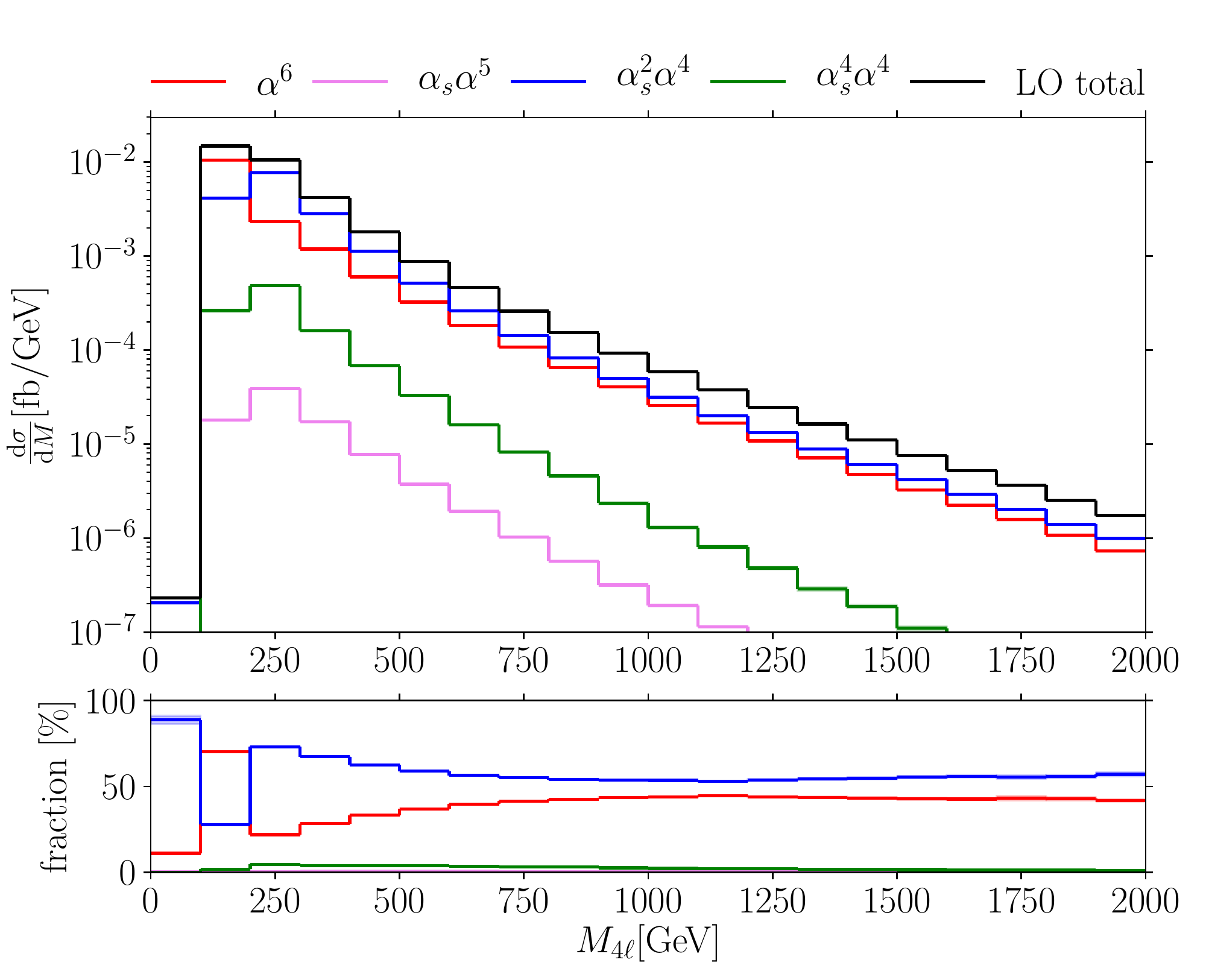}
\end{subfigure}
\begin{subfigure}{0.49\textwidth}
\centering
\includegraphics[width=1.\linewidth]{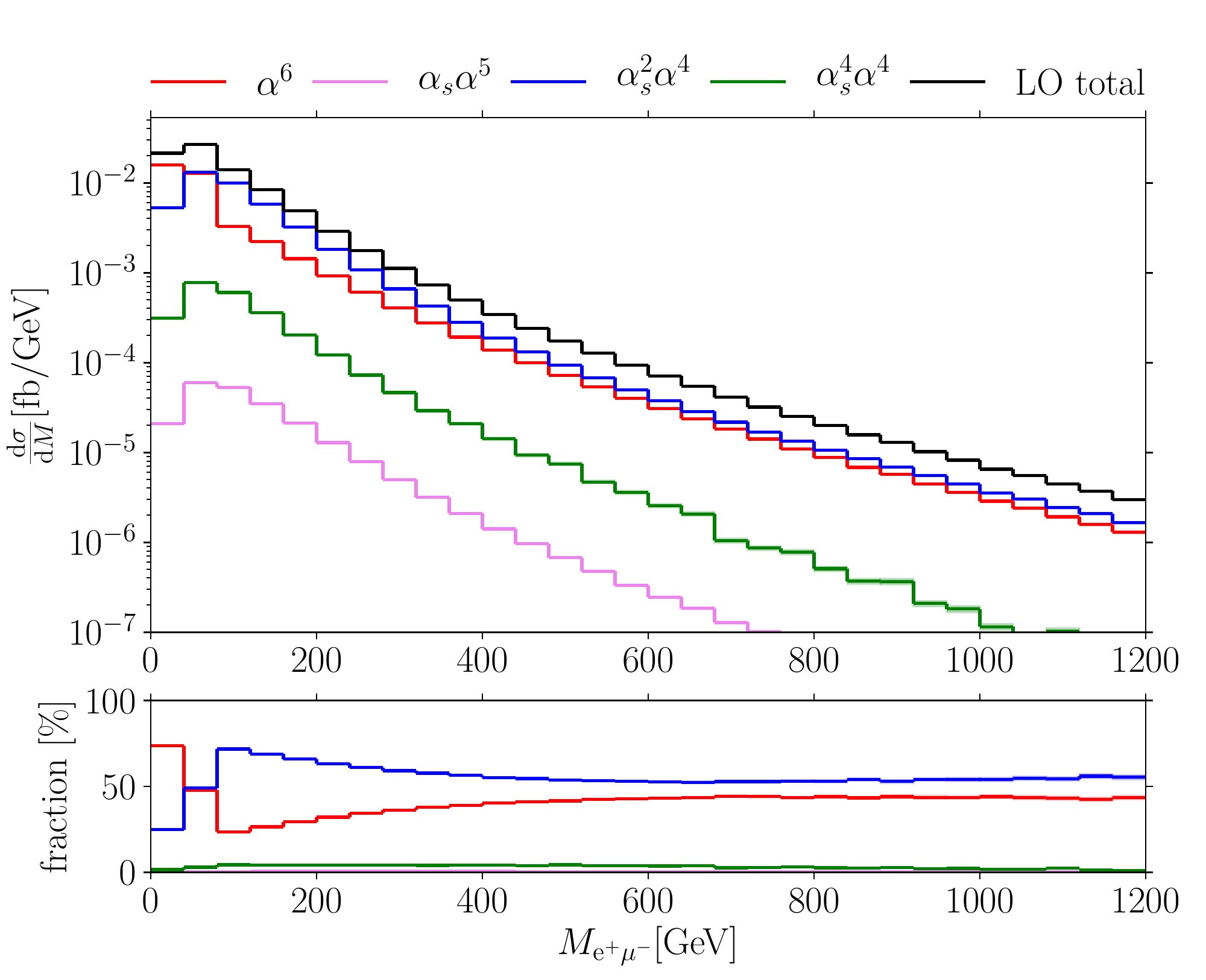}
\end{subfigure}
\caption{Differential distributions at LO in the invariant mass of the
  four final-state leptons (left) and the invariant mass of charged leptons (right) in the VBS setup (top) and the Higgs setup (bottom). The upper panels show the absolute EW contribution at $\order{\alpha^6}$, the interference at $\order{\alphas\alpha^5}$, the QCD contribution at $\order{\alphas^2\alpha^4}$, the loop-induced contribution at $\order{\alphas^4\alpha^4}$, and the sum of all contributions. The lower panels show the relative contribution normalised to the sum. Shaded bands denote integration errors.}
\label{fig:LO_III}
\end{figure}

We present distributions for the VBS and the Higgs setup in
\reffis{fig:LO_I}, \ref{fig:LO_II}, and \ref{fig:LO_III} for all
leading orders $\order{\alpha^6}$, $\order{\alphas\alpha^5}$,
$\order{\alphas^2\alpha^4}$, and $\order{\alphas^4\alpha^4}$ and their
sum as absolute predictions in the upper panels and their relative
contribution to the sum in the lower ones. The upper two diagrams
always show the results for the VBS and the lower ones those for the
Higgs setup. We note that the interference contribution of order
$\order{\alphas\alpha^5}$ is not only negligible in the total cross
section but also in most of the distributions for both setups.

In the left panels of \reffi{fig:LO_I} we show the distribution in the
invariant mass of the two jets. 
While the QCD contribution is dominant
at low invariant masses, the EW contribution starts to exceed its QCD
background at $M_{\Pj_1\Pj_2} \approx 1500\GeV$ in the VBS and at
$M_{\Pj_1\Pj_2} \approx 1000\GeV$ in the Higgs setup. 
For $M_{\Pj_1\Pj_2}> 3000\GeV$, the EW signal contributes more than $80\%$
and  $90\%$ in the VBS and Higgs setups, respectively. The loop-induced
contribution is at the level of $2.5\%$ at low invariant masses for
the VBS and $5\%$ for the Higgs setup, but drops steeply even below
the interference 
contribution for $M_{\Pj_1\Pj_2} > 1500\GeV$ in the VBS setup and
$M_{\Pj_1\Pj_2} > 2000\GeV$ in the Higgs setup.

Another variable to separate EW contributions from QCD contributions
is the rapidity difference between the jets. We show the corresponding
distribution in the right panels of \reffi{fig:LO_I}.  Jets with high
rapidity separation are a typical VBS signature, and we observe the
dominance of the EW contribution in the expected phase-space region
with a rapidity difference larger than $\Delta y_{\Pj_1\Pj_2} \approx
6$ in the VBS and $\Delta y_{\Pj_1\Pj_2} \approx 5$ in the Higgs
setup. The loop-induced contribution shows a maximum with about
$4.5\%$ at $\Delta y_{\Pj_1\Pj_2} \approx 6$ and $\Delta y_{\Pj_1\Pj_2} \approx 5.5$ 
in the VBS setup and Higgs setup, respectively.

A third distribution that is regularly used in VBS physics is shown in the left panels of
\reffi{fig:LO_II}. 
It uses the Zeppenfeld variable of the positron,
$z_{\Pe^+\Pj_1\Pj_2}$, defined in Eq.~\refeq{eq:def_zepp}, which gives
information about the rapidity of the leptons compared to those of the jets.
In agreement with the expectation that the lepton rapidity is between
the jet rapidities in the EW contribution, we recognise a maximum in
this contribution at $z_{\Pe^+\Pj_1\Pj_2} \approx 0$, which corresponds to $y_{\Pe^+}
\approx ({y_{\Pj_1} + y_{\Pj_2}})/{2}$, \ie the rapidity of the lepton
being the arithmetical mean of the two jet rapidities. The QCD
contribution, on the other hand, exhibits a maximum near $z_{\Pe^+\Pj_1\Pj_2} \approx
0.3$.

The distribution in the $\Delta R_{\Pj_1\mu^-}$ separation of the
hardest jet and the muon, in which the difference between the VBS and
the Higgs setup becomes visible as well, is presented in the right panels of
\reffi{fig:LO_II}. In the Higgs setup, the
relative EW contribution grows constantly and exceeds the QCD
contribution at $\Delta R_{\Pj_1\mu^-} \approx 3.5$, whereas in the VBS setup, the
relative EW contribution has a
maximum at $\Delta R_{\Pj_1\mu^-} \approx 4$ and decreases again. The enhancement
of the EW contribution leads to a more symmetric distribution about
$\pi$ in the Higgs setup. We verified that the main difference in the shape of the
$\order{\alphas^2\alpha^4}$ contributions between the VBS and Higgs setups is due to
the cut \refeq{eq:cut_zepp} on the Zeppenfeld variable. Since the QCD
contributions tend to cause more central jets, they favour events with
small rapidity differences between the leptons and the jets, which
explains the enhancement of these contributions for small $\Delta
R_{\Pj_1\mu^-}$.

In \reffi{fig:LO_III} (left panels) we present the distributions in the
(non-measurable) four-lepton invariant mass $M_{4\Pl}$ at LO for the VBS (top)
and the Higgs setup (bottom) to stress the impact of the Higgs
resonance. 
We find a single bin at $100 \GeV < M_{4\Pl} < 200\GeV$, \ie
the one containing the Higgs resonance, in
which the EW contribution is comparable to the QCD one (in the VBS
setup) or exceeds the QCD one (in the Higgs setup). A similar but
smaller effect of the Higgs resonance can
also be seen in the measurable invariant mass of the
two-charged-lepton system (right panels), where we find two bins at
very low two-lepton invariant masses in which the EW contribution is
with $50\%$ almost comparable to the QCD contribution in the VBS setup
and becomes even dominant with approximately $75\%$ in the Higgs
setup, until it drops very quickly to $20\%$ above $M_{\Pe^+\mu^-} =
80\GeV$, from which it rises again to a very flat second maximum near
$M_{\Pe^+\mu^-} \approx 1000\GeV.$ We note that the Higgs resonance does
not have a visible impact on the loop-induced contribution,
although it is present therein.

\subsubsection{NLO distributions}
\begin{figure}
\begin{subfigure}{0.49\textwidth}
\centering
\includegraphics[width=1.\linewidth]{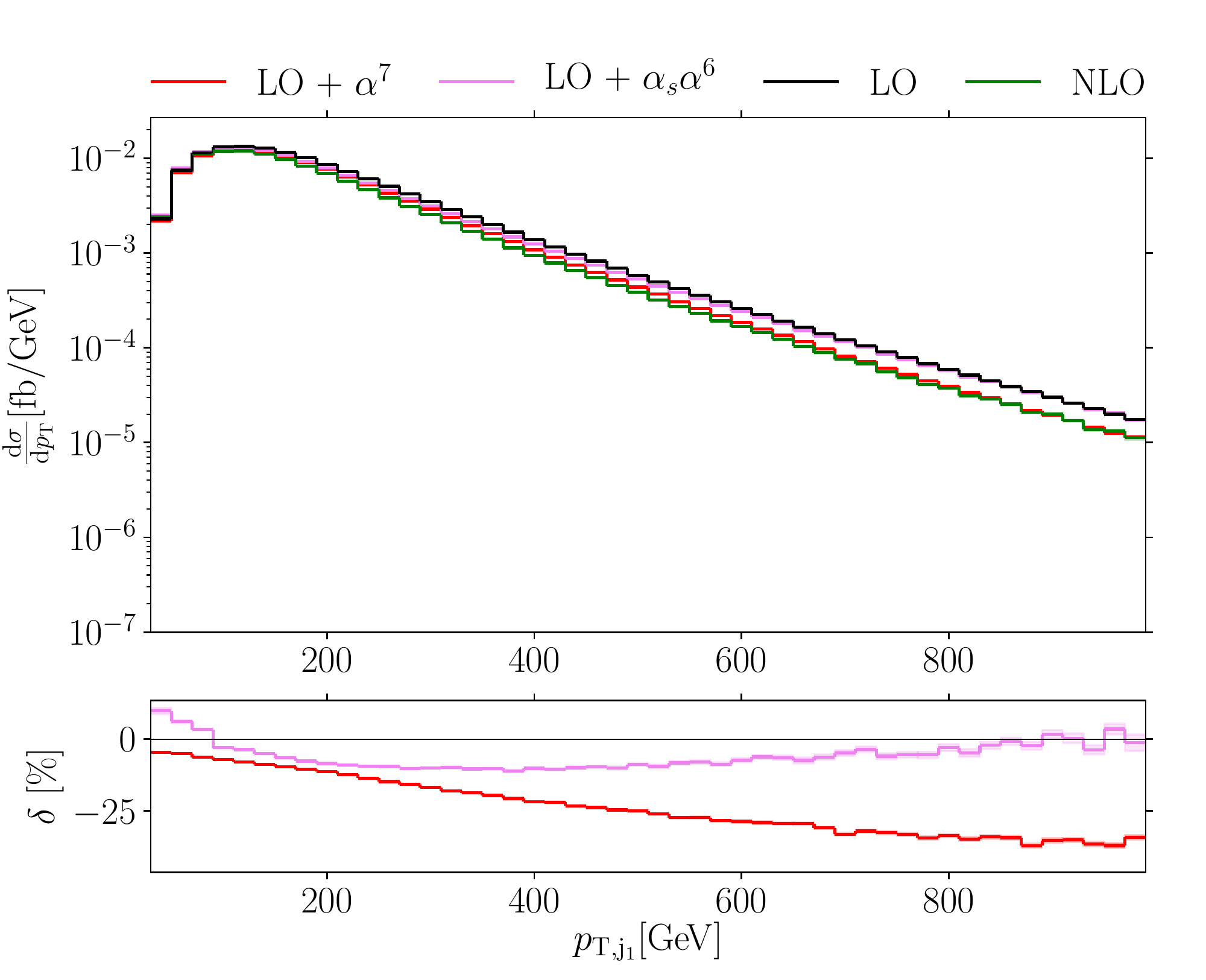}
\end{subfigure}
\begin{subfigure}{0.49\textwidth}
\centering
\includegraphics[width=1.\linewidth]{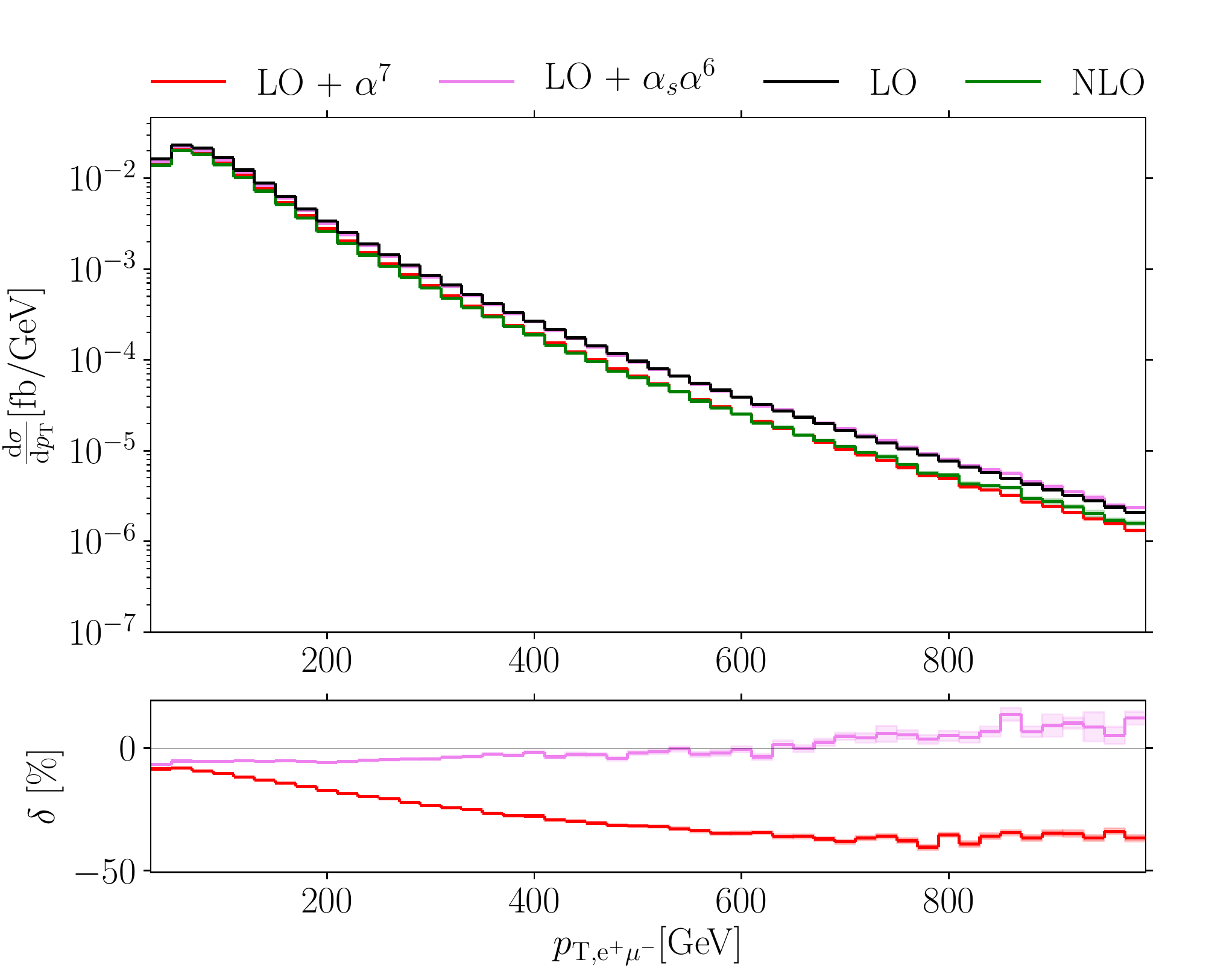}
\end{subfigure}%
\par
\begin{subfigure}{0.49\textwidth}
\centering
\includegraphics[width=1.\linewidth]{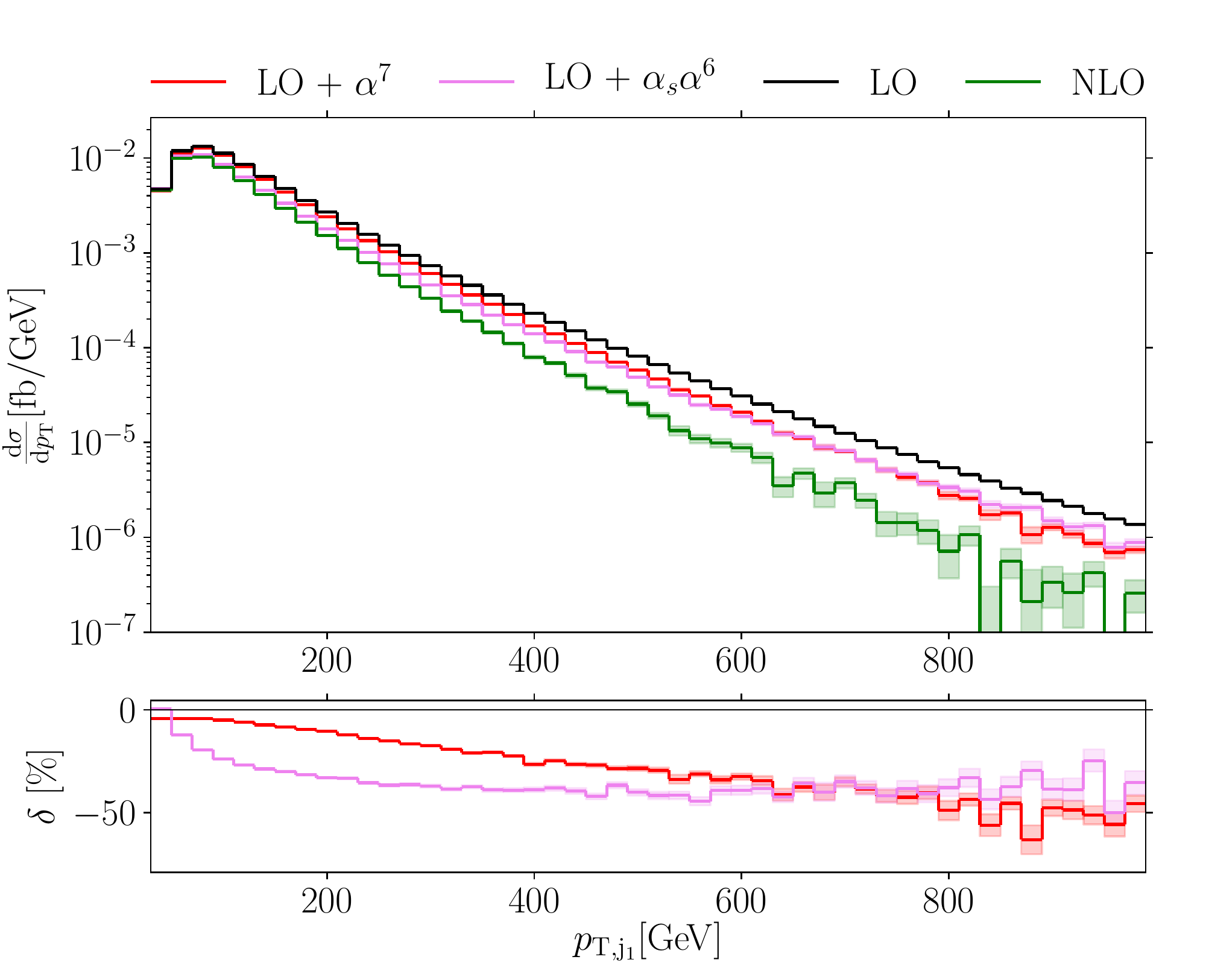}
\end{subfigure}
\begin{subfigure}{0.49\textwidth}
\centering
\includegraphics[width=1.\linewidth]{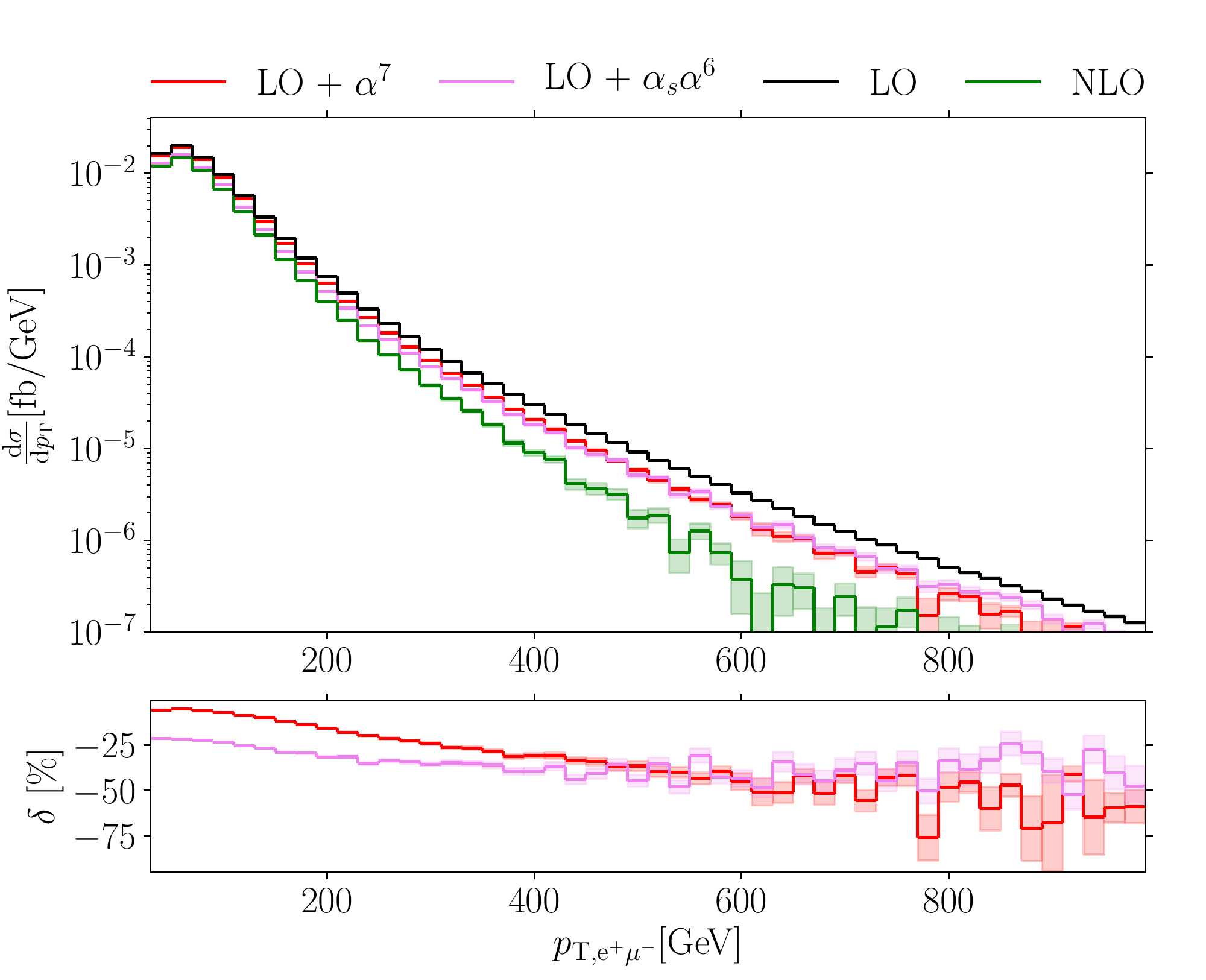}
\end{subfigure}
\caption{Differential distributions at NLO in the transverse momentum
  of the hardest jet (left) and the two charged leptons (right) in the
  VBS (top) and the Higgs setup (bottom). The upper panels show the
  absolute EW contribution at $\order{\alpha^6}$, the sum of the EW
  contribution and the EW corrections of $\order{\alpha^7}$, the sum
  of the EW contribution and the QCD corrections of
  $\order{\alphas\alpha^6}$, and the sum of all three contributions. The
  lower panels show the relative EW and QCD corrections normalised to the LO EW contribution. Shaded bands denote integration errors.}
\label{fig:NLO_I}
\end{figure}

\begin{figure}
\begin{subfigure}{0.49\textwidth}
\centering
\includegraphics[width=1.\linewidth]{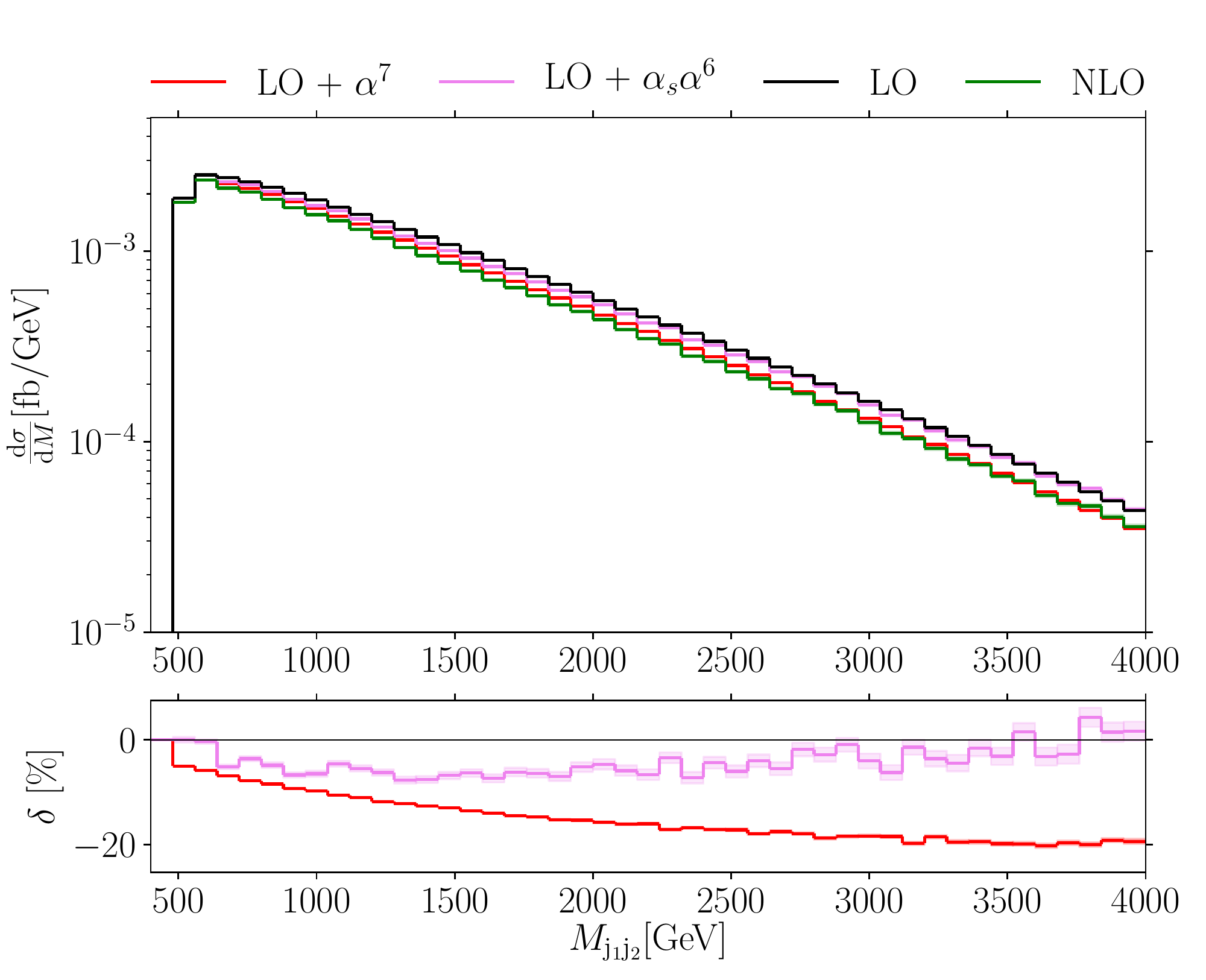}
\end{subfigure}
\begin{subfigure}{0.49\textwidth}
\centering
\includegraphics[width=1.\linewidth]{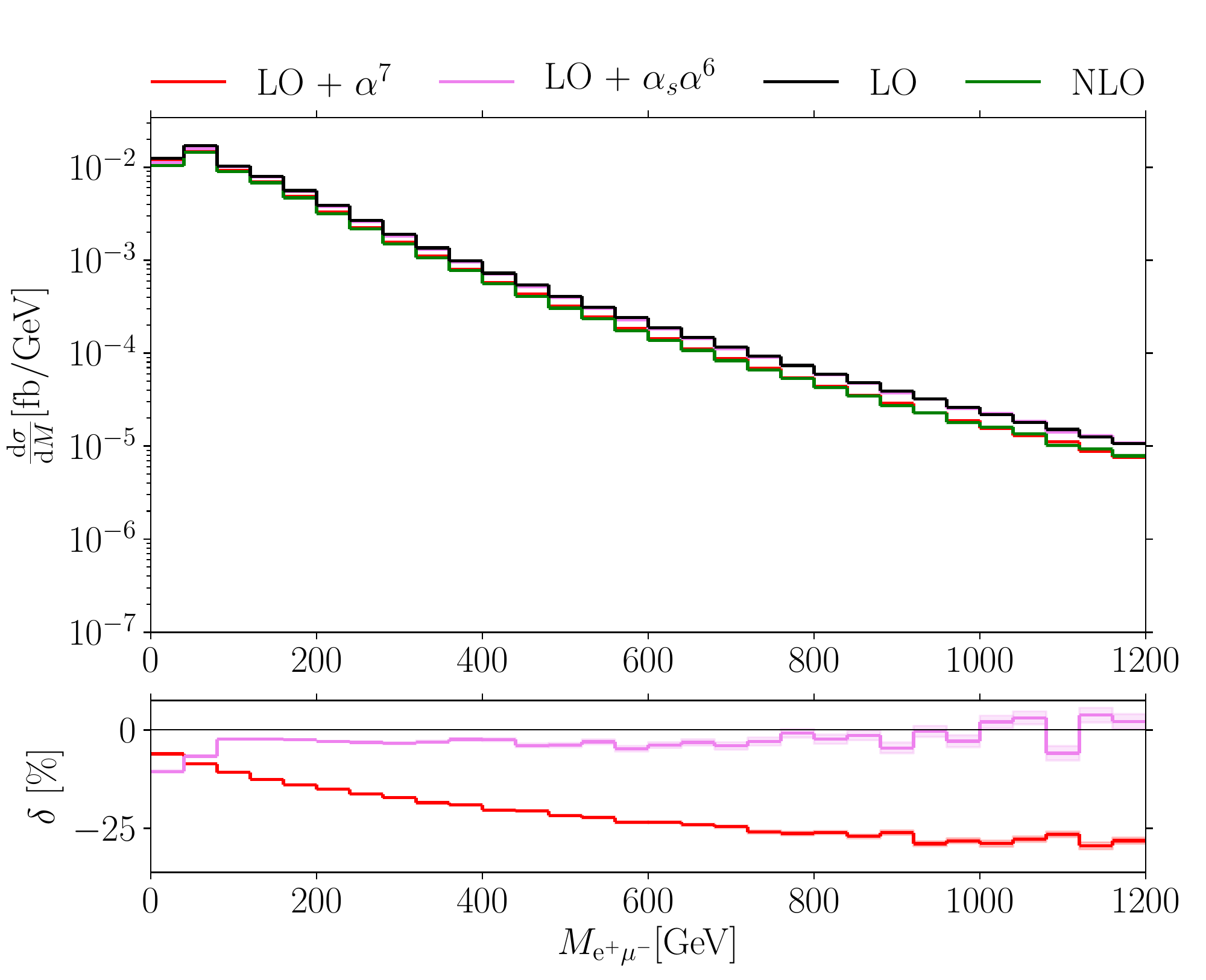}
\end{subfigure}
\par
\begin{subfigure}{0.49\textwidth}
\centering
\includegraphics[width=1.\linewidth]{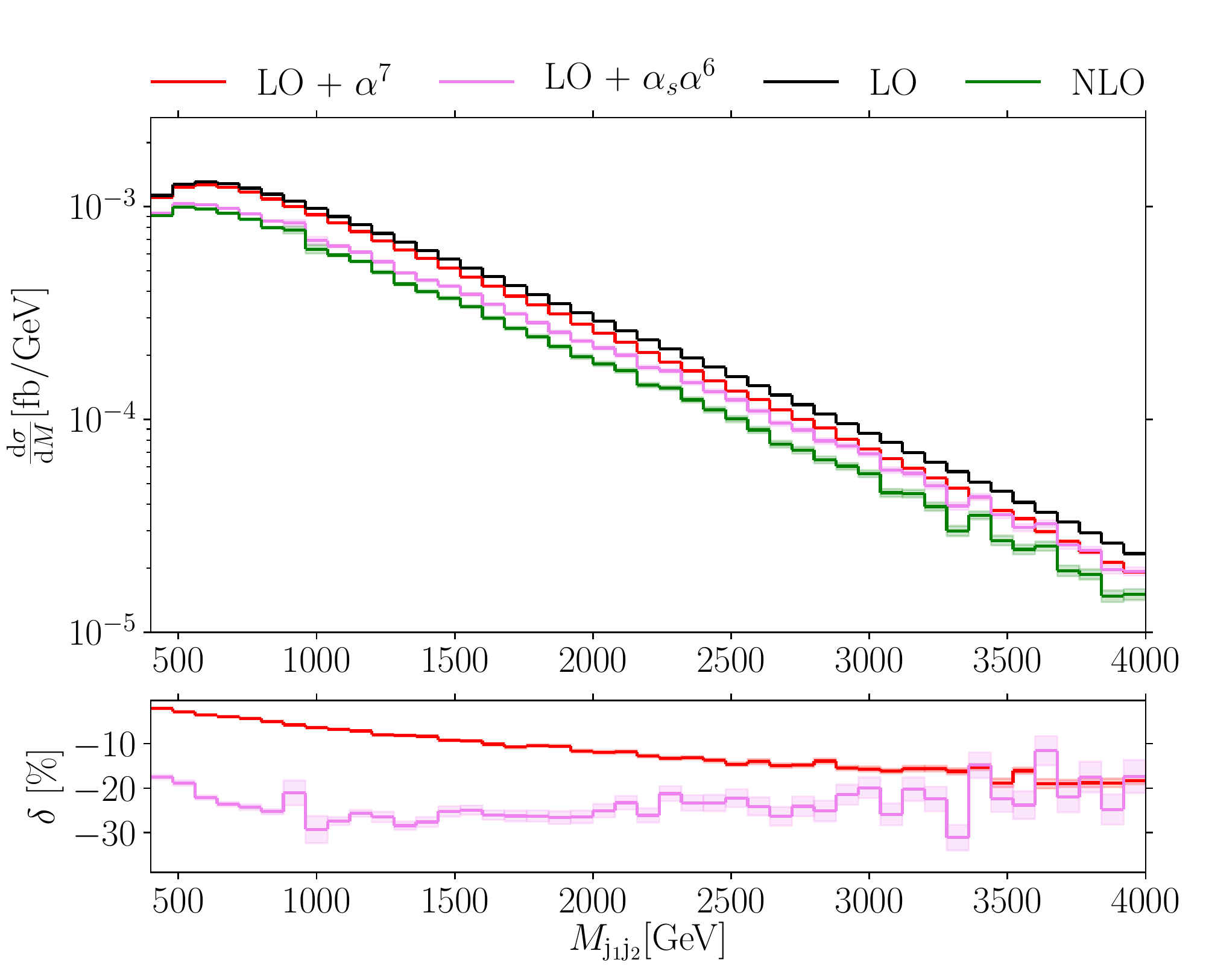}
\end{subfigure}
\begin{subfigure}{0.49\textwidth}
\centering
\includegraphics[width=1.\linewidth]{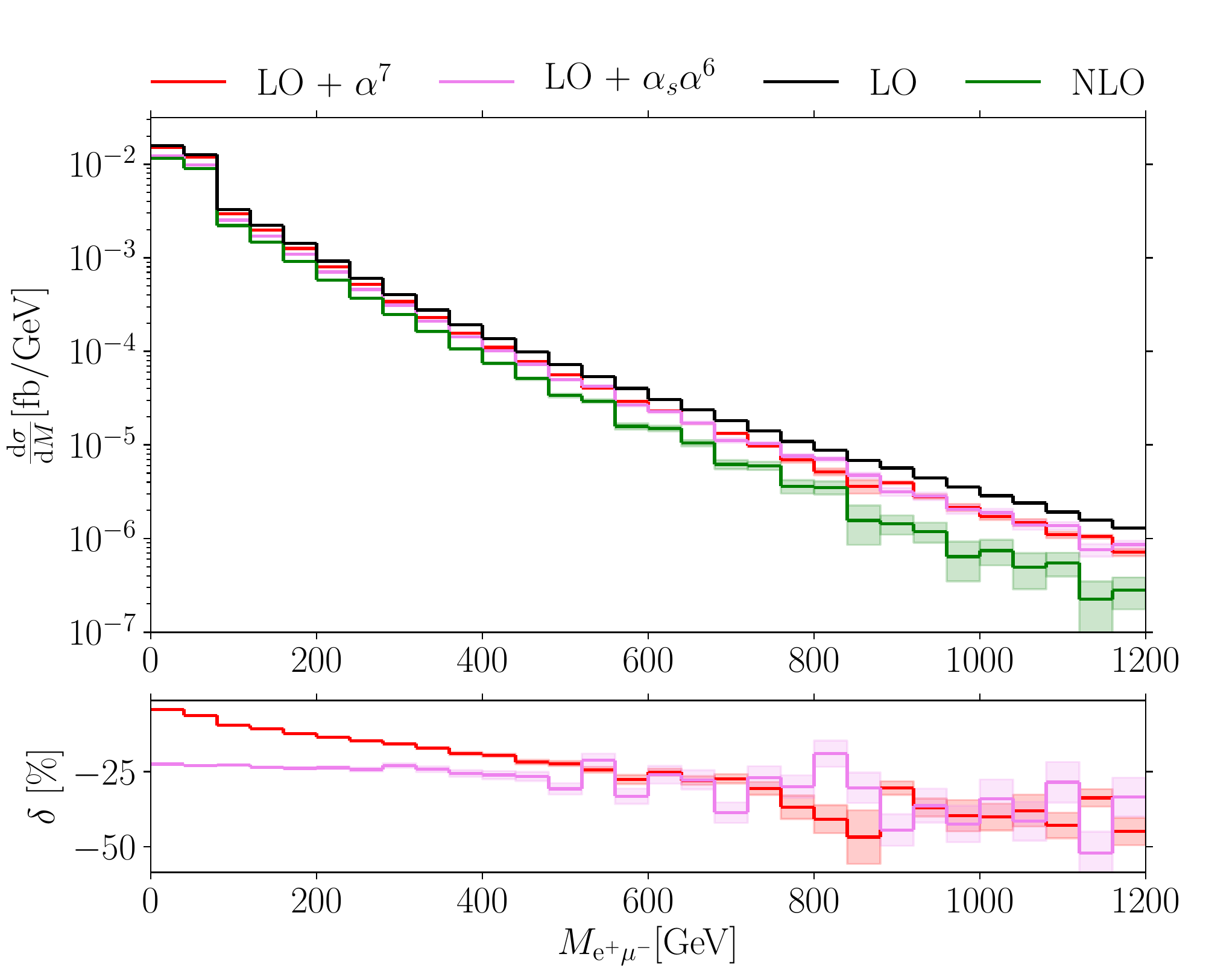}
\end{subfigure}
\caption{Differential distributions at NLO in the invariant mass of
  the two tagging jets (left) and the charged leptons (right) in the
  VBS (top) and the Higgs setup (bottom). The upper panels show the
  absolute EW contribution at $\order{\alpha^6}$, the sum of the EW
  contribution and the EW corrections of $\order{\alpha^7}$, the sum
  of the EW contribution and the QCD corrections of
  $\order{\alphas\alpha^6}$, and the sum of all three contributions. The
  lower panels show the relative EW and QCD corrections normalised to the LO EW contribution. Shaded bands denote integration errors.}
\label{fig:NLO_II}
\end{figure}

Next, we present NLO distributions for the VBS setup and for the Higgs
setup in \reffis{fig:NLO_I} and \ref{fig:NLO_II}. In those plots, we
use $\order{\alpha^6}$ as LO baseline and show results for
$\order{\alpha^6} + \order{\alpha^7}$, $\order{\alpha^6} +
\order{\alphas\alpha^6}$, and $\order{\alpha^6} + \order{\alpha^7} +
\order{\alphas\alpha^6}$ as absolute quantities in the upper panels
and as corrections relative to the $\order{\alpha^6}$ in the lower
ones.  We note that the $\order{\alpha^6}$ contribution is not
  dominant in most regions of phase space. Consequently, the impact of
  the relative corrections shown in this section on the distributions
  for the complete processes $\Pp\Pp \to \Pe^+ \nu_\Pe \mu^- \bar
  \nu_\mu \Pj\Pj+X$ should be gauged by the relative fraction presented
  in \refse{sec:LO_distributions}. On the other hand, the relative
  corrections shown below are directly relevant for the EW production processes
  typically isolated in experimental measurements.

We begin our discussion by considering distributions in
energy-dependent variables.  Specifically, we present distributions in
the transverse momenta of the hardest jet and the two-charged-lepton
system in \reffi{fig:NLO_I}, as well as distributions in the invariant
mass of the two hardest jets and the charged-lepton pair in
\reffi{fig:NLO_II}. All of these distributions in both setups show the
typical behaviour of EW corrections that grow with the energy scale
and reach $-40\%$ to $-50\%$ in the tails of the distributions in our
depicted regions. We again emphasise that the EW LO cross section
peaks at two-lepton invariant masses below $\MH$, where the EW corrections
contribute only a few percent.
This becomes especially clear in the Higgs setup, being in accordance
with our findings for the Higgs-resonance contribution to the fiducial
cross section in \refta{tab:resonance}, where we also found small EW
corrections in this phase-space region. While the distribution in the
invariant mass of the positron--muon system drops by almost an order
of magnitude above the Higgs resonance, the relative corrections
behave smoothly there.

The behaviour of the QCD corrections differs in the VBS and the Higgs
setup. In the VBS setup, the QCD corrections for the distribution in
the transverse momentum of the hardest jet (\reffi{fig:NLO_I} left)
are positive at low values $\ptsub{\Pj_1} < 100 \GeV$, reach a minimum
at $\ptsub{\Pj_1} \approx 440\GeV$, and start growing again towards
zero for large $\ptsub{\Pj_1}$. In contrast, the
QCD corrections in the Higgs setup are monotonically falling very fast
and stay roughly constant above $\ptsub{\Pj_1} \approx 400\GeV$ at a level of
$-40\%$.  This can be explained by the presence of the jet veto, which
allows only for radiation of gluons that are soft or parallel to the
beam axis. If already the hardest jet is very soft, the jet veto
does not apply to any other jet. On the other hand, if the hardest jet
is very hard, other final-state particles are also likely to have high
transverse momenta and the jet veto cuts away a fraction of events
independently of the actual transverse momentum of the hardest jet.

The QCD corrections to the distribution in the transverse momentum of
the two leptons (\reffi{fig:NLO_I} right) show a different behaviour
for the VBS and the Higgs setup. While in the VBS setup, the
corrections are increasing and even turn positive with higher values
of $\ptsub{\Pe^+\mu^-}$, in the Higgs setup they grow negatively
and fluctuate around $-40\%$ for large $\ptsub{\Pe^+\mu^-}$.
Again, the behaviour in the Higgs setup can be
explained with the jet veto, but the transverse momenta of the leptons
are not that tightly correlated to the one of the softest jet as the
one of the hardest jet. NLO corrections to other transverse-momentum
distributions of jets and leptons are qualitatively similar.

For the distributions in the invariant mass of the two tagging jets
(\reffi{fig:NLO_II} left), the relative QCD corrections in the VBS and
the Higgs setup are only quantitatively different; they are almost
constant over a wide range of $M_{\Pj_1\Pj_2}$, before statistical
fluctuations in the tails do not allow any further statement in these
regions. The jet veto leads merely to a constant shift of the
corrections, since transverse momentum and invariant mass are
uncorrelated observables. 

For invariant masses above $100\GeV$, the NLO corrections to the
distributions in the invariant mass of the charged-lepton pair
(\reffi{fig:NLO_II} right) show a similar behaviour to those in the
corresponding transverse-momentum distribution. While the NLO QCD
corrections reach $-10\%$ for $M_{\Pe^+\mu^-}<100\GeV$, they are close
to zero for higher invariant masses in the VBS setup.  

\begin{figure}
\begin{subfigure}{0.49\textwidth}
\centering
\includegraphics[width=1.\linewidth]{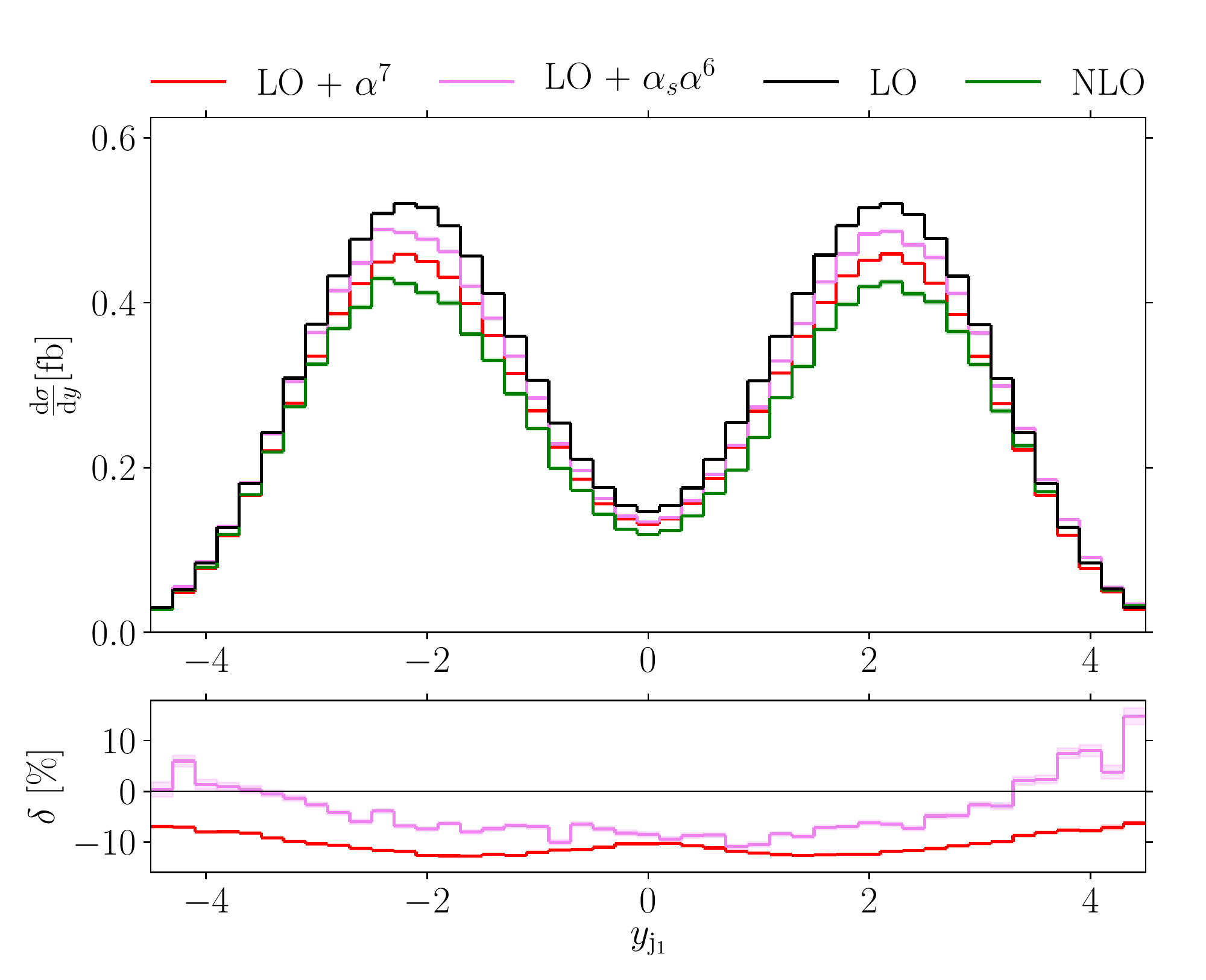}
\end{subfigure}
\begin{subfigure}{0.49\textwidth}
\centering
\includegraphics[width=1.\linewidth]{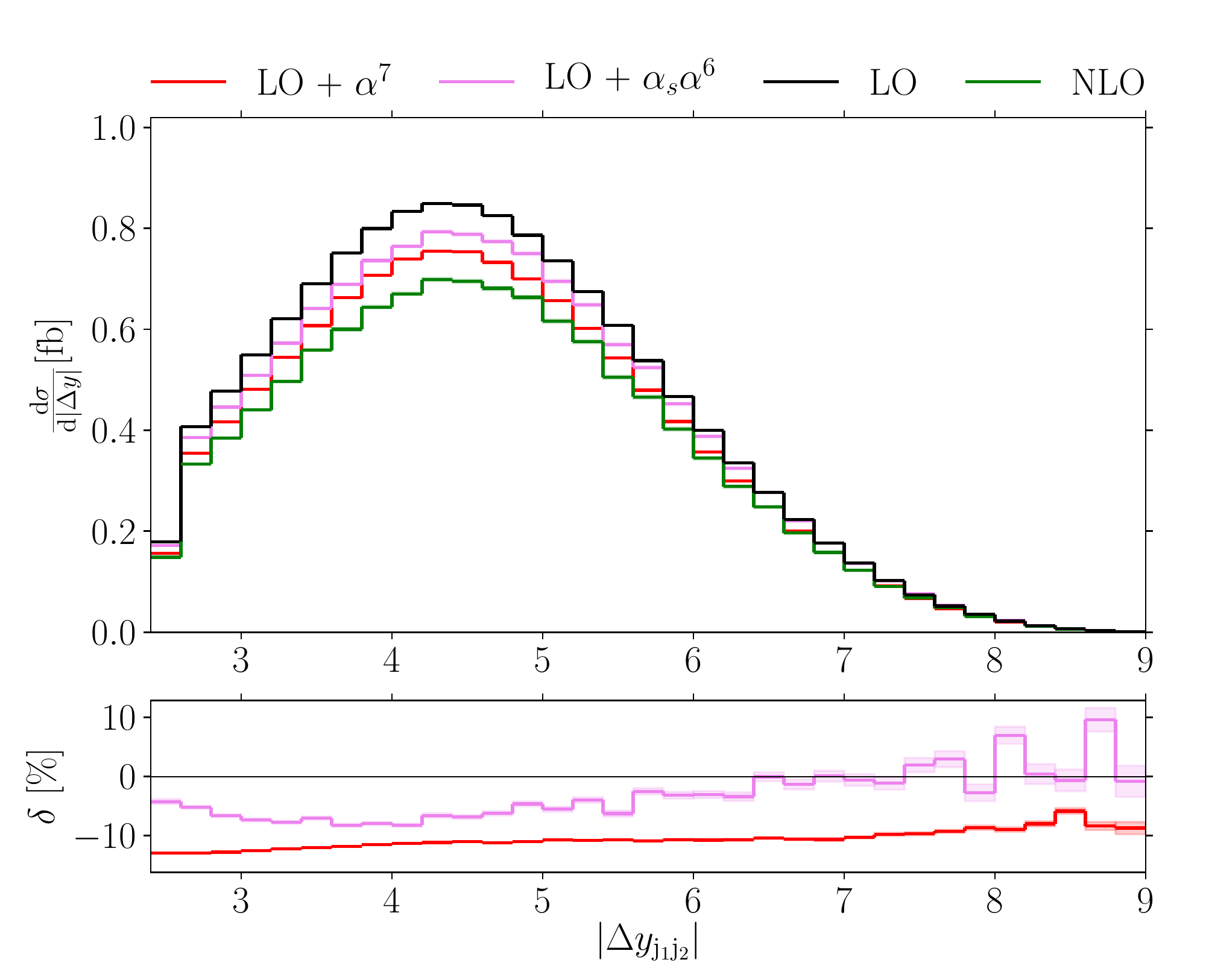}
\end{subfigure}%
\par
\begin{subfigure}{0.49\textwidth}
\centering
\includegraphics[width=1.\linewidth]{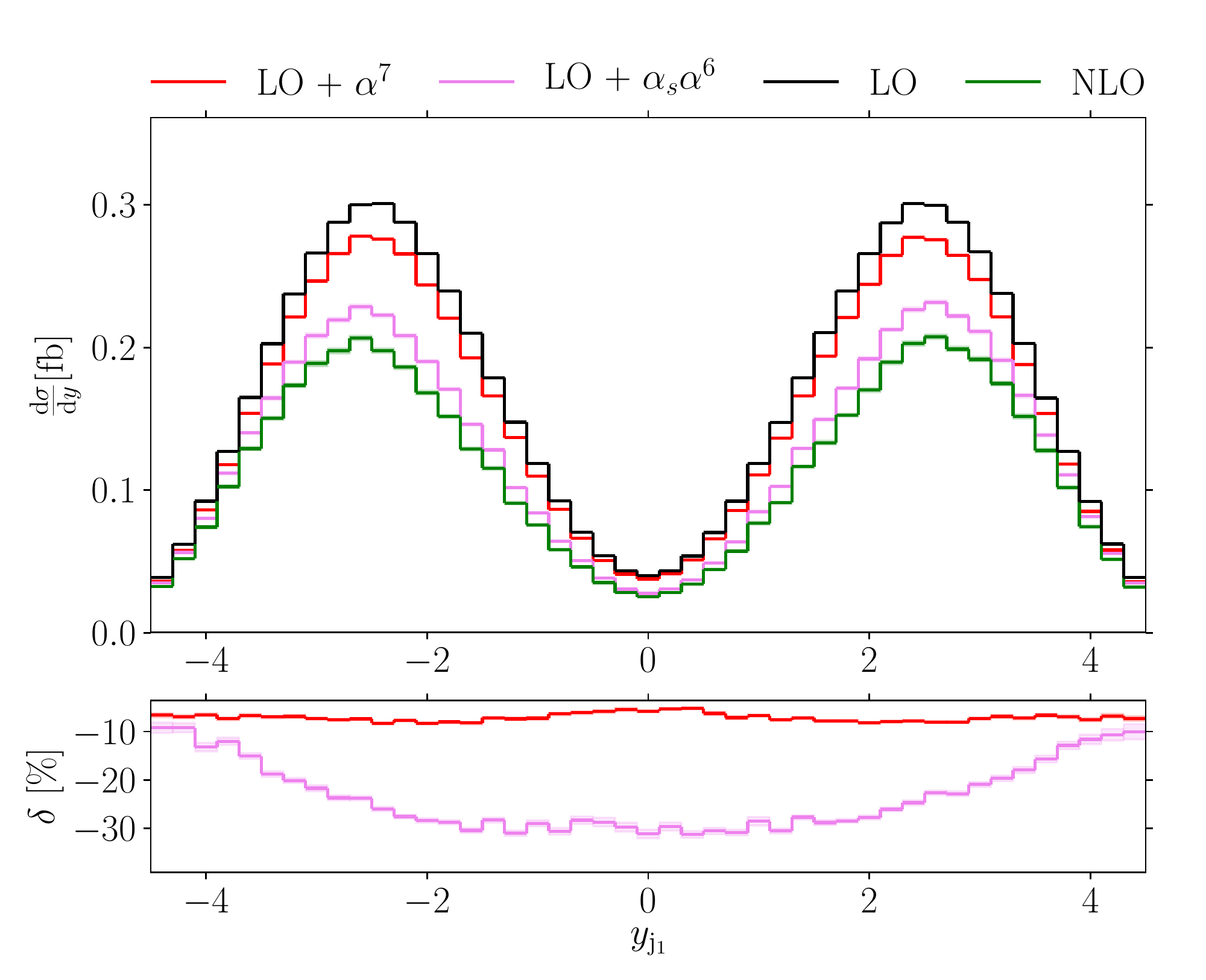}
\end{subfigure}
\begin{subfigure}{0.49\textwidth}
\centering
\includegraphics[width=1.\linewidth]{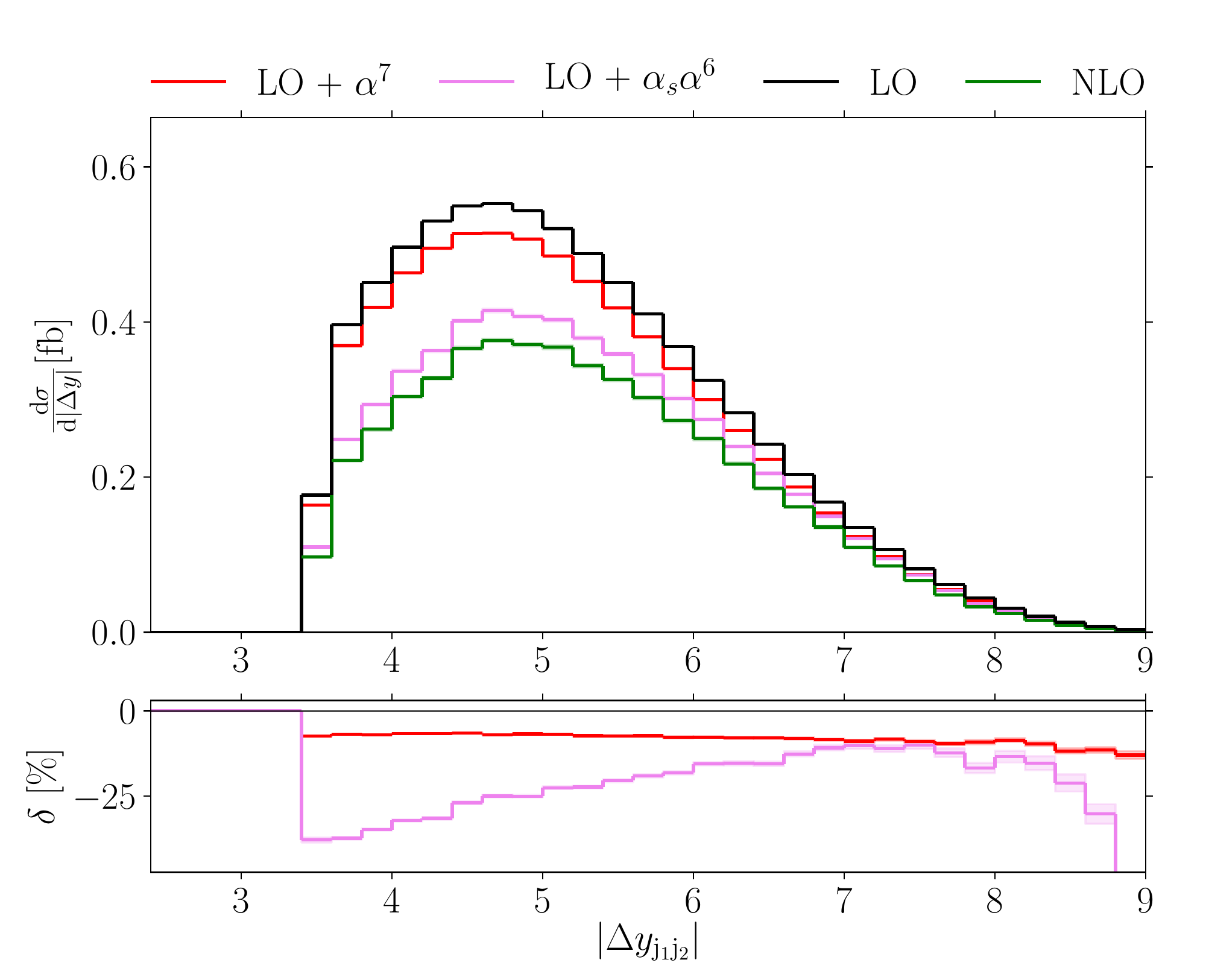}
\end{subfigure}
\caption{Differential distributions at NLO in the rapidity of the
  hardest jet (left) and the rapidity difference between the two
  tagging jets (right) in the VBS (top) and the Higgs setup
  (bottom). The upper panels show the absolute EW contribution at
  $\order{\alpha^6}$, the sum of the EW contribution and the EW
  corrections of $\order{\alpha^7}$, the sum of the EW contribution
  and the QCD corrections of $\order{\alphas\alpha^6}$, and the sum of
  all three contributions. The lower panels show the relative EW and QCD
  corrections normalised to the LO EW contribution. Shaded bands denote integration errors.}
\label{fig:non-E-dependent_NLO}
\end{figure}

Next we turn to distributions in energy-independent variables.
Specifically, we discuss the distributions in the rapidity of the
hardest jet and the rapidity difference between the two tagging jets
in \reffi{fig:non-E-dependent_NLO}. The EW corrections to the rapidity
distribution of the leading jet (\reffi{fig:non-E-dependent_NLO} left)
exhibit a flat (negative) maximum at moderate rapidities around
$|y_{\Pj_1}| \approx 1.5$ and $|y_{\Pj_1}| \approx 2$ in the VBS and
Higgs setup, respectively. The QCD corrections show qualitatively the
same variation in both setups: They peak at $y_{\Pj_1} \approx 0$ and
become smaller at high rapidities in the Higgs setup or even change
 sign in the VBS setup.

The distribution in the rapidity difference of the tagging jets
(\reffi{fig:non-E-dependent_NLO} right), a typical VBS observable,
shows a qualitative difference in the two setups for both EW and QCD
corrections. While the EW corrections are nearly flat in the VBS setup
and decrease slightly with large rapidity difference, they increase
for the Higgs setup. The QCD corrections also decrease in the VBS
setup and and become compatible with zero for high rapidity differences, while they
have a minimum at $|\Delta y_{\Pj_1\Pj_2}| \approx 7.5$ in the Higgs
setup. This behaviour points to a correlation of large rapidity
differences with high energies.

We note that the EW and QCD corrections to the distributions in the
rapidities of the leptons are flat, \ie the corresponding corrections
are roughly given by those to the fiducial cross section. This holds
also for the EW corrections to the distribution in the Zeppenfeld
variable.

\subsubsection{Scale dependence}
\begin{figure}
\begin{subfigure}{0.49\textwidth}
\centering
\includegraphics[width=1.\linewidth]{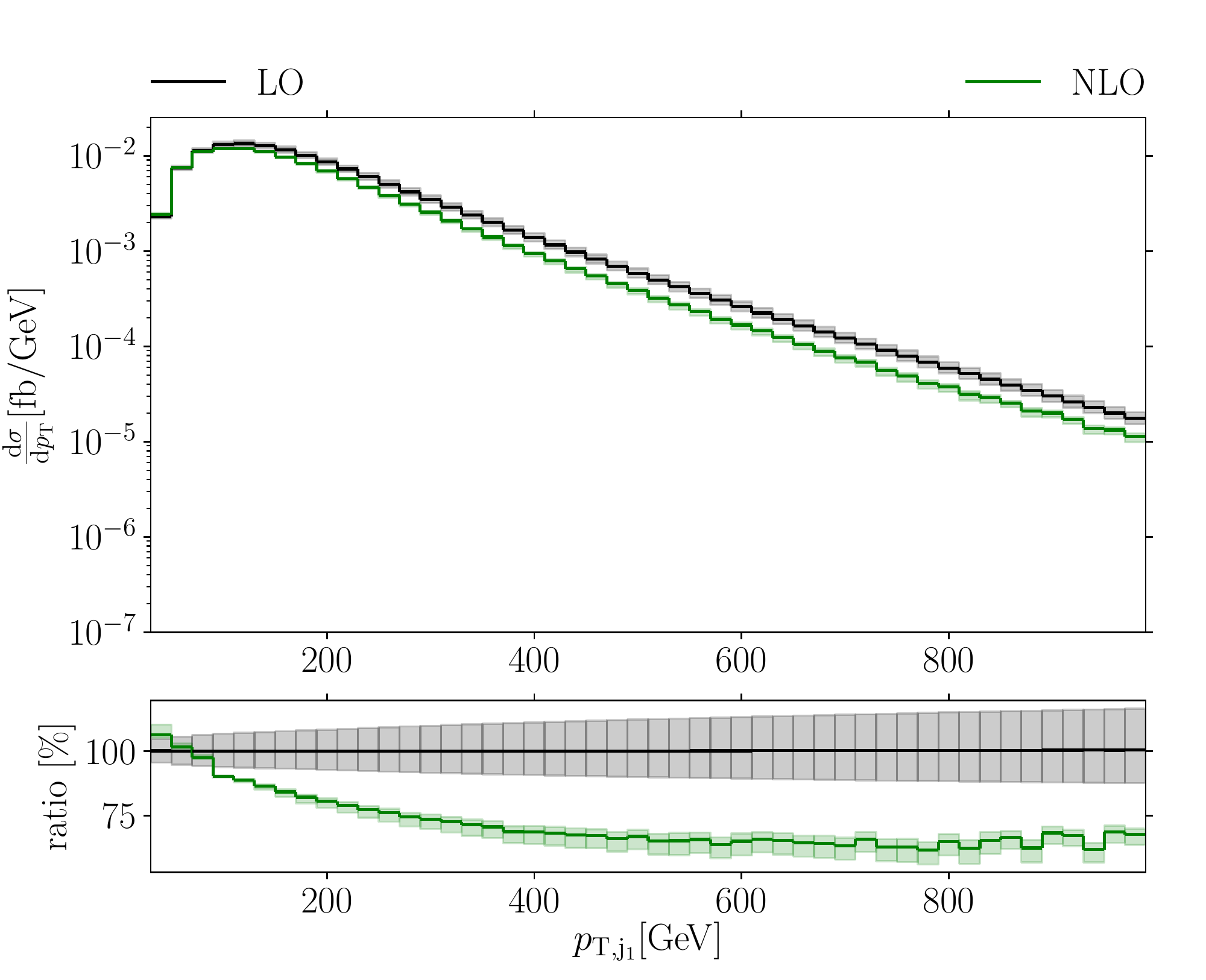}
\end{subfigure}
\begin{subfigure}{0.49\textwidth}
\centering
\includegraphics[width=1.\linewidth]{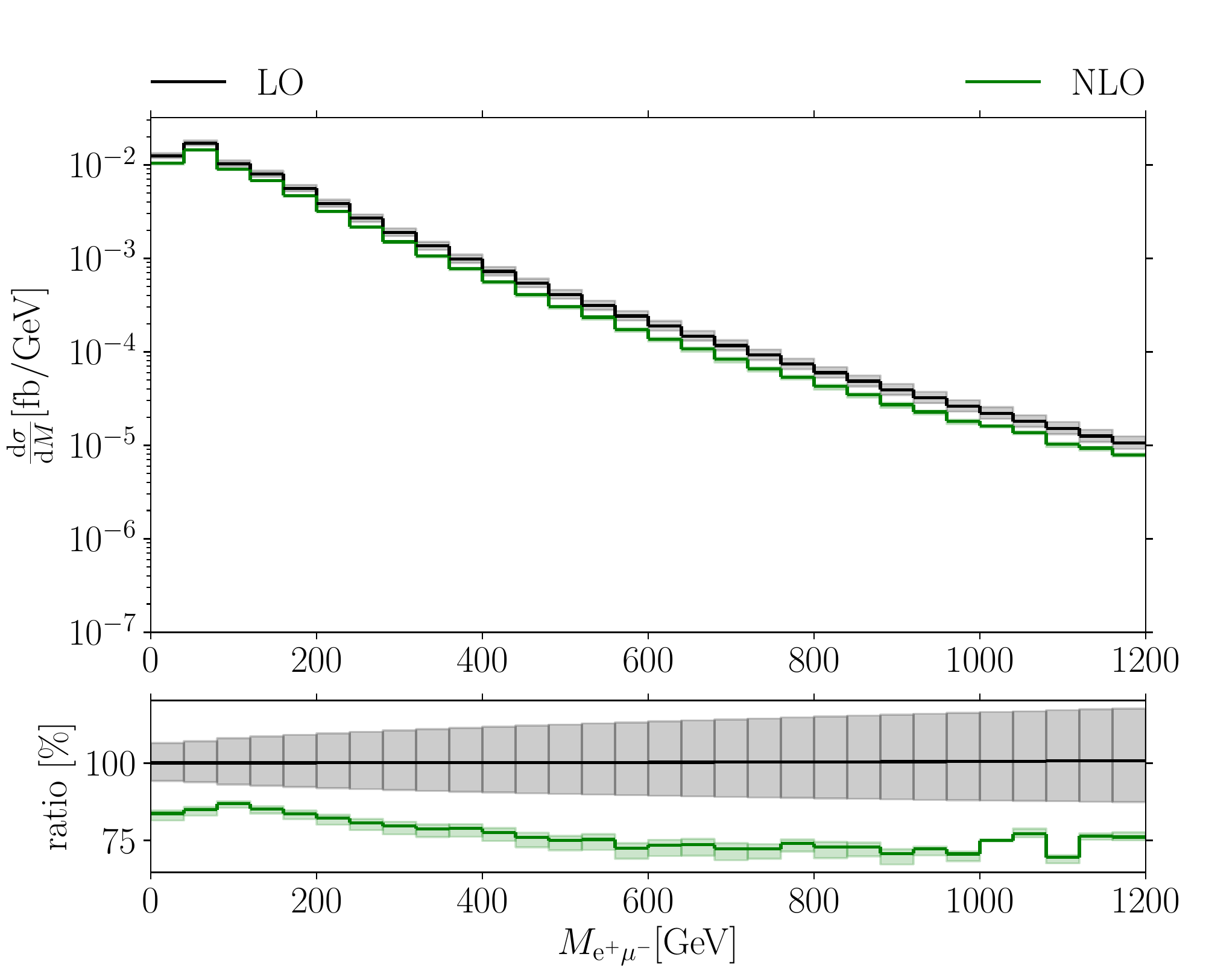}
\end{subfigure}%
\par
\begin{subfigure}{0.49\textwidth}
\centering
\includegraphics[width=1.\linewidth]{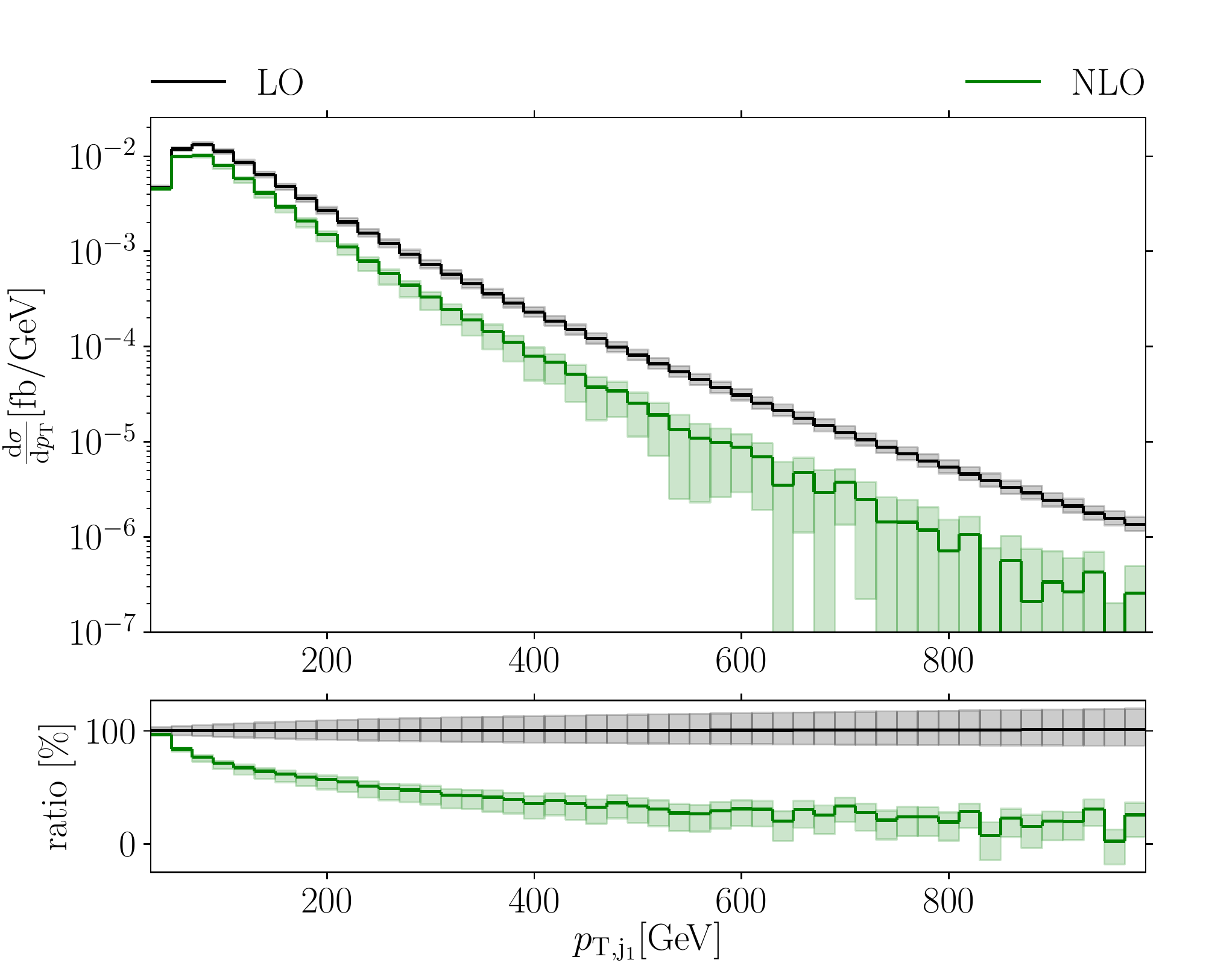}
\end{subfigure}
\begin{subfigure}{0.49\textwidth}
\centering
\includegraphics[width=1.\linewidth]{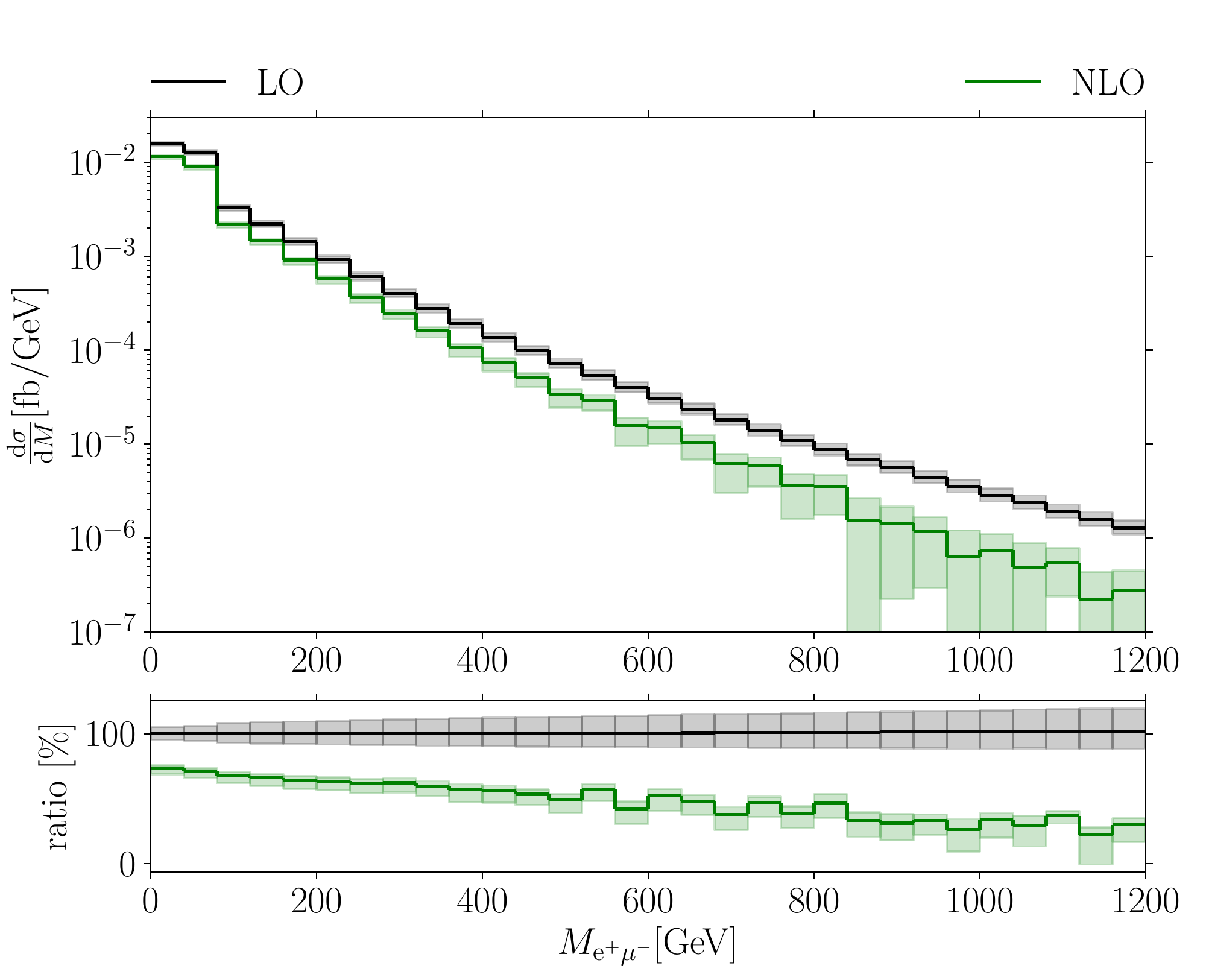}
\end{subfigure}
\caption{Differential distributions at LO and NLO in the transverse
  momentum of the leading jet (left) and the invariant mass of the
  charged-lepton pair (right) in the VBS setup (top) and the Higgs
  setup (bottom). The upper panels show the absolute EW contribution
  at $\order{\alpha^6}$ and the total NLO contribution
  $\order{\alpha^6} + \order{\alpha^7} + \order{\alphas\alpha^6}$, the
  lower panels the relative corrections. Shaded bands indicate the
  7-point scale uncertainty.}
\label{fig:scale_I}
\end{figure}

\begin{figure}
\begin{subfigure}{0.49\textwidth}
\centering
\includegraphics[width=1.\linewidth]{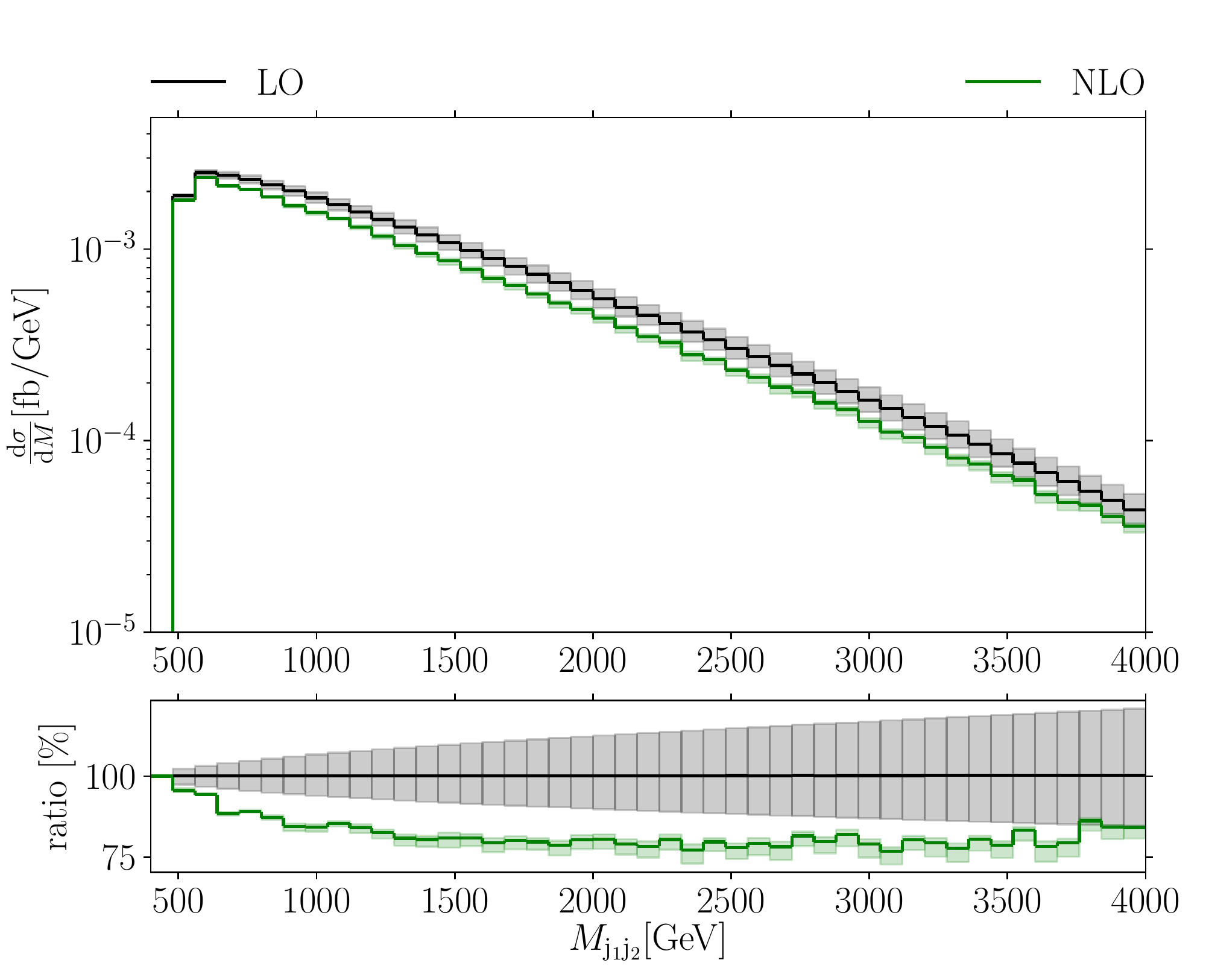}
\end{subfigure}
\begin{subfigure}{0.49\textwidth}
\centering
\includegraphics[width=1.\linewidth]{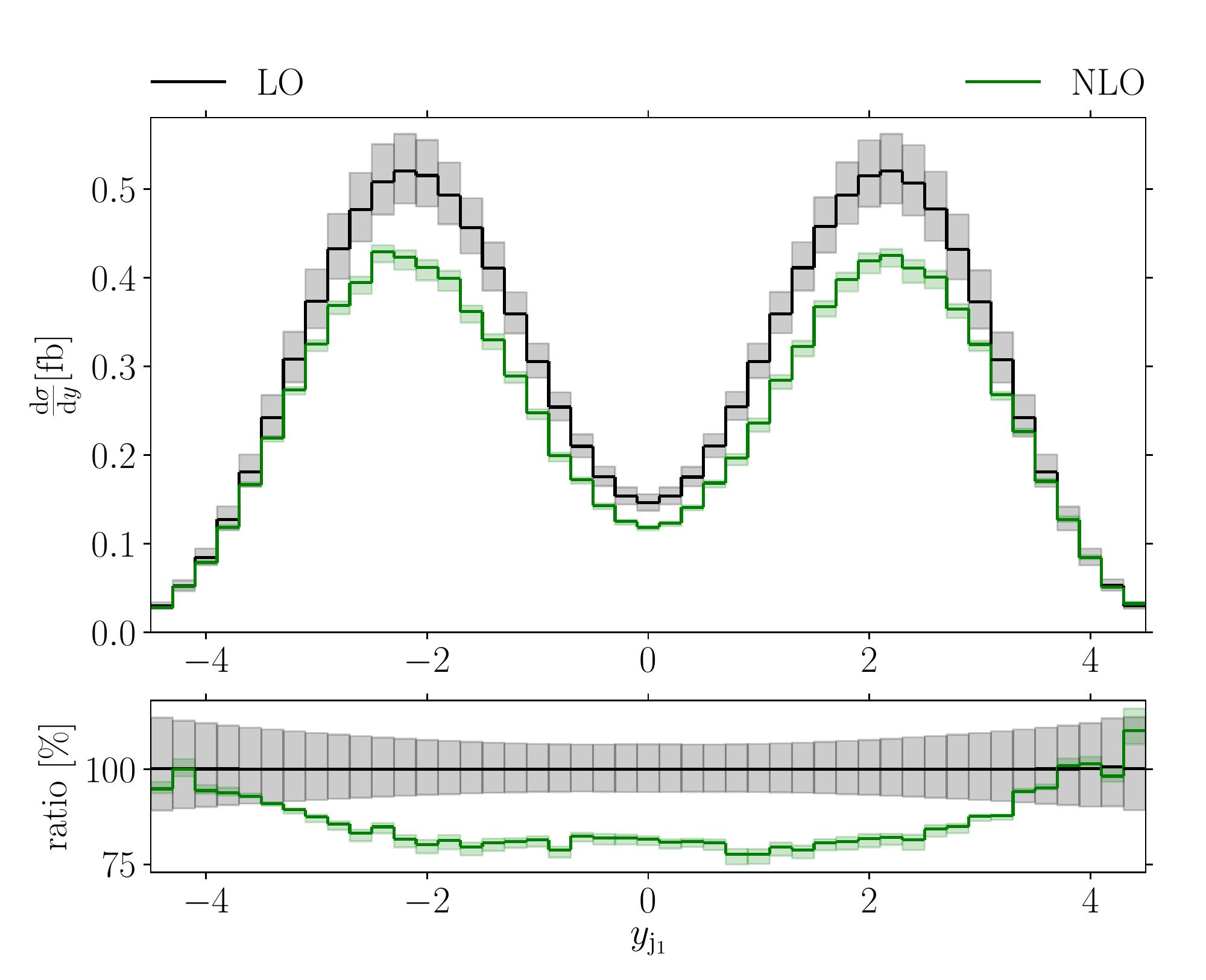}
\end{subfigure}%
\par
\begin{subfigure}{0.49\textwidth}
\centering
\includegraphics[width=1.\linewidth]{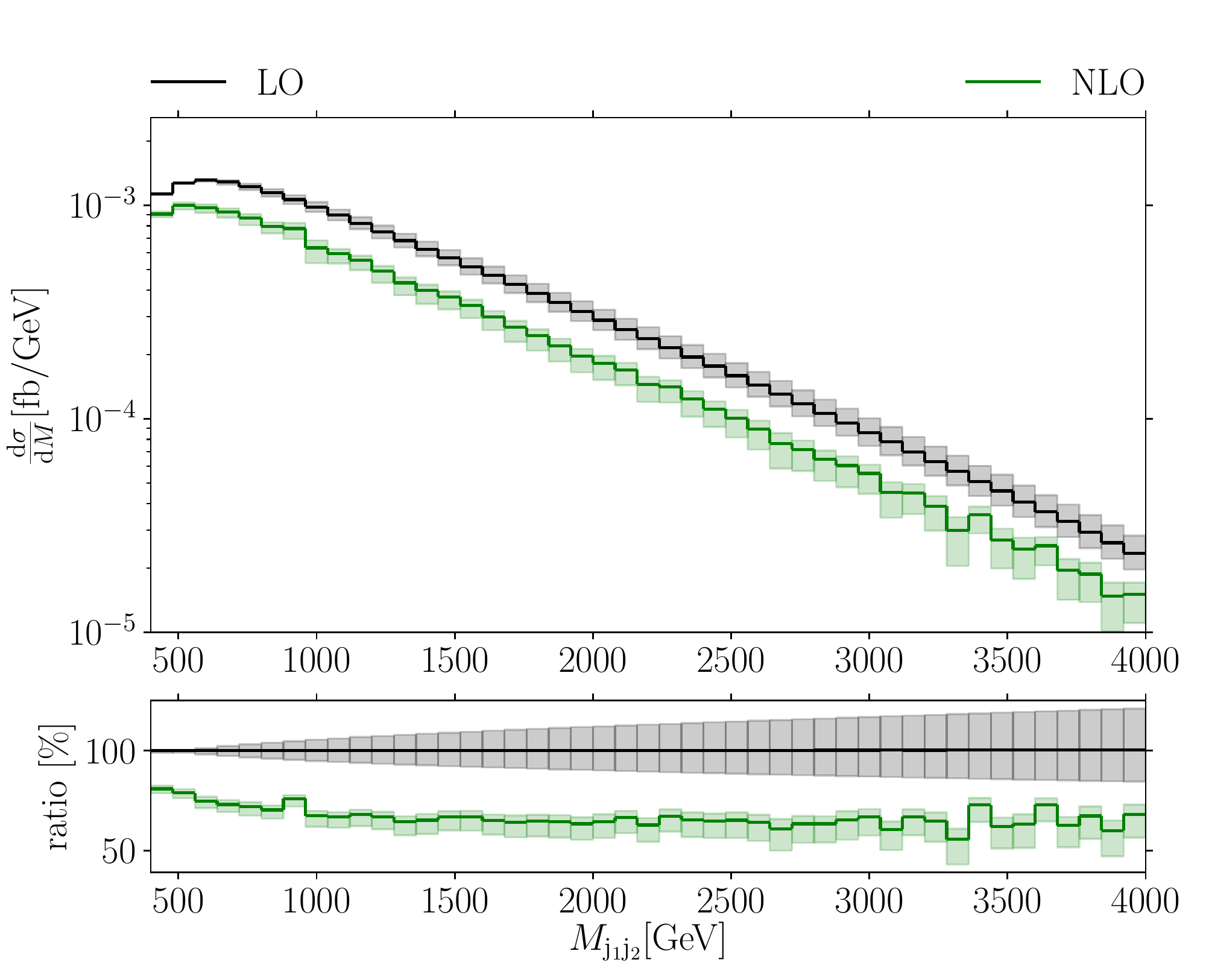}
\end{subfigure}
\begin{subfigure}{0.49\textwidth}
\centering
\includegraphics[width=1.\linewidth]{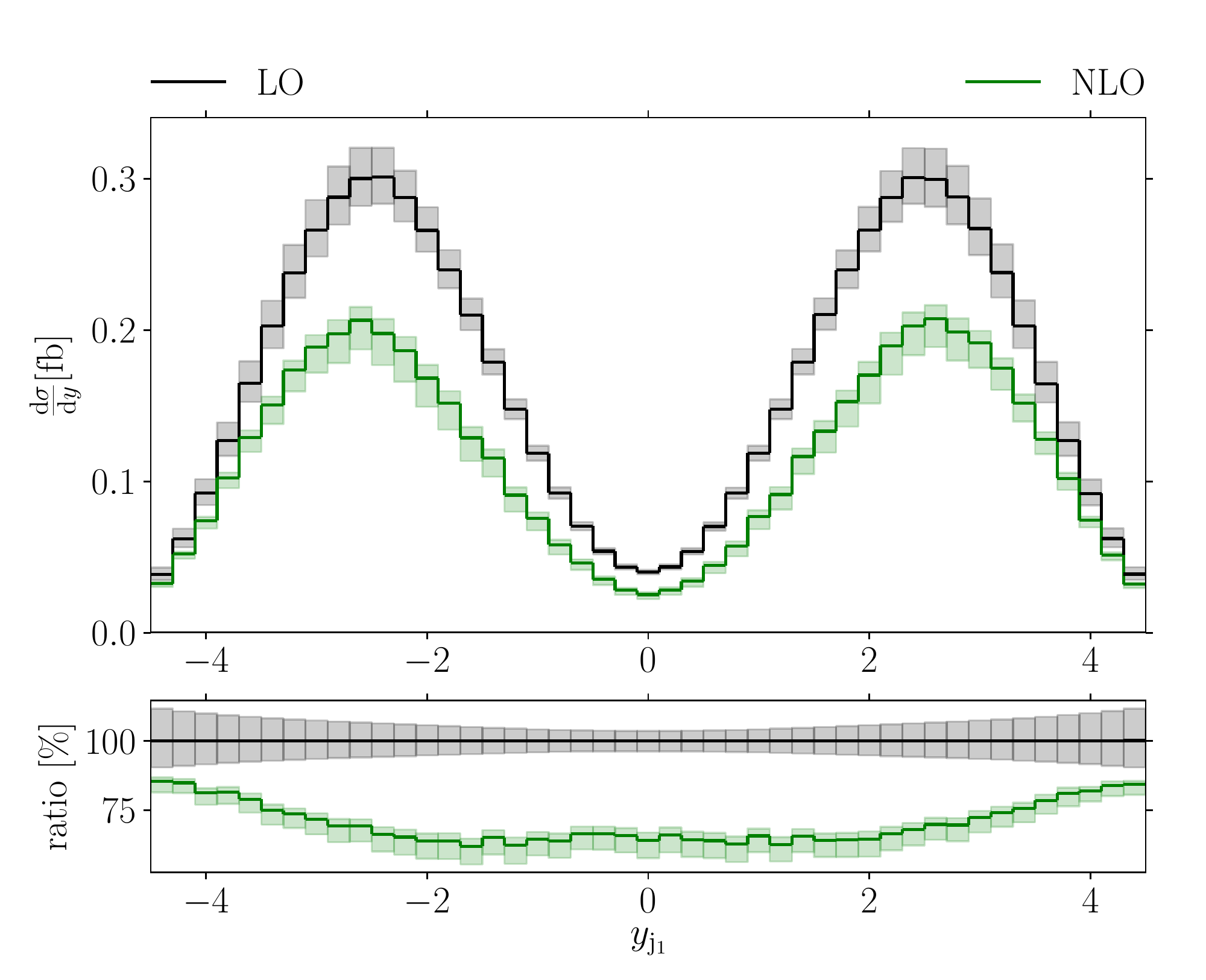}
\end{subfigure}
\caption{Differential distributions at LO and NLO in the invariant
  mass of the two tagging jets (left) and the invariant mass of the
  charged-lepton pair (right) in the VBS setup (top) and the Higgs
  setup (bottom). The upper panels show the absolute EW contribution
  at $\order{\alpha^6}$ and the total NLO contribution
  $\order{\alpha^6} + \order{\alpha^7} + \order{\alphas\alpha^6}$, the
  lower panels the relative corrections. Shaded bands indicate the
  7-point scale uncertainty.}
\label{fig:scale_II}
\end{figure}

Finally, we investigate the 7-point scale dependence of some
distributions. In \reffi{fig:scale_I}, we present those in the
transverse momentum of the hardest jet (left) and in the invariant
mass of the charged-lepton pair (right) in the VBS setup (top) and the
Higgs setup (bottom), and in \reffi{fig:scale_II} those in the
invariant mass of the two hardest jets (left) and the rapidity of the
hardest jet (right).  We show the absolute LO, \ie $\order{\alpha^6}$,
and NLO, \ie $\order{\alpha^6} + \order{\alpha^7} +
\order{\alphas\alpha^6}$, predictions including their scale dependence
in the upper panels. The lower panels display the scale dependence,
defined as the ratio of the envelope (shown as the band) of
scale-varied LO and NLO cross sections and the corresponding LO cross
section at the central scale.
We stress that the NLO corrections are outside the LO scale-uncertainty band
in essentially all cases.  

In the VBS setup, the NLO scale dependence  does not
show peculiarities and stays at a modest level for all values of
transverse momenta or invariant masses. 
In the Higgs setup, we observe a qualitatively different behaviour in the
scale dependence of variables that are strongly affected by the jet veto,
\eg $p_{\rT,\Pj_1}$ and $M_{\Pe^+\mu^-}$,
and those that are not, \eg $M_{\Pj_1\Pj_2}$ and $y_{\Pj_1}$.
The scale variation in the Higgs setup grows significantly with
transverse momenta  or invariant masses correlated to the third jet
transverse momentum. This is related to the fact
that the relative NLO QCD corrections become rather large and
even exceed $-100\%$ in the tails, leading to unphysical results.
This behaviour can be attributed to the jet veto and indicates that
some resummation is required to obtain decent predictions 
in this part of phase space.
The scale dependence of the distributions in variables that are not
correlated to the transverse momentum of the third jet, as the jet-pair invariant mass
or rapidities, is larger in the Higgs setup than in the VBS setup.
However, for those variables, varying the scale does not lead to
corrections larger than 50\%.

This brief analysis of the scale dependence demonstrates that the use
of a jet veto requires additional efforts to provide stable NLO QCD
predictions already at $\order{\alphas\alpha^6}$, \ie for the QCD
corrections to a purely EW process. This will become an even more
pressing issue if the QCD corrections to the QCD background at
$\order{\alphas^3\alpha^4}$ are included.

\section{Conclusion}
\label{sec:conclusion}
In this article, we have presented results for the NLO EW and QCD
corrections of the orders $\order{\alpha^7}$ and
$\order{\alphas\alpha^6}$ for the process $\Pp \Pp \to \Pe^+ \nu_\Pe
\mu^- \bar \nu_\mu \Pj\Pj + X$ excluding external bottom quarks in two
different setups. We have taken the full matrix elements into account
without relying on any approximations. At orders $\order{\alpha^6}$, 
$\order{\alpha^7}$, and $\order{\alphas\alpha^6}$ within the
considered setups, the process is dominated by vector-boson scattering
(VBS) into a pair of $\PW$ bosons, which contains vector-boson fusion
into a Higgs boson as a subprocess, but receives also contributions of
triple-vector-boson production. We have included all partonic channels
in the considered orders apart from those involving bottom quarks,
which are dominated by contributions from top-pair production.

VBS into oppositely-charged W~bosons offers the largest cross section
for VBS at the LHC in the range of several femto-barns. Within VBS
cuts, about 70\% of the cross section is due to the QCD-induced
background, while in a dedicated Higgs setup the fraction is reduced
to 50\%. The contribution of interferences is below one percent and
the one of the loop-induced gluon channels below few percent.

Compared to other VBS processes, the EW corrections to the cross section are
relatively small owing to the presence of the Higgs resonance in the
fiducial phase space, which leads to a sizeable contribution to the EW
cross section around a four-lepton invariant mass of $M_{4\Pl} \approx
\MH$. 
In the tails of the distributions, the EW NLO corrections reach values
of $-40\%$ to $-50\%$ in accordance with findings for other  VBS processes.

The overall QCD corrections in the VBS setup
are, as for other VBS processes, relatively small, whereas they are
large and negative in the Higgs setup owing to the veto on a third jet.
An investigation of the scale dependence of the cross section shows
that jet vetoes should be used with care in VBS processes.  The EW and
QCD corrections depend on the fiducial volume and on the distribution.
While both QCD and EW corrections exhibit a considerable energy
dependence, also for energy-independent distributions the QCD and EW
corrections show variations of $20\%$ and $5\%$,
respectively.

A further investigation in a setup in which we cut out the
Higgs resonance validated our assumptions on its impact on the
cross section. We found that the Higgs resonance contributes a large
fraction of the cross section for VBS processes. After removing it
from the fiducial phase space, NLO EW corrections are comparable to
those of other VBS processes at the level of $-15\%$.

After a detailed investigation of the corrections to partonic channels
that have been characterised by their subprocesses, we found a
correlation between the size of the corrections and the type of
processes: The relative EW corrections to non-VBS processes turned out
to be large, since they are not affected by the Higgs resonance, but
these contributions are negligible due to their small absolute size.
As already known from other VBS processes, the relative QCD corrections
for non-VBS processes are also large, since the additional emission of
a gluon from an initial-state quark changes the action of the VBS cuts
on the process, when not all tagging jets have to be produced via
a decay of a vector boson in the $s$ channel. The full calculation of
all photon-induced corrections leads to a contribution at the percent
level at NLO, because new partonic channels with VBS content open up.

This article concludes the calculation of EW corrections to massive
VBS processes. In the future, it will be relevant to calculate the
full NLO corrections to the corresponding irreducible backgrounds and the
EW corrections to polarised VBS.

\section*{Acknowledgements}
We are grateful to Jean-Nicolas Lang and Sandro Uccirati for
continuously supporting and improving \recola. Many thanks go to
Mathieu Pellen for his support of \mocanlo. We
acknowledge financial support by the German Federal Ministry for
Education and Research (BMBF) under contract no. 05H18WWCA1 and the
German Research Foundation (DFG) under reference numbers DE 623/6-1
and DE 623/6-2.

\bibliographystyle{JHEPmod}
\bibliography{vbs_ww}{}

\end{document}